\documentclass[reprint,superscriptaddress,amsmath,amssymb,aps,prb,floatfix]{revtex4-2}
\usepackage{comment}
\usepackage{bm}
\usepackage{hyperref}
\usepackage{graphics}
\usepackage{graphicx}
\usepackage{braket}
\usepackage{MnSymbol}
\usepackage{amssymb}
\usepackage{amsmath,empheq}
\usepackage{mathrsfs}
\usepackage{textgreek}
\usepackage{xcolor}
\usepackage{enumitem}
\usepackage{dcolumn}
\usepackage[normalem]{ulem}
\usepackage{amsfonts}
\usepackage{mathrsfs}
\usepackage{nicefrac}
\usepackage{simpler-wick}
\usepackage{dsfont}
\hypersetup{colorlinks,linkcolor=blue,urlcolor=blue,citecolor=blue}

\usepackage[export]{adjustbox} 

\newcommand{\hc}{\text{H.c.}}
\newcommand{\sign}{\text{sign}}
\newcommand{\hphi}{\hat{\phi}}
\newcommand{\hvarphi}{\hat{\varphi}}
\newcommand{\htheta}{\hat{\theta}}
\newcommand{\expp}[1]{\text{e}^{#1}}
\newcommand{\phd}{{\phantom{\dagger}}}

\begin{document}
 
\title{Topological Kondo effect with spinful Majorana fermions}
\author{Steffen Bollmann}
\affiliation{Max-Planck-Institut f\"{u}r Festk\"{o}rperforschung, Heisenbergstra{\ss}e 1, 70569 Stuttgart, Germany}

\author{Jukka I. V\"ayrynen}
\affiliation{
Department of Physics and Astronomy, Purdue University, West Lafayette, Indiana 47907, USA.
}

\author{Elio J. K\"{o}nig}
\affiliation{Max-Planck-Institut f\"{u}r Festk\"{o}rperforschung, Heisenbergstra{\ss}e 1, 70569 Stuttgart, Germany}

\begin{abstract}
 Motivated by the importance of studying topological superconductors beyond the mean-field approximation, we here investigate mesoscopic islands of time reversal invariant topological superconductors (TRITOPS). We characterize the spectrum in the presence of strong order parameter fluctuations in the presence of an arbitrary number of Kramers pairs of Majorana edge states and study the effect of coupling the Coulomb blockaded island to external leads. In the case of an odd 
fermionic parity on the island, we derive an unconventional Kondo Hamiltonian in which metallic leads couple to both topological Majorana degrees of freedom (which keep track of the parity in different leads) and the overall spin-1/2 in the island. For the simplest case of a single wire (two pairs of Majorana edge states), we demonstrate that anisotropies are irrelevant in the weak coupling renormalization group flow. This permits us to solve the Kondo problem in the vicinity of a Toulouse-like point using Abelian Bosonization. We demonstrate a residual ground state entropy of $\ln(2)$, which is protected by spin-rotation symmetry, but reduced to $\ln(\sqrt{2})$ (as in the spinless topological Kondo effect) by symmetry breaking perturbations. In the symmetric case, we further demonstrate the simultaneous presence of both Fermi liquid and non-Fermi liquid like thermodynamics (depending on the observable) and derive charge and spin transport signatures of the Coulomb blockaded island.
\end{abstract}

\maketitle

\section{Introduction}
Band structure topology has become a pillar of modern condensed matter physics, with implications for both quantum technologies and quantum materials~\cite{ChiuRyu2016}. In particular, unconventional spin triplet superfluids and superconductors, as realized in $^3$He~\cite{Volovik2003,VollhardtWoelfle2013}, Uranium based heavy fermion superconductors~\cite{JoyntTaillefer2002,JiaoMadhavan2020,AokiYanase2022} and possibly in twisted van der Waals multilayers~\cite{LakeSenthil2022}, are important candidates to host topological fermionic boundary and low energy states. While the emergent single particle band structure of these phases is by now theoretically well understood, the additional complexity of strong electronic correlations leads to much richer physics~\cite{FidkowskiKitaev2011, Rachel2018} and is the object of ongoing research. This particularly concerns the interplay of fermionic boundary states (Majorana fermions) with quantum fluctuations of the order parameter.

\begin{figure}
    \centering
    \includegraphics{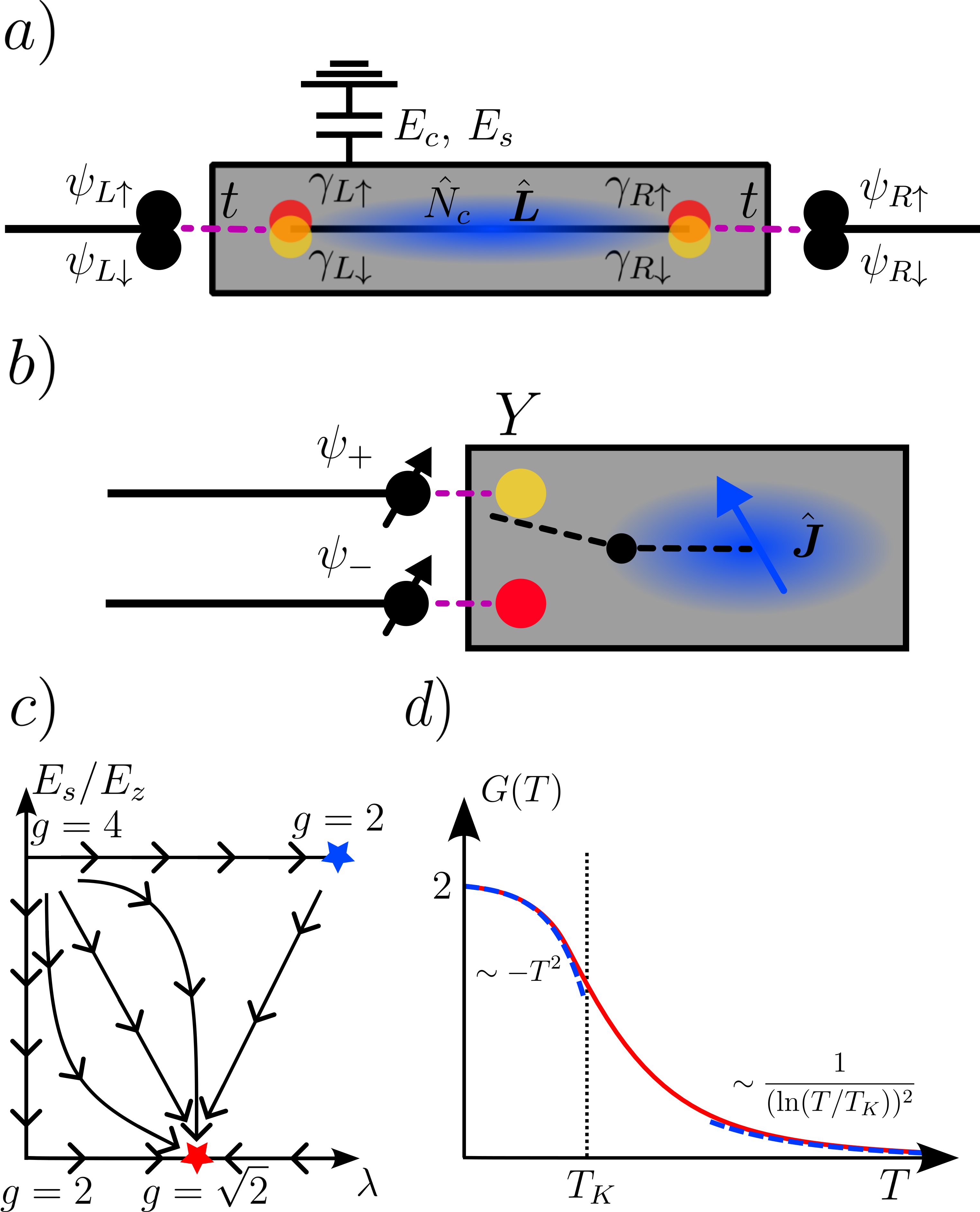}
    \caption{ a) Schematic of physical setup. The grey box represents the Majorana Cooper pair box, which harbors a one-dimensional time-reversal invariant SC (black line) that hosts four Majorana zero modes (yellow and red dots). The device is coupled to
    leads (blue lines). b) Schematic representation of the effective Kondo Hamiltonian \eqref{eq:KondoHamEzSmall}. The even $\psi_+$ and odd $\psi_-$ superpositions of the lead electrons are coupled to an effective quantum impurity with four internal states 
    (square). The angular momentum $\hat{\bm{J}}$ of the superconducting condensate (blue ellipse) acts as an effective spin 1/2. The spin of the lead electrons (black arrows) is coupled to the impurity via the orbitals formed by the MZMs (red and yellow ellipses). The Kondo coupling to $\hat{\bm{J}}$ is effective only when the orbital next to the lead electrons is occupied. c) Schematic Renormalization Group flow diagram based in the plane of effective Kondo coupling $\lambda$ and relative strength of Cooper pair spin fluctuations $E_s$ and polarizing field $E_z$. The parameter $g = e^{\mathcal S_{\rm imp}}$ is defined through the zero temperature impurity entropy $\mathcal S_{\rm imp}$. 
     d) Absolute value of the charge conductance between the left and right wire in the limit $E_s/E_z \gg 1$.}
    \label{fig:Summary}
\end{figure}

As a prime example, strong quantum fluctuations of the superconducting phase in mesoscopic effectively \textit{spinless} Majorana Cooper-pair boxes (MCPB) can be induced by a large charging energy~\cite{Fu2010}. These devices are floating islands of Josephson-connected spinless (i.e., spin-polarized) p-wave superconducting chains~\cite{Kitaev2001,Alicea2012} and allow to implement a paradigmatic topological qubit, the \textit{tetron}~\cite{OregvonOppen2020}. While arrays of such tetrons emulate exotic fractionalized many-body phases with topological order\cite{TerhalDiVincenzo2012,SagiOreg2019,ZiesenRoy2019,KoenigTsvelik2020b,OregvonOppen2020}, a single spinless MCPB coupled to metallic leads realizes a topological quantum impurity problem, the \textit{topological Kondo effect}~\cite{BeriCooper2012,AltlandTsvelik2014,AltlandTsvelik2014b,Buccheri2015,GauEgger2018,PapajFu2019,LiVayrynen2023b,WautersBurrello2023}, hosting non-Fermi liquid like low-energy characteristics and an irrational residual entropy reminiscent of a non-trivial anyonic quantum dimension.

Exotic Kondo impurity problems of SU($N$) symmetry, most notably for $N = 2$, have been studied extensively for the past forty years both theoretically~\cite{Andrei1980,Vigman1980,AndreiDestri1984,TsvelickWiegmann1985,AffleckLudwig1991a,AffleckLudwig1991b,AffleckLudwig1993,JerezZarand1998,KellerGoldhaberGordon2014,KellerGoldhaberGordon2015,PouseGoldhaberGordon2021}, most recently in the prospect of quantum information~\cite{LopesSela2020,Komijani2020,GabaySela2022,LotemGoldstein2022,LotemGoldstein2022b}, and experimentally, in particular in mesoscopic quantum electronics devices~\cite{PotokGoldhaberGordon2007,IftikharPierre2015,IftikharPierre2018}. For this symmetry group, non-Fermi liquid behavior occurs only in the overscreened multichannel case~\cite{NozieresBlandin1980}. The Majorana based topological Kondo effect is special, inasmuch it realizes an impurity spin transforming under O($M$) (where $M$ denotes the number of Majorana zero modes coupled to external leads). Very recently, Majorana-free setups of orthogonal~\cite{MitchellAffleck2021,LibermanSela2021} and symplectic symmetry~\cite{LiVayrynen2023,KoenigTsvelik2023,RenTsvelik2023} were proposed, mathematically completing the set of possible classical Lie groups. Interestingly, in the single channel case, both orthogonal and symplectic Kondo models display exotic physics, including fingerprints of bound anyons, already~\cite{Kimura2021}. 

Within the superconducting Altland-Zirnbauer classes, two non-trivial topological phases exist in one dimension. Apart from spinless p-wave superconductors (e.g., the ``Kitaev chain'' \cite{Kitaev2001}, class D), there is a spinful p-wave time reversal invariant topological superconductor (TRITOPS~\cite{HaimOreg2019}, class DIII). In this work, we study TRITOPS islands in the Coulomb blockade regime and, when the fermionic parity on the island is odd, uncover an unconventional topological Kondo effect of spinful Majorana fermions. Using Abelian Bosonization, we solve this problem in the simplest and most relevant case of two pairs of Majorana edge states coupled to two spinful leads, Fig.~\ref{fig:Summary}. We characterize the phase diagram and demonstrate that the topological Kondo effect of spinful Majorana fermions is protected by
spin-rotation 
symmetry, but flows to the fixed point of the spinless topological Kondo effect in the presence of symmetry breaking perturbations. Both fixed points display exotic hallmarks of non-Fermi liquids. We make numerically and (potentially) experimentally verifiable predictions for thermodynamic and transport signatures of the non-trivial low-energy fixed point.

Single-particle physics in one-dimensional TRITOPS gained substantial attention over the years~\cite{WangLaw2012,NakoseiNagaosa2013,KeselmanBerg2013,ZhangMele2013}, in particular regarding its unconventional transport through Josephson junctions~\cite{SchradeLoss2015,KimLutchyn2016,CamjayivonOppen2017,SchradeFu2018,AligiaArrachea2018,ArracheaGruneiro2019,KnappAlicea2020}. Strong order parameter fluctuations of the superconducting phase, i.e. spinful MCPBs in the Coulomb blockade regime, also gained some attention~\cite{NakoseiNagaosa2013,SchradeFu2018}, both in their context as topological qubits\cite{SchradeFu2022}, topological Josephson junction arrays~\cite{FatemehKargarian2022}, and mesoscopic Kondo impurities~\cite{BaoZhang2017}. At the same time, to the best of our knowledge, the impact of strong order parameter fluctuations of the Cooper pair orientation $\hat{\bm d}$ in the spin sector was considered only in Ref.~\cite{RamppSchmalian2022}, where it was uncovered that the bandstructure topology induces a theta term\cite{Else2021} in the effective non-linear sigma model of the $\hat{\bm d}$ vector, thereby stabilizing 2e-paired-superconductivity. Here, similar fluctuations will be studied in zero-dimensional mesoscopic islands.

The paper is organized as follows: In section \ref{sec:SpinfulMCPB} we introduce the spinful Majorana Cooper-pair box and present the solution, finding the eigenfunctions and spectrum. Building upon that result, we discuss in section \ref{sec:spinfulTopologicalKondoEffect} how a novel topological Kondo effect can arise from coupling a Coulomb blockaded MCPB to spinful normal metal leads. This is done in the cases of dominant and sub-dominant spin interactions within the MCPB. In section \ref{sec:singleWireAtTheToulosePoint}, we discuss the topological Kondo effect in more depth by means of Bosonization and poor man's scaling. This allows for constructing a schematic RG-flow diagram in the parameter space of Kondo coupling and spin interaction strength, see Fig.~\ref{fig:Summary}c. This is followed by presenting several observables, such as transport coefficients and thermodynamic properties, in section \ref{sec:Observables}. We end the paper with a conclusion and give an outlook to further research concerning MCPBs. 

\section{Spinful MCPB}
\label{sec:SpinfulMCPB}

\subsection{Setup}

We consider a floating mesoscopic quantum device consisting of a one-dimensional time-reversal invariant topological triplet superconductor in the Cartan-Altland-Zirnbauer class DIII. 
We focus on the Coulomb blockade regime where the charging energy $E_c$ is large enough that (thermal) fluctuations of the total charge of the island can be neglected. Also, $E_c$ will be the largest energy scale we assume in the system. 
In analogy to the charging energy, we also consider an interaction ($E_s$) that punishes the formation of a 
large total angular momentum within the island. The full Hamiltonian reads
\begin{subequations}
\begin{align}
    \begin{split}
    \hat{H} & = E_c(2\hat{N}_c + \hat{n}_f - 2N^c_g)^2 + E_s(\hat{\bm{L}} +
    \hat{\bm{S}})^2 \\ & + \hat{H}_{\text{BdG}} + \hat{H}_z
    \end{split}
    \label{eq:model}
\end{align}
where 
\begin{equation}
    \hat H_z = -E_z \hat{d}_y
\end{equation}
and
\begin{equation}
    \hat{H}_{\text{BdG}} = \int \text{d}x\left\{ \psi(x)^\dagger\epsilon(\hat{p})\psi(x) + [\psi(x)^{\text{T}}\hat\Delta \psi(x) + \hc ] \right\},
\end{equation}
with 
\begin{equation}
    \hat\Delta = u\sigma_y \hat{\bm{d}}\cdot\bm{\sigma}e^{i\hat{\varphi}}\partial_x .
\end{equation}
\end{subequations}
This model can be thought of as a two-fluid model with the condensate of Cooper-pairs and the electrons, described by spinor fields $\psi=(\psi_\uparrow, \psi_\downarrow)^\text{T}$, where $\psi^{(\dagger)}_\sigma(x)$ annihilates (creates) an electron of spin $\sigma$ at position $x$; 
similar models have been discussed in the context of one-dimensional superconductors~\cite{KaneHalperin2017,Lapa2020,LapaLevin2020,RamppSchmalian2022}. We chose the topological triplet superconductor large enough to host independent zero energy edge states 
(larger than the superconducting coherence length) but sufficiently small that, in view of long-range interactions,
the condensate incorporates only quantum (i.e., temporal) fluctuations and no spatial fluctuation. The assumptions of long-range interactions in the charge channel is realized by Coulomb interactions, while the limit of long-range interactions in the spin channel is met in the vicinity of magnetic phase transition.
The problem will thus eventually turn out to be 0+1 dimensional. 

The first term in Eq.~(\ref{eq:model}) corresponds to the previously mentioned charging energy where $\hat{N}_c  = - i \partial_\varphi$ is the total number operator of Cooper-pairs, and $\hat{n}_f=\int \text{d}x \,\psi^\dagger(x) \psi(x)$ is the total number operator of electrons. Hence the total charge operator on the island is~\cite{LapaLevin2020} $\hat{N}_\text{tot}=2\hat{N}_c + \hat{n}_f$ and commutes with $\hat H$. The constant $N^c_g = \frac{e V_g}{2}$ ($e$ being the electron charge) can be tuned by changing the gate voltage $V_g$ and determines the expected value of $\hat{N}_\text{tot}$. 

The second term describes the spin interactions on the island. The vectors $\hat{\bm{L}}=(\hat{L}_x, \hat{L}_y, \hat{L}_z)$ and $\hat{\bm{S}}=(\hat{S}_x, \hat{S}_y, \hat{S}_z)$ are the canonically conjugate 
operator 
to order parameter $\hat{\bm{d}}  \in \mathbb S^2$ and the total spin operator of the fermions $\hat S_i = \frac{1}{2}\int \text{d}x \,\psi^\dagger(x) \sigma_i \psi(x)$, respectively. 
In the following, we will refer to $\hat{\bm{L}}$ as the angular momentum of the superconducting order parameter, which should not be confused with the angular momentum of the individual Cooper-pair wavefunction (which is obviously absent in the present problem of one-dimensional TRITOPS).

The third term, $\hat{H}_\text{BdG}$ describes the triplet superconductor. The free dynamics of the electrons is governed by the dispersion relation $\epsilon(\hat{p})=\frac{\hat{p}^2}{2m}-\mu$ where $\hat{p}$ is the momentum operator. The pairing operator $\hat{\Delta}$ consists of the superconducting phase $\hat{\varphi}$, the Cooper-pair orientation $\hat{\bm{d}}$ in spin space, the pairing strength $u$ (units of velocity) and the partial derivative $\partial_x$ reflecting the p-wave nature of the superconductor.
Furthermore, the Cooper-pair orientation can be parameterized by the two operators $\htheta$ and $\hphi$, which eigenvalues can take the values $\theta\in[0, \pi)$ and $\phi\in[0, 2\pi)$. A possible parametrization in terms of $\htheta$ and $\hphi$ is given as follows:
\begin{subequations}
\begin{equation}
    \hat{\bm{d}} = \begin{pmatrix} \cos(\hphi)\sin(\htheta) \\ \cos(\htheta) \\ \sin(\hphi)\sin(\htheta) \end{pmatrix}.
\end{equation}
Another representation of $\hat{\bm{d}}$ that will be heavily used in this work can be given in terms of a unitary operator, that is
\begin{equation}
    \hat{U}^\dagger \sigma_y \hat{U} = \hat{\bm{d}}\cdot\bm{\sigma},
\end{equation}
where 
\begin{equation}
    \hat{U} = e^{i\frac{\htheta}{2}\sigma_x}e^{i\frac{\hphi}{2}\sigma_y}.
\end{equation}
\end{subequations}

Finally, the term $\hat H_z$  in Eq.~(\ref{eq:model})  is a 
{spin-rotation symmetry breaking} 
perturbation, which will be used to quantify the dominance of spin fluctuations within the MCPB. This term can be understood as a Zeeman coupling that favors a polarization of the $\hat{\bm{d}}$ vector in the y-direction. Thus, there are two competing terms in the Hamiltonian. The first term, $E_s \hat{\bm{J}}^2$, attempts to fix the angular momentum, resulting in strong fluctuations in $\hat{\bm{d}}$, while $\hat{H}_z$ tries to localize $\hat{\bm{d}}$, leading to strong fluctuations in $\hat{\bm{J}}$. 
The two aforementioned limiting cases correspond to situations where either $E_s \gg E_z$ or $E_s \ll E_z$. In the first case,  $\hat{\bm{d}}$ is pinned 
while $\hat{\bm J}$ fluctuates strongly, while in 
the other case, $\hat{\bm{J}}$ is pinned, and $\hat{\bm{d}}$ freely fluctuates. 

\subsection{Spectrum of the spinful MCPB}

In order to find the spectrum of Hamiltonian \eqref{eq:model}, we transform the electrons into a co-moving basis, which follows the fluctuations of the condensate. Thus, the transformed electrons become essentially chargeless, and their spin will be fixed to either point up or down without any fluctuation. The canonical transformation that achieves this goal reads
\begin{equation}
    \mathscr{U} = \exp(-i\frac{\hvarphi}{2} \hat{n}_f)\exp(-i\hphi \hat{S}_y)\exp(-i\htheta \hat{S}_x).
\end{equation}
This transformation applied to the Hamiltonian \eqref{eq:model} yields
\begin{align}
     \, & \mathscr{U}^\dagger  \hat{H} \mathscr{U}|_{E_z=0}  = 4 E_c(\hat{N}_c - N^c_g)^2 + E_s \hat{\bm{J}}^2 \notag \\
    & + \int \text{d}x\left\{ \psi(x)^\dagger\epsilon(\hat{p})\psi(x) + u[\psi(x)^{\text{T}}\partial_x \psi(x) + \hc ] \right\}, 
    \label{eq:modelTransformed}
\end{align}
where $\hat{\bm{J}} = \hat{\bm{L}} +  \frac{\hat{S}_y}{\sin(\htheta)} (\hat{\bm{d}}+\cot(\htheta)  {\bm{e}}_{\theta})$ with ${\bm{e}}_{\theta}$ being the unit vector in $\theta$ direction.

A few comments are in order: First, the transformation $\mathscr{U}$ decouples the spin up and down sector of the fermions and diagonalized $\hat{H}_\text{BdG}$ in spin space. 
Now, the superconducting part consists of two copies of a one-dimensional p-wave superconductor of spinless fermions per wire. Each system is a Kitaev topological superconductor and is known to host Majorana zero modes (MZMs) at the edges. Thus, the MCPB  can host up to four MZMs, which we will call $\gamma_{L\uparrow}, \gamma_{L\downarrow}, \gamma_{R\uparrow}$ and $\gamma_{R\downarrow}$. Note that the symbols $\uparrow$ and $\downarrow$ should not be taken too seriously since the MZMs do not carry a spin anymore, and the reader may think of them as mere labels. The labels $R$ and $L$ denote whether an MZM is localized at the right or left edge of the wire (see Fig.~\ref{fig:Summary}a). Furthermore, we define non-local fermions 
    \begin{equation}
        \Gamma_\sigma = \frac{1}{2}(\gamma_{L\sigma} + i\gamma_{R\sigma}).
    \end{equation}
    
Second, due to the canonical transformation, the total electron charge vanished from the Hamiltonian.  However, it enters implicitly through $\hat{N}_c$, whose quantization properties have changed. The total number operator of the Cooper-pair operator is usually integer quantized. However, after the transformation $\hat{N}_c$ is half integer quantized, as it physically describes the total charge on the island, including the contribution of unpaired electrons. A technical explanation of the change in the quantization condition is contained in appendix \ref{secApx:BoundaryCondition}. 

Third, in contrast to the total charge, the spin operator $\hat{\bm{S}}$ did not fully disappear since the spin is a vector quantity. Instead, the fermion spin survives as a background field for the angular momentum  $\hat{\bm{L}}$ of the order parameter $\hat{\bm{d}}$. Suppose one understands $\hat{\bm{d}}$ as the coordinate of a particle on a sphere. Then, 
due to the fermions, this particle 
feels a magnetic monopole of charge $s_y$ (being an eigenvalue the $\hat S_y$ operator) as 
can straightforwardly be calculated by taking the rotation of $\hat{\bm{J}}-\hat{\bm{L}}$. Analogously, 
one can show that the action corresponding to the Hamiltonian $E_s \hat{\bm{J}}^2$ takes the shape of a non-linear sigma model with a topological Wess-Zumino-Witten term. As a consequence of the background field, and somewhat in 
analogy to the Cooper-pair charge $\hat{N}_c$, the angular momentum $\hat{\bm{L}}$ is integer quantized while its transformed version $\hat{\bm{J}}$ is half odd integer quantized if the fermion parity is odd.

Finally, we can find the spectrum of the Hamiltonian \eqref{eq:modelTransformed}. The total Hilbert space $\mathcal{H}$ of the MCPB can be decomposed as a tensor product of three subspaces. That is, 
\begin{equation}
    \mathcal{H} = \mathcal{H}_n \otimes \mathcal{H}_j \otimes \mathcal{H}_f,
    \label{eq:hilbertspace}
\end{equation}
where $\mathcal{H}_n$ is the Hilbert space spanned by the eigenstates of $\hat{N}_c$, $\mathcal{H}_j$ is spanned by the eigenstates of $\hat{J}_y$ and $\mathcal{H}_f$ is the Fock space of the fermionic neutral excitations in the wire.
Since we are interested in the low-temperature behavior of the system, we assume that Bogoliubov  quasi-particles in the superconductor can not be created. However, the non-local electrons $\Gamma_\sigma$ formed from the MZMs are accessible even at zero temperature. Thus, we project $\mathcal{H}_f$ down to the Fock space, which in the case of a single wire $w = 1$ is
\begin{equation}
    P_{T=0}\mathcal{H}_f P_{T=0} = \text{span} (\ket{0}, \Gamma_{\sigma}^\dagger\ket{0}, \Gamma ^\dagger_\uparrow \Gamma^\dagger_\downarrow \ket{0}) ,
\end{equation}
where the state $\ket{0}$ is the vacuum destroyed by the operators $\Gamma_\sigma$. 

\begin{figure}
    \centering
    \includegraphics{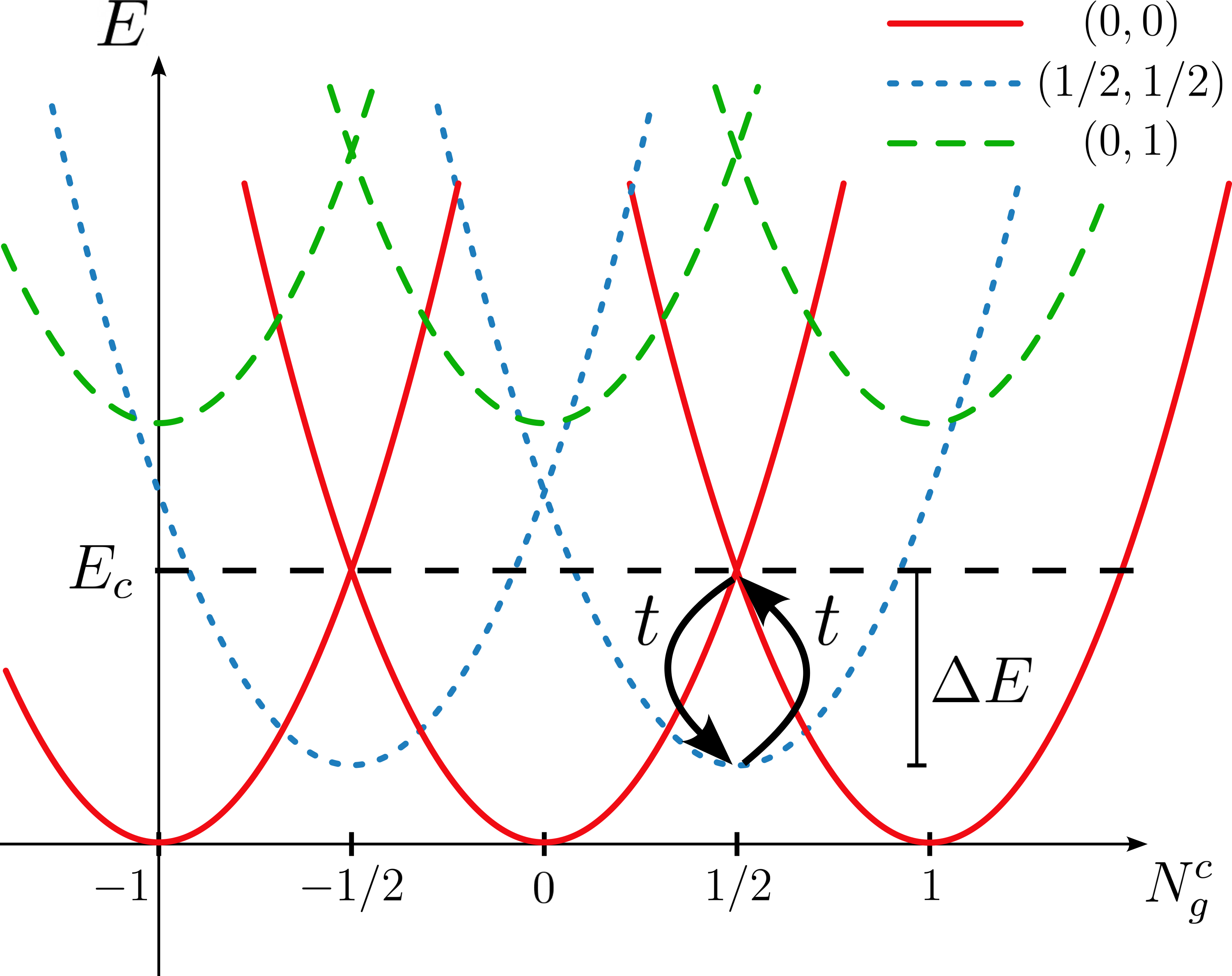}
    \caption{The spectrum of Hamiltonian \eqref{eq:modelTransformed} 
    as a function of $N^c_g$ in the case $E_c > 3E_s/4$. 
     The labeling of parabolas corresponds to eigenvalues $(N_c, J)$ of total charge and total angular momentum as defined in Tab.~\ref{tab:spectrum}.
    The crossing point of two red, solid parabolas denotes the value of $N^c_g$ where it becomes favorable to add one unit of charge (one Cooper pair) onto the island and corresponds to $E_c$ where $E_c$ is the charging energy. The two black arrows denote a second-order process of virtual transitions from a state with odd fermion parity to a state with even fermion parity and reverse and determine the superexchange interaction 
    The energy gap $\Delta E = E_c - 3E_s/4$ must be overcome for such a process.
    }
    \label{fig:spectrum}
\end{figure}

Furthermore, the Hamiltonian \eqref{eq:modelTransformed} commutes with the mutually commuting operators $\hat{N}_c$ and $\hat{\bm{J}}$. Therefore, we can construct the energies and energy eigenstates by solving the eigenvalue equation for $\hat{N}_c$ and $\hat{\bm{J}}^2$ separately. The low energy spectrum of the MCPB is summarized in the Tab.~\ref{tab:spectrum} for the case $w = 1$ and plotted in Fig.~\ref{fig:spectrum}. In the case of multiple wires, $w >1$, the low-energy spectrum of Tab.~\ref{tab:spectrum} is the same and characterized by the very same quantum numbers as displayed in the first four columns of Tab.~\ref{tab:spectrum}. However, at $w >1$, the entries for $s_y$ and for the degeneracy of the various states may be higher. In particular, the ground state degeneracy is $4,16,204, \dots$ for $w = 1,2,3, \dots$.

\begin{table}[]
    \centering
    \begin{tabular}{c|c|c|c|c||c|c}
        ${N}_c$ & ${j}_y$ & $J$ & ${P}_f$ & ${s}_y$ & $E$ & $\#$ \\
        \hline
        0 & 0 & $0$ & 1 & 0 & $4E_c(N^c_g)^2$ & 2  \\
    \hline
    $\nicefrac{1}{2}$ & $\pm \nicefrac{1}{2}$ & $\nicefrac{1}{2}$ & $-1$ & $\pm\nicefrac{1}{2}$ & $4E_c(\nicefrac{1}{2} - N_g^c)^2 + E_s\nicefrac{3}{4}$ & 4 \\
            \hline
         0 & $0,\pm 1$  & 1 & 1 & 0 & $4E_c(N^c_g)^2 + 2 E_s$ & 6 
    \end{tabular}
    \caption{Classification of the three lowest energy eigenstates of Hamiltonian \eqref{eq:modelTransformed} 
    by their quantum numbers: Charge ${N}_c$ (in units of $2e$), magnetic quantum number ${j}_y$, total angular momentum $J$ as determind by the $\hat {\bm{J}}^2$ eigenvalue $J(J+1)$, fermion parity ${P}_f=(-1)^{n_f}$ and magnetic quantum number of the fermions ${s}_y$.  The second last column shows the total energy of a particular state as a function of $N^c_g$ while the last column 
    shows the corresponding degeneracy of each state.}
    \label{tab:spectrum}
\end{table}

\section{Spinful Topological Kondo Effect}
\label{sec:spinfulTopologicalKondoEffect}

This section explains the setup that gives rise to a topological Kondo-effect by coupling leads to an MCPB. We model the system with two free electron gases that are brought into the vicinity of an MCPB, where electrons from a left and right lead can tunnel via the tunneling amplitude $t$ into the Majorana edge states on the left and right sides of the superconductor, respectively. A schematic of the setup for $w = 1$ can be found in Fig.~\ref{fig:Summary}a. The Hamiltonian of the full system reads:

\begin{equation}
\hat{H} = \hat{H}_{\text{dot}}(\gamma) + \hat{H}_0(\psi) - t \,\hat{H}_{t}(\psi, \gamma) +\hat H_z.
\label{eq:fullBareHamiltonian}
\end{equation}

Here, $\hat{H}_\text{dot}$ denotes the Hamiltonian \eqref{eq:modelTransformed}, and $\hat{H}_0$ is the Hamiltonian of free electrons $\psi_{R/L}$ of the left and right leads. The Hamiltonian $\hat{H}_t$ couples the electrons to the MZMs and is given by:

\begin{equation}
\hat{H}_t = \sum_{i=R/L} \left(\psi^\dagger_{i,\sigma}e^{-i\varphi/2}(U^\dagger)^{\phantom{\dagger}}_{\sigma\sigma'}\gamma^{\phantom{\dagger}}_{i\sigma'} + \text{H.c.}\right),
\end{equation}
Here, the operator $\psi_{i,\sigma}^\dagger(x)$ creates an electron in the $i$th lead with spin $\sigma$. To straighten up the notation, we define $\psi^{(\dagger)}_{i\sigma}\equiv\psi^{(\dagger)}_{i\sigma}(0)$ where we assume that the MCPB is located at $x=0$. This setup is readily generalized to $w>1$ where we assume a spinful lead coupled to each of the $2w$ Kramers pairs of Majoranas on the island.

We choose the gate charge $N^c_g=\frac{1}{2}$ such that the ground state of the quantum dot has an odd fermion parity, i.e., one non-local fermion $\Gamma_\sigma$ is occupied. Therefore, the ground state corresponds to the blue parabola in Fig.~\ref{fig:spectrum} and the second row in Table \ref{tab:spectrum}. Thus, the ground state is fourfold degenerate in the $w = 1$ wire case. In this exemplary case,
the restriction to the odd fermion parity operator imposes a constraint on the MZMs of the form:

\begin{equation}
\gamma_{L\downarrow}\gamma_{L\uparrow}\gamma_{R\downarrow}\gamma_{R\uparrow} = -1
\label{eq:parityConstraint}
\end{equation}

The objective is to develop an effective low-temperature theory for the ground state manifold spanned by the four degenerate states. Processes in which fermions tunnel into and away from the MCPB change the fermion parity from $P_f=-1$ to $P_f=1$. The first mentioned process adds charge, while the second process removes charge from the box. These tunneling events correspond to transitions in higher energy eigenstates (see Fig.~\ref{fig:spectrum}).

We account for these processes in the low energy regime by treating them as virtual and integrating out the higher excited states using a Schrieffer-Wolf transformation \cite{SchriefferWolff1966}. To illustrate this procedure and as a preliminary problem, we first apply it to the limiting case where $E_c \gg E_s$. Thus, the timescale on which the superconducting phase $\hat\varphi$ fluctuates is much smaller compared to the timescale of the fluctuations of $\hat{\bm{d}}$ ($\hat U$). Therefore, we first only perform a Schrieffer-Wolf transformation in the charge sector $\mathcal{H}_n$ of the condensate Hilbert space. 

While in all of the previous discussions, a single TRITOPS wire on the island is assumed, see Fig.~\ref{fig:Summary}, the problem is readily generalized to a spinful MCPB with $w$ wires ($w >1$), all of which are coupled to two leads and to the same order parameter field by introducing a wire index to the fermions and assuming the mean field Hamiltonian is diagonal in wire-space (see \cite{BeriCooper2012} for the discussion of the analogous situation without spin fluctuations). We will suppress the wire index in what follows but occasionally comment on the case $w>1$. 
The such derived, effective Hamiltonian reads

\begin{equation}
    \hat H_K = - \frac{\lambda}{2} \sum_{i,j} \sum_{\sigma, \rho} [\psi^\dagger_{i}\hat U^\dagger]_{\sigma} \gamma_{i,\sigma} \gamma_{j, \rho} [\hat U\psi_{j}]_{\rho}, \label{eq:KondoGeneralw}
\end{equation}
where $\lambda = \frac{4t^2}{E_c}$. This expression holds for an arbitrary number of wires (in which case  $i,j \in \{ 1, \dots, 2w \}$). In the simplest case $w = 1$ (in this case we use labels $i,j \in \{ L,R \}$) this expression can be further simplified as follows

\begin{equation}
    \hat{H}_K = \lambda \left(\psi_+^\dagger \hat U^\dagger \boldsymbol
{\sigma} \hat U \psi_+^{\phantom{\dagger}} + \psi^\dagger_- \sigma_y \hat U^\dagger \sigma_y \boldsymbol{\sigma}\sigma_y \hat U \sigma_y  \psi_-^{\phantom{\dagger}}\right)\cdot\boldsymbol{\hat{S}},
\label{eq:KondoEcggEs}
\end{equation}
where 
\begin{equation}
    \begin{split}
        \psi_+ & =\frac{1}{\sqrt{2}}\left(\psi_L +i \psi_R\right) \\
        \psi_- & = \frac{i\sigma_y}{\sqrt{2}}\left(\psi_L -i \psi_R\right)
    \end{split}
\end{equation}
with $\psi_{L/R}=(\psi_{L/R\uparrow},\psi_{L/R\downarrow})^\text{T}$ being spinors in spin space. The impurity spin $\hat{\bm{S}}=(\hat{S}_x, \hat{S}_y, \hat{S}_z)^\text{T}$ contains the fermion spin operators $\hat{S}_i = \frac{1}{2}\Gamma^\dagger\sigma_i\Gamma$, where $\Gamma=(\Gamma_{\uparrow},\Gamma_{\downarrow})^\text{T}$. These spin operators can be expressed in terms of the MZMs. Using the parity constraint \eqref{eq:parityConstraint}, we obtain
\begin{align}
\,& \hat{S}_x = \frac{X}{2}, \quad\quad X = i\gamma_{L\downarrow}\gamma_{R\uparrow}=i\gamma_{L\uparrow}\gamma_{R\downarrow}, \\
& \hat{S}_y = \frac{Y}{2}, \quad\quad Y = i\gamma_{L\downarrow}\gamma_{L\uparrow}=i\gamma_{R\downarrow}\gamma_{R\uparrow}, \\
& \hat{S}_z = \frac{Z}{2}, \quad\quad Z = i\gamma_{L\uparrow}\gamma_{R\uparrow}=i\gamma_{R\downarrow}\gamma_{L\downarrow},
\end{align}
where the operators $X$, $Y$, and $Z$ fulfill an SU(2) algebra. Since the spin quantum number of the MZMs behaves more like a static quantum number than an actual spin, we should rather think of $\hat{\bm{S}}$ as an impurity that acts on the orbital (charge parity) space spanned by $\gamma_{\uparrow/\downarrow}$  than an actual spin. 

\subsection{Limit $E_z\gg E_s$}

We take $E_s$ and $E_c$ of the same order of magnitude but also take the perturbation $\hat{H}_z$ into account where we work in the limit $E_z\gg E_s$. Thus, the $\hat{\bm d}$ vector is polarized and, again, fluctuates on a much larger time scale than the superconducting phase $\hat \varphi$. Therefore, we can continue working with the equation \eqref{eq:KondoEcggEs}.

In the case where $E_s/E_z\rightarrow 0$, $\hat U$ does not display any quantum fluctuations (it is a constant of motion). We can thus 
choose $U=\mathds{1}$, and the Hamiltonian \eqref{eq:KondoGeneralw} becomes an O(4$w$) topological Kondo model~\cite{BeriCooper2012}. Moreover, in view of a well established equivalence between O(4) topological Kondo effect and two-channel SU(2) Kondo effect (cf. App.~\ref{app:Superex}), the $w = 1$ Hamiltonian \eqref{eq:KondoEcggEs} becomes a two-channel Kondo Hamiltonian
\begin{equation}
    \hat{H}_{2CK} = \lambda \sum_{a=\pm} \psi^\dagger_a \bm{\sigma} \psi^\phd_a \cdot \hat{\bm{J}}. \label{eq:H2CK}
\end{equation}

It is well known that the two-channel Kondo model has a quantum critical point at a finite Kondo coupling $\lambda$~\cite{AndreiDestri1984,TsvelickWiegmann1985,AffleckLudwig1991a}. Emery and Kivelson~\cite{EmeryKivelson1992} already formulated an exact solution to this problem with Bosonization methods and found that fractionalized (Majorana fermions) excitations govern the system. 

However, in the case of fluctuating $\hat U$ (i.e.~finite $E_s/E_z$), the situation becomes more complex. As discussed previously, in the limit $E_s/E_z \rightarrow 0$ we expect two-channel Kondo physics, Eq.~\eqref{eq:H2CK}, for the $w = 1$ spinful topological Kondo effect. This is illustrated as a red star in Fig.~\ref{fig:Summary}c. Here, we consider small $E_s/E_z$ corrections and find that the two-channel Kondo fixed point is stable concerning this perturbation.

We assume that the vector $\hat{\bm{d}}$ is predominantly polarized in the y-direction of the ``north-pole'' and weakly fluctuates around this state.
Mathematically, this can be expressed as follows:

\begin{multline}
\hat{\bm{d}} = \bm{e}_y + \delta\hat{\bm{d}}, \\ \quad \delta\hat{\bm{d}} = \begin{pmatrix}
\delta \hat{d}_x, \ 0, \ \delta \hat{d}_z
\end{pmatrix}
= \begin{pmatrix}
\hat{\theta} \sin(\hat{\phi}), \ 0, \ \hat{\theta} \cos(\hat{\phi})
\end{pmatrix},
\end{multline}

The $\hat{\bm{d}}$ vector has been linearized with respect to the deviation angle $\hat{\theta}$. This linearization yields the effective two-dimensional vector $\delta \hat{\bm{d}}$ that characterizes the spin sector of the condensate, confined to a tangent plane attached to the north pole of the sphere encompassing all possible configurations of $\hat{\bm{d}}$.

To determine the spectrum of the vector $\hat{\bm{d}}$, we expand the Hamiltonian of the MCPB (including the $E_z$ term) to the first non-trivial order in $\theta$. This leads to a Hamiltonian $\hat{H}_{\theta\ll 1}$, in which the Schrödinger equation resembles the radial part of a harmonic oscillator in two dimensions (see App.~\ref{AppSubSecZeeman}). The level spacing of the energy eigenstates is $\Delta E_n = \sqrt{2E_s E_z}$. The four-fold degeneracy of the MCPB ground state for $E_z=0$ gets lifted by the presence of a finite $E_z$ with a new two-fold degenerate ground state manifold where $s_y=j_y=\pm 1/2$. Furthermore, the ground state expectation value of $\delta \hat{\bm d}$ is 0. A comprehensive derivation of the spectrum can be found in appendix \ref{AppSubSecZeeman}. 

To derive the effective Kondo Hamiltonian in the limiting case $\theta \ll 1$ (i.e. $E_z \gg E_s$), we use equation \eqref{eq:KondoEcggEs} as a starting point and, similarly, linearize $U$. Initially, we introduce a new parametrization $\hat U = \exp(i\frac{\hat{W}}{2})$, where $\hat{W} = \partial_\phi \delta\hat{\bm d}\cdot\bm{\sigma}$. Expanding $\hat U$ to first order in $\hat{W}$, we arrive at a linearized version of the Hamiltonian \eqref{eq:KondoEcggEs}, that is 

\begin{equation}
\hat{H}_K = \lambda \sum_{a=\pm} \psi^\dagger_a \bm{\sigma} \psi_a \cdot \left(1 - a\,\partial_\phi\delta\hat{\bm d}\times\right)\hat{\bm{S}}.
\label{eq:KondoEzLarg}
\end{equation}
From this expression, it becomes evident that even at the first order in $\hat{\bm{d}}$ fluctuations, the effective Hamiltonian exhibits a more intricate structure than a two-channel Kondo Hamiltonian. 

However, if $E_z \gg \left(\frac{\lambda^4}{E_s}\right)^{\frac{1}{3}}$ the system effectively stays in the ground state manifold. Consequently, we can replace the operator $\delta\hat{\bm d}$ with its ground-state expectation value, which is zero. {Second-order processes in the first excited state and back into the ground state modify the Hamiltonian such that the isotropic 2CK Hamiltonian becomes anisotropic {in SU(2) spin space, but still preserves channel isotropy}. Furthermore, they add an interaction term coupling the relative spin densities between the two types of lead electrons. However, the interaction term has a scaling dimension of two and is deemed irrelevant in the Renormalization Group (RG) {sense}. The anisotropic 2CK model flows towards isotropy and, consequently, converges to the same fixed point as the isotropic 2CK model~\cite{EmeryKivelson1992}. Therefore, the two-channel Kondo effect remains stable against small fluctuations in $\hat{\bm{d}}$.}
\subsection{Limit $E_z \ll E_s$}

If, conversely, the magnitude of $E_s$ greatly surpasses that of $E_z$, the phase $\hvarphi$ and $\hat U$ experiences rapid fluctuations and, necessarily, need to be treated on equal footing. Within the low-energy domain, the system is confined to the fourfold degenerate ground state manifold in Tab.~\ref{tab:spectrum} (blue parabola in Fig.~\ref{fig:spectrum}). Details concerning the precise computation are presented in appendix~\ref{app:Superex}. Employing the Schrieffer-Wolf transformation, we arrive at the effective Hamiltonian given by
\begin{equation}
\hat{H}_\text{eff} = \lambda \sum_{a=\pm} \psi^\dagger_a \bm{\sigma} \psi^{\phantom{\dagger}}_a \cdot \frac{1-a\,Y}{2}\hat{\bm{J}} - \frac{E_z}{3}Z.
\label{eq:KondoHamEzSmall}
\end{equation}

The vector $\hat{\bm{J}}$ is projected onto the ground state manifold and can be expressed as $\hat{\bm{J}}=(\frac{J_x}{2}, \frac{J_y}{2}, \frac{J_z}{2})$, with $J_i$ denoting the Pauli matrices acting within the space spanned by the states $\ket{j=\pm\nicefrac{1}{2}, J=\nicefrac{1}{2}}$ in the Hilbert space $\mathcal{H}_j$. The Kondo coupling parameter is now defined as $\lambda = \frac{4t^2}{E_c - \nicefrac{3}{4}E_s}$. The disparity from the Kondo coupling discussed in the previous section arises from including virtual processes in the angular momentum sector. Of course, in the limit $E_c \gg E_s$, which was assumed to derive Eq.~\eqref{eq:KondoEcggEs}, this disparity vanishes.

At first glance, one might mistake the Hamiltonian \eqref{eq:KondoHamEzSmall} for twice a conventional SU(2) Kondo Hamiltonian, given that the projector $P_a = \frac{1-aY}{2}$ projects the impurity $\hat{\bm{J}}$ onto two independent sectors. In other words, the two effective impurities $\hat{\bm{J}}_a = P_a \hat{\bm{J}}$ commute with each other, i.e., $\left[\hat{J}_+^i, \hat{J}_-^j\right]=0$. However, the distinction to two SU(2) Kondo effects lies in the fact that the entire impurity transforms under SU(2) $\oplus$ SU(2) rather than SU(2) $\otimes$ SU(2), as it is the case for two instances of a standard SU(2) Kondo effect. To give Hamiltonian \eqref{eq:KondoHamEzSmall} a physical meaning, one can understand the projector $\frac{1-aY}{2}$ as an operator that distinguishes between the occupation of the two orbitals formed by the four MZMs. The lead electrons $\psi_a$ can only interact with the impurity if the orbital assigned by the projection operator is occupied. Therefore, the MZMs act as gatekeepers, limiting access to the impurity for the lead electrons, see Fig.~\ref{fig:Summary} b) for an illustration.

\begin{figure}
    \centering
    \includegraphics{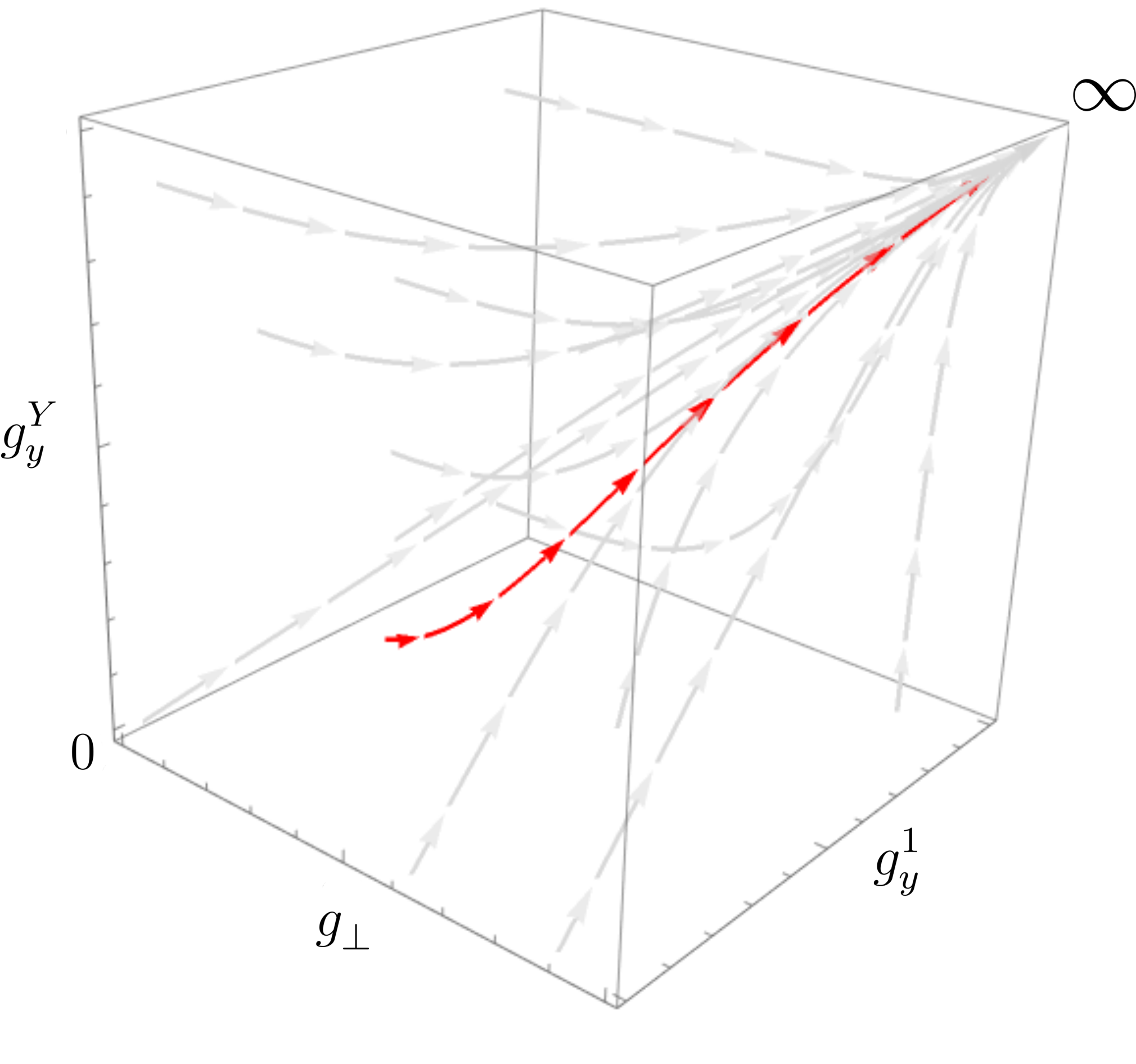}
    \caption{{RG flow for anisotropic coupling constants. In the lower-left corner, all couplings are zero. In contrast, the upper left corner symbolizes the point where all coupling constants are infinity, which is the strong coupling fixed point. The red trajectory reflects the RG flow for which the Toulouse point is the starting point. A more detailed discussion on how this figure is created can be found in App.~\ref{app:PoorMan}}}
    \label{fig:RGflow}
\end{figure}

For further use in the remainder of the paper, we generalize the isotropic $E_z = 0$ Hamiltonian to an anisotropic model as follows: 
\begin{align}
\hat{H}_{K} & = \sum_{a=\pm} \left( \sum_{i\in{x, z}} \psi^\dagger_a \sigma_i \psi^{\phantom{\dagger}}_a(\lambda^1_\perp - a\, \lambda^Y_\perp Y)\hat{J}_i \right. \notag\\ & \left. + \phantom{\sum_a} \psi^\dagger_y \sigma_i \psi^{\phantom{\dagger}}_a (\lambda^{1}_y - a\, \lambda^{Y}_y)Y\hat{J}_y \right). \label{eq:anisotropicKondoHam}
\end{align}
We employ a poor man's scaling \cite{Anderson1970} analysis (see App.~\ref{app:PoorMan}) for Eq.~\eqref{eq:anisotropicKondoHam} and obtain the following flow equations: {
\begin{subequations}
\begin{align}
    \frac{\text d g_\perp}{\text d l} & = g_\perp\left( g^1_y + g^Y_y \right), \\
    \frac{\text d g^1_y}{\text d l} & = 2g_\perp^2, \\
    \frac{\text d g^Y_y}{\text d l} & = 2g_\perp^2,
\end{align}
\label{eq:RGequations}
\end{subequations}
where $l=\log\left(\frac{D_0}{D}\right)$ is the RG time and $D_0$ being the {bare} 
bandwidth. We introduced the     dimensionless $g^{1/Y}_{\perp/y} = \nu              \lambda^{1/Y}_{\perp/y}$ coupling constants, with $\nu$ denoting the density of states at the Fermi level and set $g^1_\perp=g^Y_\perp=g_\perp$. Fig.~\ref{fig:RGflow} displays the RG flow created by Eq.~\eqref{eq:RGequations}, where it is clearly visible that the couplings flow towards strong coupling and isotropy. In particular, this is true for the red trajectory, which starts at the {parameter values corresponding to the} Toulouse point {introduced in the next section}.} This result suggests that the Toulouse point and weak isotropic coupling points in parameter space reside in the basin of attraction of the same strong coupling fixed point. This observation will have significance in the next section, where we present an exact solution of the model at the Toulouse point: It can be expected that the Toulouse point solution correctly describes the strong coupling physics of Eq.~\eqref{eq:KondoHamEzSmall} and Eq.~\eqref{eq:anisotropicKondoHam}, alike. 

\section{Single wire at the Toulouse point}
\label{sec:singleWireAtTheToulosePoint}

Generalizing the protocol of Emery and Kivelson \cite{EmeryKivelson1992}, we can exactly solve
Hamiltonian Eq.~\eqref{eq:anisotropicKondoHam} at a specific hyperplane in parameter space called ``Toulouse point''.
For this, we bosonize the lead fermions, apply a non-local canonical transformation $U_{\text{E.K.}}$, and refermionize the lead electrons as well as expressing the impurity in terms of Majorana fermions.

\subsection{Bosonization and Refermionization}
\label{sec:BosonizationAndRefermionization}

We take the usual Bosonization approach and decompose the fields $\psi_{a \sigma}$, with respect to the lattice constant $a_0$, into slow-varying right and left-moving fields:
\begin{equation}
\psi_{a\sigma}(x) = \expp{ik_F x}R_{a\sigma}(x) + \expp{-ik_F x}L_{a\sigma}(x),
\end{equation}
where $k_F$ is the Fermi wavevector.
Note that $\psi_{+\sigma}(x)$ and $\psi_{-\sigma}(x)$ only have support on one half-axis. Thus, one can represent the two fields by only one chiral (right-moving) field that extends over the whole real axis \cite{Gogolin2004}. This new chiral fermion, written in terms of the chiral Bose fields $\phi_{a\sigma}(x)$, read
\begin{equation}
\tilde{\psi}_{a\sigma}(x) = \frac{F_{a\sigma}}{\sqrt{2\pi a_0}}\expp{i\sqrt{4\pi}\phi_{a\sigma}(x)},
\label{eq:fermionToBoson}
\end{equation}
where $F_{a\sigma}$ is a Klein factor that ensures the correct fermionic statistics. { The bosonic field follows the commutation relation 
\begin{equation}
    [\partial_x \phi_i(x), \phi_j(y)] = \frac{i}{2}\delta(x-y)\delta_{ij},
\end{equation}
where $i,j$ are multindices of the shape $(a,\sigma)$}. Again, we define $\tilde{\psi}_{a\sigma}(0)\equiv\tilde{\psi}_{a\sigma}$ and $\phi_{a\sigma}(0) \equiv \phi_{a\sigma}$.

It will be convenient to introduce a new basis of Bose fields that are related by
\begin{equation}
\begin{pmatrix}
\phi_{s} \\ \phi_{sf} \\ \phi_{c} \\ \phi_{cf}
\end{pmatrix}
= \frac{1}{2}
\begin{pmatrix}
1 & -1 & 1 & -1 \\
1 & -1 & -1 & 1 \\
1 & 1 & 1 & 1 \\
1 & 1 & -1 & -1 \\
\end{pmatrix}
\begin{pmatrix}
\phi_{+\uparrow} \\ \phi_{+\downarrow} \\ \phi_{-\uparrow} \\ \phi_{-\downarrow}
\end{pmatrix}.
\label{eq:BosonMatrix}
\end{equation}
A similar basis transformation can also be done for the Klein factors with the identities \cite{LibermanSela2021, Zarnd2000, vonDelft1998}
\begin{align}
& F^\dagger_{sf} F^\dagger_s = F^\dagger_{+\uparrow}F_{+\downarrow}, \quad F_{sf} F_s^\dagger = F^\dagger_{-\uparrow}F_{-\downarrow}, \notag \\
& F^\dagger_{sf} F^\dagger_{cf} = F^\dagger_{+\uparrow}F_{-\uparrow}, \quad F^\dagger_c F^\dagger_s = F^\dagger_{+\uparrow}F^\dagger_{-\uparrow}.
\label{eq:basisChangeKleinFactors}
\end{align}
Also, the Kondo Hamiltonian can be rewritten in the following form
\begin{align}
\hat{H}_K & = \frac{\lambda_\perp}{2}\left[(I^+_+ + I^+_-)\hat J_- + (I^+_- - I^+_+)Y\hat J_- + \text{H.c.}\right] \nonumber \\
& \quad + \lambda^1_y(I^y_+ + I^y_-)\hat J_y + \lambda_y^Y(I^y_- - I^y_+)Y \hat J_y,
\end{align}
where
\begin{equation}
I^{+/-}_a = \tilde{\psi}^\dagger_{a\uparrow/\downarrow}\tilde{\psi}^{\phantom{\dagger}}_{a\downarrow/\uparrow} \quad \text{and} \quad I^y_a = \frac{1}{2}(\tilde{\psi}^\dagger_{a\uparrow}\tilde{\psi}^\phd_{a\uparrow} - \tilde{\psi}^\dagger_{a\downarrow}\tilde{\psi}^\phd_{a\downarrow}).
\end{equation}
Note that we work in a basis in which $\sigma_y$ is diagonal and, thus, the spin indices $\uparrow/\downarrow$ refer to spins quantized in the $y$-direction. We used the notation $\hat J_\pm = \hat  J_z \pm i \hat J_x$. We next bosonize according to equation \eqref{eq:fermionToBoson} and perform a non-local unitary rotation (see App.~\ref{app:EK}) to simplify the Hamiltonian. It is convenient to define
the new fermions 
\begin{equation}
\chi_{r} = \frac{F_{r}}{\sqrt{2\pi a_0}}\expp{i\sqrt{4\pi}\phi_r} \quad \text{and} \quad {d} = F^\dagger_s \hat J_-,
\end{equation}
$(r=s,sf,c,cf)$ one can show that the effective Kondo Hamiltonian becomes
\begin{align}
\hat{H}'_K & = \frac{\lambda\perp}{\sqrt{8\pi a_0}}(\chi_{sf}^\phd + \chi_{sf}^\dagger)(d - d^\dagger) \nonumber \\
& \quad - \frac{\lambda_\perp}{\sqrt{8\pi a_0}}(\chi_{sf}^\phd - \chi_{sf}^\dagger)(d + d^\dagger) Y \nonumber \\
& \quad + \delta\lambda_y :\chi_s^\dagger\chi_s^\phd: (d^\dagger d - \frac{1}{2}) \nonumber \\
& \quad + \lambda_y^Y :\chi_{sf}^\dagger\chi^\phd_{sf}:Y (d^\dagger d - \frac{1}{2}),
\label{eq:fermionizedHam}
\end{align}
where $\delta\lambda_y = \lambda^1_y - 2\pi v_F$ and $v_F$ is the Fermi velocity.

The former Hamiltonian takes a more convenient shape if one rewrites the fermions in terms of real Majorana fermions. Let us consider the following decomposition
\begin{equation}
\chi_r = \frac{1}{2}(\xi_r + i\zeta_r), \quad \text{and} \quad d = \frac{1}{2}(\alpha+i\beta),
\end{equation}
of the complex fermions into real Majorana fermions.
Note that in the present basis, the number of Majorana fermions that describe the four-dimensional impurity is represented by the six Majorana fermions and one constraint. 
Namely, the four MZMs (i.e. $\gamma_{L\uparrow}, \gamma_{L\downarrow}, \gamma_{R\uparrow},$ and $\gamma_{R\downarrow}$) and the two 
Majorana fermions $\alpha$ and $\beta$ that describe 
${\hat{\bm{J}}}$.
It is useful to represent every term in Hamiltonian \eqref{eq:fermionizedHam} in terms of four composite Majorana fermions without constraint, i.e.
\begin{equation}
\eta = \beta, \quad \eta_x = \alpha X, \quad \eta_y = \alpha Y, \quad \text{and} \quad \eta_z = \alpha Z.
\end{equation}
These four fermions allow for a faithful representation of the Hamiltonian \eqref{eq:fermionizedHam} that preserves all mutual commutation relations of the different operators appearing in the Hamiltonian. Using this representation, the Hamiltonian becomes
\begin{align}
\hat{H}'_\text{eff} & = i\frac{\lambda_\perp}{\sqrt{8\pi a_0}}\xi_{sf} \eta + i\frac{\lambda_\perp}{\sqrt{8\pi a_0}}\zeta_{sf}\eta_y \nonumber \\
& \quad + i \frac{\delta \lambda_y}{4} \xi_s \zeta_s \eta_x\eta_y\eta_z\eta + \frac{\lambda^Y_y}{4}\xi_{sf}\zeta_{sf}\eta_y\eta + i\frac{E_z}{3}\eta_y \eta_x.
\label{eq:HamMajoranas}
\end{align}
Thus, the spinful topological Kondo effect for a single wire maps to a special interacting resonant level model of Majorana fermions, see Fig.~\ref{fig:SchematicMajoHam} for illustration.

\subsection{Toulouse point and strong coupling fixed point}

In this section, we investigate the Hamiltonian at the Toulouse point and the stability of the emergent strong coupling fixed point by means of the symmetry-breaking Zeeman term as well as small perturbations away from the Toulouse point. The Toulouse point~\cite{ToulouseSeances1969} is defined as the point in parameter space where interactions in the resonant level Hamiltonian \eqref{eq:HamMajoranas} are absent, i.e., $\delta\lambda_y=\lambda^Y_y = 0$. As was already discussed in the previous section, the poor man's scaling reveals that the different coupling constants flow towards strong coupling and isotropy. This is also true if we choose the Toulouse point as the starting point for the RG flow, see Fig.~\ref{fig:RGflow}.
Thus, we expect that the physics that is present in the Hamiltonian at the exact solvable Toulouse point extends towards the strong coupling fixed point.

\begin{figure}
\centering
\includegraphics{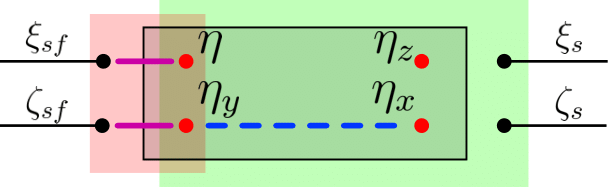}
\caption{Schematic representation of Hamiltonian \eqref{eq:HamMajoranas}. The red dots denote the impurity Majorana fermions while the black symbols represent the lead Majorana fermions. The hybridization terms are represented by colored lines (solid magenta for $\lambda_\perp$, dashed blue for $E_z$). The interaction terms are given by transparent rectangles (red for $\lambda^Y_y$, green for $\delta\lambda_y$).}
\label{fig:SchematicMajoHam}
\end{figure}

Fig.~\ref{fig:SchematicMajoHam} shows a schematic of the Hamiltonian \eqref{eq:HamMajoranas}. At the Toulouse point and in the case of $E_z=0$, only two impurity Majorana fermions hybridize with the lead electrons, namely $\eta_y$ and $\eta$. Physically, that means that the two sectors spanned by the MZMs are degenerate. Thus, there are two dangling Majorana fermions.
This translates into a twofold ground state degeneracy. However, the moment one has a finite $E_z$, regardless of how small, $\eta_x$ hybridizes with $\eta_y$ and there is only one Majorana fermion dangling anymore. Thus, the twofold ground state degeneracy reduces to a non-integer degeneracy of $\sqrt{2}$. 

\subsection{Renormalization group flow}

In the following, it is useful to invoke the
impurity entropy defined by
\begin{equation}
S_{\text{imp}} = S_\text{tot} - S_\text{bulk} \equiv \ln(g),
\end{equation}
where the
quantity $g$ can be interpreted as a generalized ground state degeneracy which can take non-integer values. We remind the reader that, generally under RG 
unstable fixed points flow towards stable fixed points with lower $g$ (``$g$-theorem'' \cite{FriedanKonechy2004}).

We are now in the position to discuss the stability of the Toulouse point solution in a renormalization group sense and to construct the schematic
RG flow diagram, Fig.~\ref{fig:Summary}c). For the case $E_z=0$ and at 
the Toulouse point, we found a twofold degeneracy $g = 2$ (represented by a blue star). We repeatedly argued that anisotropy (i.e. unequal $\lambda_{y, \perp}^{1,Y}$) is irrelevant within poor man's scaling. The effective interacting resonant level model, Eq.~\eqref{eq:HamMajoranas} corroborates this statement from the strong-coupling perspective, since $\lambda_y^Y$, $\delta \lambda_y$ terms are RG irrelevant (the corresponding operators have scaling dimension 4 and 2 respectively).

While the Toulouse point is stable towards restoring isotropy, it 
is unstable towards the inclusion of 
$E_z$, which couples to an operator of scaling dimension $1/2$ ($\eta_y$ acquires the scaling dimension of lead electrons through $\lambda_\perp$ hybridization) and generates a state of $g = \sqrt{2}$. 
We have thus demonstrated that the Kondo fixed point at infinite $E_s/E_z$, blue star in Fig.~\ref{fig:Summary}c), is unstable and, given that our Bosonization solution is valid for any $E_z$, flows towards a fixed point with $g=\sqrt{2}$. As we had previously argued, this fixed point may be interpreted as an O(4) topological Kondo effect (or equivalently a two-channel Kondo effect).

\section{Observables}
\label{sec:Observables}

Within this section, we present a comprehensive overview of specific observables derived from the application of the Hamiltonian \eqref{eq:HamMajoranas} at the Toulouse point, with a fixed value of $E_z=0$. Details are relegated to App.~\ref{app:Observables}.

\subsection{Thermodynamics}

We first focus on
the correction to the free energy denoted by $\delta F$. To determine the scaling dimension of the interaction terms, we analyzed the operators $O_s$ and $O_{sf}$ defined as follows:
\begin{align}
O_s & = i \frac{\delta \lambda_y}{4} \xi_s \zeta_s \eta_x\eta_y\eta_z\eta, \\
O_{sf} & = \frac{\lambda^Y_y}{4}\xi_{sf}\zeta_{sf}\eta_y\eta.
\end{align}
By calculating the expectation value 
\begin{equation}
    \braket{O_r(\tau)O_r(0)}\sim\left(\frac{1}{\tau}\right)^{2\Delta_x} ,
\end{equation}
where $\Delta_r$ represents the scaling dimension, we found that the scaling dimensions of $O_s$ and $O_{sf}$ are $\Delta_{s}=2$ and $\Delta_{sf}=4$, respectively. Following \cite{Gogolin2004}, we inferred that the correction to the free energy scales as:
\begin{equation}
\delta F \sim \mathscr{O}(T^2),
\end{equation}
which relates to the thermodynamic entropy and specific heat, respectively:
\begin{align}
S = -\frac{\partial}{\partial T} F \sim \kappa T, \quad\text{and}\quad
C_v = T\frac{\partial}{\partial T} S \sim \kappa T,
\end{align}
with $\kappa$ being some constant. 
This behavior resembles Fermi-liquid characteristics.

Furthermore, we evaluated the susceptibilities at zero temperature for the impurity, of which we have six different kinds,  three in the orbital space and three in the angular momentum space. The susceptibilities are as follows:
\begin{align}
\chi_X & \sim \ln\left(\nicefrac{\Gamma}{T}\right), \quad\quad \chi_{J_x} \sim \text{const.}, \\
\chi_Y & \sim \nicefrac{1}{T}, \quad \quad \quad \quad \chi_{J_y} \sim \text{const.}, \\
\chi_Z & \sim \ln\left(\nicefrac{\Gamma}{T}\right), \quad\quad \chi_{J_z} \sim \text{const.},
\end{align}
where $\Gamma = \frac{\lambda_\perp^2\nu}{2 a_0}$. In the angular momentum sector, we observed Pauli susceptibility with constant values akin to those in a Fermi liquid. However, in the orbital space, the susceptibilities diverge as $T\rightarrow 0$. Specifically, $\chi_{X, Z}$ diverges logarithmically, while $\chi_Y$ exhibits an algebraic divergence. This anisotropic behavior is 
 possibly due to the explicit breaking of the SU(2) symmetry in the orbital space by the Kondo Hamiltonian \eqref{eq:KondoHamEzSmall}. These divergences signal non-Fermi liquid behavior akin to the physics of the two-channel Kondo effect.

\subsection{Transport Properties}

Next, we consider various transport coefficients of our mesoscopic setup, notably the DC conductance denoted as $G_{ij}^c$ and the spin current conductance represented by $G_{ij}^s$. We remind the reader that the indices $i,j \in \{L,R\}$ correspond to the left and right leads. At the smallest 
temperatures, the DC conductance can be expressed as:
\begin{equation}
    G^c_{ij} = (2\delta_{ij}-1) \underbrace{(2+\mathscr{O}(T^2))}_{G(T)}G_0^c,
\end{equation}
where $ G_0^c = e^2/h$ is the perfect conductance. 
This equation reveals that both channels within the impurity contribute to the charge transfer. 
In parallel, when examining the spin conductance, we arrive at a comparable outcome:
\begin{equation}
G_{ij}^s = (2\delta_{ij}-1)\frac{1}{h} G(T). 
\end{equation}

In the small coupling regime where the impurity is not screened (i.e. above the Kondo temperature $T_K$) we find for both conductances a dependence on the square of the Kondo coupling $\lambda$, which scales as 
\begin{equation}
    \lambda \sim \frac{1}{\ln(\nicefrac{T}{T_K})}.
\end{equation}
The temperature dependence $G(T)$ of the transport coefficients is illustrated in Fig.~\ref{fig:Summary}d).
 
\section{Outlook and Conclusions}
\label{sec:outlookAndConclusions}

In summary, we have studied a floating mesoscopic topological superconductor of symmetry class DIII, which realizes the spinful Majorana Cooper pair box. In contrast to the more prominent spinless Majorana Cooper pair boxes, the present system is subject to strong quantum fluctuations of the non-Abelian order parameter describing the Cooper pair orientation $\hat{\bm d}$ in spin space. We carefully characterized the spectrum of such a box. After coupling the device to external leads, we uncovered a spinful topological Kondo problem in the Coulomb blockade regime. This problem has an SU(2) $\oplus$ SU(2) symmetry in the simplest situation of just two external leads.

We study the spinful topological Kondo problem in the isotropic case, but also in the presence of anisotropies and in the presence of a perturbation polarizing the $\hat{\bm d}$-vector. At weak coupling, we find that the unperturbed anisotropic model flows to isotropy. We argue that this justifies solving the problem at an anisotropic Toulouse point. We use the Abelian Bosonization technique to solve the problem and demonstrate that the unperturbed spinful topological Kondo problem realizes a non-Fermi liquid fixed point and determines its thermodynamic and transport observables. We also determine that this fixed point is unstable to the $\hat{\bm d}$-polarizing  perturbation, i.e. it relies on symmetry protection.

Beyond its apparent relevance in the context of realizing strongly correlated phases in mesoscopic topological devices, we hope that this work with further help understanding the non-trivial interplay of topology and strong correlations in triplet superconductors. Future directions of research could involve the spinful topological Kondo effect with more leads, multichannel versions thereof~\cite{LiVayrynen2023b}, and arrays of spinful Majorana Cooper pair boxes. In the long run, the latter could be valuable emulators of the phases of matter in quantum materials with triplet pairing tendency.

\acknowledgments
It is a pleasure to thank Erez Berg, Matan Lotem, Pavel Ostrovsky, Aline Ramires, Pietro M. Bonetti, Nikolaos Parthenios, Raffaele Mazzilli, Lukas Debbeler, Kirill Alpin und Robin Scholle for useful discussions on the problem. 
JIV thanks the Max Planck Institute for Solid State Research for hospitality.   
JIV was supported by the U.S. Department of Energy, Office of Science, National Quantum Information Science Research Centers, Quantum Science Center. 
EJK acknowledges hospitality by the Kavli Institute for Theoretical Physics, where part of this work was completed. This research was supported in part by the National Science Foundation under Grants No. NSF PHY-1748958 and PHY-2309135.

\appendix

\section{Spectrum of the spinful MCPB}

In this appendix, we present the calculation of the spectrum of the spinful MCPB. The starting point is Hamiltonian \eqref{eq:modelTransformed}, where the model has already been transformed into the co-moving frame of the condensate
\begin{align}
    & \hat{H}_\text{dot} = 4 E_c(\hat{N}_c - N^c_g)^2 + E_s \hat{\bm{J}}^2 \notag \\
    & + \int \text{d}x\left\{ \psi(x)^\dagger\epsilon(\hat{p})\psi(x) + u[\psi(x)^{\text{T}}\partial_x \psi(x) + \hc ] \right\}.
\end{align}

\subsection{Boundary Condition and Quantization of Condensate Operators}
\label{secApx:BoundaryCondition}

First, we demonstrate how the quantization condition of the operator $\hat{N}_c$ and $\hat{\bm{J}}$ change with respect to their transformed counterparts. Let be the Cooper-pair number operator before the transformation that is 
\begin{equation}
    \mathscr{U}^\dagger \left(2\hat{N}_c + \hat{n}_f\right) \mathscr{U} = 2\hat{N}_c.
\end{equation}
Before the transformation, the operator $\hat{N}_c$ is integer quantized and has the eigenstates $\ket{\Phi}_N$ with eigenvalue $N$, which denotes the number of Cooper-pairs. We can project the state into the basis of the superconducting phase, that is $\braket{\varphi|\Phi}_N=\Phi_N(\varphi)$. Note that, due to the missing hat, $\varphi$ is a quantum number and not an operator. Since the phase is only defined up to $2\pi$, the wavefunction has to fulfill the boundary condition 
\begin{equation}
    \Phi_N(\varphi) = \Phi_N(\varphi + 2\pi).
\end{equation}
The transformed wavefunction reads 
\begin{equation}
    \mathscr{U}^\dagger \Phi_N(\varphi) = \text{e}^{
i\frac{\varphi}{2}\hat{n}_f}\Phi_N(\varphi).
\end{equation}
There, we can see that for an even number of electrons, the wave function is $2\pi$ periodic, and for an odd number of electrons $4\pi$ periodic. Thus, the operator $\hat{N}_c$ after the transformation is half-integer quantized since we changed the boundary condition by $\mathscr{U}$.

A similar effect can also be observed in $\hat{\bm{J}}$. The operator $\hat{L}_y = -i\partial_\phi$ has, since it is an angular momentum operator, integer eigenvalues $l_y$ with eigenfunctions $\Psi_{l_y}(\phi)$. These eigenfunctions are also $2\pi$ periodic. After the transformation, the new eigenfunctions read
\begin{equation}
    \mathscr{U}^\dagger \Psi_{l_y}(\phi) = \expp{i\phi\hat{S}_y}\Psi_{l_y}(\phi),
\end{equation}
where these are eigenfunctions of $\hat{J}_y=-i\partial_\phi$ and $\hat{S}_y$ is the spin operator of the fermion excitation in the y-direction. Similar to the superconducting phase and the Cooper-pair number, in the case of an odd fermion number (i.e. $\hat{S}_y$ has half-integer eigenvalues), the eigenfunction of $\hat{J}_y$ are $4\pi$ periodic and have half-integer eigenvalues. 

\subsection{Spectrum and Eigenfunctions}

As already discussed in the main text, the fermion sector does not contribute to the energy and is spanned by a four-dimensional Hilbert space in the case of a single wire $w = 1$.

In the condensate sector, we note that the two operators $\hat{N}_c$ and $\hat{\bm{J}}$ commute with the Hamiltonian, and hence, the spectrum can be constructed using their eigenvalues and eigenfunctions. The charge eigenfunctions are straightforwardly constructed by the eigenvalue equation:
\begin{equation}
\hat{N}_c \ket{\Phi}_N = N \ket{\Phi}N \iff -i\partial_\varphi \Phi_N(\varphi) = N \Phi_N(\varphi),
\end{equation}
where the solutions are $\Phi_N(\varphi) = \expp{iN\varphi}$, and $N$ is a half-integer.

For the angular momentum sector, the energy eigenstates correspond to the eigenstates of the operator $\hat{\bm{J}}^2$. Since $\hat{\bm{J}}$ is an angular momentum/spin operator, we can construct the spectrum and the eigenfunctions using standard methods. That is, we choose the operator $\hat{J}_y$ to commute with $\hat{\bm{J}}^2$, and the eigenstates are given by the following two eigenvalue equations:

\begin{align}
\hat{J}_y \ket{\Xi} & = j_y \ket{\Xi}, \label{eqApx:eigenvalueEquationAngularMomentumY} \\
\hat{\bm{J}}^2 \ket{\Xi} & = J(J+1) \ket{\Xi}. \label{eqApx:eigenvalueEquationsAngularMomentum}
\end{align}

To construct the corresponding differential equation, we note that $\hat{\bm{L}}$ is the angular momentum operator that is canonically conjugate to $\hat{\bm d}$. Thus, in terms of coordinates, the $\hat{L}$ operator can be written as:

\begin{equation}
\hat{\bm{L}} = -i\hat{\bm{d}}\times \nabla = -i\hat{\bm{d}}\times \left(\bm{e}_\theta \partial_\theta + \frac{\bm{e}_\phi}{\sin(\theta)}\partial_\phi\right),
\end{equation}
where $\bm{e}_i$ are unit vectors. Note that in coordinate space, $\hat{\bm{d}} \equiv \bm{e}_r$. From this, we can calculate:
\begin{equation}
\mathscr{U}^\dagger \hat{\bm{L}} \mathscr{U} = \hat{\bm{L}} - \bm{e}_\phi \hat{S}_x + \cot(\phi)\bm{e}_\theta \hat{S}_y - \bm{e}_\theta \hat{S}_z.
\end{equation}
On the other hand, the fermion spin transforms as:
\begin{equation}
\mathscr{U}^\dagger \hat{\bm{S}} \mathscr{U} = \bm{e}_r\hat{S}_y + \bm{e}_\phi \hat{S}_x + \bm{e}_\theta \hat{S}_z.
\end{equation}
Now, we can conclude:
\begin{equation}
\mathscr{U}^\dagger \left(\hat{\bm{L}} + \hat{\bm{S}}\right)\mathscr{U} = \hat{\bm{L}} + \frac{\hat{S}_y}{\sin(\hat\theta)}
\begin{pmatrix}
\sin(\hat\phi) \\
0 \\
\cos(\hat\phi)
\end{pmatrix}\equiv \hat{\bm{J}}.
\end{equation}

In order to solve the two eigenvalue equations \eqref{eqApx:eigenvalueEquationAngularMomentumY} and \eqref{eqApx:eigenvalueEquationsAngularMomentum}, we use the separation of variables as an Ansatz, that is, $\braket{\theta, \phi|\Xi}=\Xi(\theta,\phi)=\Theta(\theta)\Psi(\phi)$. This Ansatz leads to the equation:

\begin{equation}
-i\partial_\phi \Psi(\phi) = j_y \Psi(\phi)
\end{equation}
with the solution $\Psi(\phi) = \expp{ij_y \phi}$. Note that $\hat{S}_y$ also commutes with the Hamiltonian. Therefore, we can replace the operator $\hat{S}_y$ with its eigenvalue $s_y$. The second eigenvalue equation yields the differential equation:

\begin{equation}
\begin{split}
\, & \left[(1-x^2)\partial^2_x -  2x\partial_x - \frac{(j_y - s_y x)^2}{1-x^2}\right. \\ & + \left(J(J+1) - s_y^2\right)\biggr]P(x) = 0,
\label{eqApx:eigenvalueEquationJCoordinates}
\end{split}
\end{equation}
where $x=\cos(\theta)$ and $P(x) = P(\cos(\theta))=\Theta(\theta)$. This corresponds to the eigenvalue equation of ``monopole harmonics'', i.e. the generalization of spherical harmonics to $s_y \neq 0$.\cite{Shnir2006}

Assuming $N_g^c=\nicefrac{1}{2}$, the ground state manifold is given by the second row in table \ref{tab:spectrum}. Inserting the corresponding quantum numbers in equation \eqref{eqApx:eigenvalueEquationJCoordinates}, we find the normalizable solutions:

\begin{equation}
\Theta(\theta) = \sqrt{1 + 4j_y s_y\cos(\theta)}.
\end{equation}
Thus, the full wave function of the condensate (charge and spin part) in the ground state manifold reads:
\begin{equation}
\Phi_{\frac{1}{2}}(\varphi)\Xi_{s_y, j_y}(\phi, \theta) = \expp{i\varphi/2}\expp{ij_y\phi}\sqrt{1+4j_y s_y \cos(\theta)}.
\end{equation}
In the first excited states, the condensate wavefunction takes a trivial shape:
\begin{equation}
\Phi_{N={0, 1}}(\varphi)\Xi_{0, 0}(\phi,\theta) = \exp(iN\varphi) \cdot 1.
\end{equation}

\subsubsection{Spectrum in the limit of a strong perturbation \label{AppSubSecZeeman}}
We now consider a regime in which the perturbation $E_z$ is large and
find the spectrum in the limit $E_z \gg E_s$, and hence, $\theta \ll 1$. To do so, we expand the Hamiltonian:
\begin{equation}
\begin{split}
\hat{H} & = E_s \hat{\bm{J}}^2 - E_z \hat{d}_y \\
& = E_s \left(\frac{-\partial_\theta(\sin(\theta)\partial_\theta \cdot)}{\sin(\theta)} + \frac{j_y^2 - 2s_y j_y\cos(\theta)+s_y^2}{\sin^2(\theta)}\right) \\ & - E_z \cos(\theta)
\\ & \approx E_s\left(-\frac{\partial_\theta (\theta \partial_\theta \cdot)}{\theta} + \frac{n^2}{\theta^2}\right) + \frac{1}{2}E_z \theta^2,
\label{appEq:newThetaDiffEq}
\end{split}
\end{equation}
up to the first non-trivial order in $\theta$ and ignoring the constant shift in energy. Note that $n=j_y - s_y$. After applying the approximation $\theta\ll1$, the differential equation for the $\Theta(\theta)$ part of the condensate wavefunction changed, which, therefore, needs to be adopted. The Hamiltonian \eqref{appEq:newThetaDiffEq} resembles the radial part of a harmonic oscillator. The spectrum for the 2-dimensional harmonic oscillator in polar coordinates is well known and reads
\begin{equation}
E_n = \omega(|n| + 1 + 2n_\theta),
\end{equation}
with energy eigenstates:
\begin{equation}
    \Theta_{n,n_\theta}(\theta) = C_{n, n_\theta}\theta^{|n|}\exp\left(-\frac{m\omega}{2}\theta^2\right)L_{n_\theta}^{n}(m\omega \theta^2).
\end{equation}
In analogy to the harmonic oscillator we identify $\omega = \sqrt{2E_zE_s}$ and $m=\frac{1}{2E_s}$. The quantum number $n_\theta$ describes the quantization of radius in the $x$-$z$-plane. However, in the low-energy sector, we can choose $n_\theta=0$. $C_{n,n_\theta}$ is a normalization constant, and $L_{n_\theta}^{n}$ are the generalized Laguerre polynomials. The total condensate wave function is most conveniently expressed in terms of $n$ and $s_y$ as 
\begin{equation}
    \Xi_{n, s_y}(\theta, \phi) = \Theta_{n, 0}(\theta)\expp{i(n+s_y)\phi}.
\end{equation}
The ground state has $n=0$, from which it follows that $s_y = j_y =\pm\nicefrac{1}{2}$. Hence, the ground state manifold is twofold degenerate. The first excited states have $n=\pm 1$ and are, therefore, 4 states. 
Furthermore, it follows immediately that:
\begin{equation}
\bra{\Xi_{0,s_y}}\delta\hat{d}_{i}\ket{\Xi_{0,s_y}} = 0.
\end{equation}

\section{Second order perturbation theory and emergent Kondo effect}
\label{app:Superex}

In this section, we describe how we obtained an effective Kondo Hamiltonian in the low-energy sector. First, we demonstrate the application of the Schrieffer-Wolf transformation to the warm-up problem in the limit where $E_c \gg E_s$, and we also assume $E_z \gg E_s$. This will be followed by a discussion of how to obtain the effective Kondo Hamiltonian in the limit where $E_c \sim E_s$ and $E_z \ll E_c$.

To derive the effective low-energy Hamiltonian, we apply a Schrieffer-Wolf transformation, treating the hopping parameter $t$ of lead electrons onto the quantum dot as a perturbation, which implies that $E_c \gg t$. 

\subsection{Limit $E_c \gg E_s$}

In this limiting case, the fluctuations in the superconducting phase ($\hat{\varphi}$) are much faster than the spin fluctuations ($\hat U$). Thus, we absorb the $\hat U$ matrix into the lead electrons:
\begin{equation}
\Psi_i = \hat U\psi_i,
\end{equation}
where $\psi_i = (\psi_{i\uparrow}, \psi_{i\downarrow})$ are spinors of the lead electrons. The hopping Hamiltonian becomes:
\begin{equation}
- t \hat{H}_t = -t \sum_{i} \left(\Psi^\dagger_i \expp{-i\hat{\varphi}/2}\gamma_i + \text{H.c.}\right).
\end{equation}
The low-energy sector is mostly determined by the charge on the island. If we choose $N_g^c = \frac{1}{2}$, the ground state manifold in the charge sector can be classified by the total charge $\ket{GS} = \ket{N_c = \nicefrac{1}{2}}$. The states with the next higher energy are the states where a fermion takes some charge from the condensate and tunnels into an electron state in the wire or an electron that tunnels into the MZMs and donates its charge to the condensate. These states can be written as $\ket{0}$ and $\ket{1}$, respectively.

Applying the Schrieffer-Wolf transformation, Hamiltonian $\eqref{eq:fullBareHamiltonian}$ becomes:
\begin{align}
\hat{H} = \hat{H}_\text{dot} + \hat{H}_0 + H_z \underbrace{- \frac{t^2}{E_c}\bra{\nicefrac{1}{2}}\sum_{n=0,1}\hat{H}_{t}\ket{n}\bra{n}\hat{H}_t \ket{\nicefrac{1}{2}}}_{\Delta \hat H} .
\end{align}
The last term will become the effective Kondo Hamiltonian. Hence, we will focus on this term and drop the other parts of the Hamiltonian.
If we expand the Hamiltonian, it becomes:
\begin{widetext}
\begin{equation}
\begin{split}
\, & \Delta\hat H = -\frac{t^2}{E_c}\sum_{i, j}\bra{\frac{1}{2}}\left[\left(\Psi^\dagger_i \expp{-i\hat{\varphi}}\gamma_i + \text{H.c.}\right)\ket{0}\bra{0}\left(\Psi^\dagger_j \expp{-i\hat{\varphi}/2}\gamma_j + \text{H.c.}\right)\right. \\
& + \left.\left(\Psi^\dagger_i \expp{-i\hat{\varphi}}\gamma_i + \text{H.c.}\right)\ket{1}\bra{1}\left(\Psi^\dagger_j \expp{-i\hat{\varphi}/2}\gamma_j + \text{H.c.}\right)\right]\ket{\frac{1}{2}}.
\end{split}
\end{equation}
\end{widetext}

The operator $\expp{\pm i\hat{\varphi}/2}$ are translation operators in charge space that destroy (create) charge in the condensate. The matrix elements of the charge translation operators are straightforwardly calculated and read
\begin{equation}
    \begin{cases}
    \bra{1/2}e^{-i\hvarphi/2}\ket{0}=0, \quad\quad\quad  \bra{1/2}e^{i\hvarphi/2}\ket{0}=1 \\
    \bra{0}e^{-i\hvarphi/2}\ket{1/2} = 1, \quad\quad\quad \bra{0}e^{i\hvarphi/2}\ket{1/2}= 0 \\
    \bra{1/2}e^{-i\hvarphi/2}\ket{1} = 1, \quad\quad\quad \bra{1/2}e^{i\hvarphi/2}\ket{1} = 0 \\
    \bra{1}e^{-i\hvarphi/2}\ket{1/2} = 0, \quad\quad\quad \bra{1}e^{i\hvarphi/2}\ket{1/2} = 1.
    \end{cases}
\end{equation}
Applying these rules, the effective Hamiltonian collapses to:
\begin{align}
\Delta\hat{H} = & -\frac{t^2}{E_c}\sum_{i,j}\left(\gamma^\text{T}_i\Psi_i^{\phantom{\dagger}} \Psi_j^\dagger \gamma_j^{\phantom{T}} + \Psi^\dagger_i \gamma_i^{\phantom{\dagger}} \gamma_j^\text{T} \Psi^{\phantom{\dagger}}_j\right).
\end{align}
This expression is the origin of Eq.~\eqref{eq:KondoGeneralw} of the main text, note that $i = 1, \dots, w$.

In the case $w = 1$ we use the notation
\begin{align}
\Delta\hat{H} =  & - \frac{2t}{E_c} \Psi^\dagger (\gamma \cdot\gamma^\text{T}) \Psi,
\end{align}
where we introduced the four-component spinor $\Psi = (\Psi_{L\uparrow}, \Psi_{L\downarrow}, \Psi_{R\uparrow}, \Psi_{R\downarrow})^\text{T}$ and the symbol $(\gamma \cdot\gamma^\text{T})$ that represents the matrix:
\begin{equation}
\begin{split}
(\gamma \cdot \gamma^\text{T}) & =
\begin{pmatrix}
1 & \gamma_{L\uparrow}\gamma_{L\downarrow} & \gamma_{L\uparrow}\gamma_{R\uparrow} & \gamma_{L\uparrow}\gamma_{R\downarrow} \\ \gamma_{L\downarrow}\gamma_{L\uparrow} & 1 & \gamma_{L\downarrow}\gamma_{R\uparrow} & \gamma_{L\downarrow}\gamma_{R\downarrow} \\ \gamma_{R\uparrow}\gamma_{L\uparrow} & \gamma_{R\uparrow}\gamma_{L\downarrow} & 1 & \gamma_{R\uparrow}\gamma_{R\downarrow} \\
\gamma_{R\downarrow}\gamma_{L\uparrow} & \gamma_{R\downarrow}\gamma_{L\downarrow} & \gamma_{R\downarrow}\gamma_{R\uparrow} & 1
\end{pmatrix} \\
& = -2
\begin{pmatrix}
0 & -i\hat{S}_y & i\hat{S}_z & i\hat{S}_x \\
i\hat{S}_y & 0 & i\hat{S}_x & -i\hat{S}_z \\
-i\hat{S}_z & -i\hat{S}_x & 0 & -i\hat{S}_y \\
-i\hat{S}_x & i\hat{S}_z & i\hat{S}_y & 0 \\
\end{pmatrix}
+ \mathds{1}_{4x4}.
\end{split}
\end{equation}
Decomposing this matrix further into Pauli matrices yields:
\begin{equation}
(\gamma \cdot \gamma^\text{T}) = -2\left(\mathds{1}_\tau \sigma_y \hat{S}_y - \tau_y \sigma_z \hat{S}_z - \tau_y \sigma_x \hat{S}_x\right) +\mathds{1}_\tau \mathds{1}_{\sigma},
\end{equation}
where $\sigma_{i}$ and $\tau_i$ act in the spin and left/right space, respectively. In the next step, we change into the eigenbasis of $-\tau_y$ and obtain:

\begin{equation}
\Delta\hat{H} = \frac{4t^2}{E_c}
\begin{pmatrix}
\Psi_+^\dagger & \Psi_-^\dagger
\end{pmatrix}
\begin{pmatrix}
\bm{\sigma} & 0 \\
0 & \bm{\sigma}
\end{pmatrix}
\begin{pmatrix}
\Psi_+ \\ \Psi_-
\end{pmatrix}
\cdot \boldsymbol{\hat{S}} - \frac{2t^2}{E_c}\Psi^\dagger \Psi,
\end{equation}
where $\Psi_+ = \frac{1}{\sqrt{2}}(\Psi_L + i\Psi_R)$ and $\Psi_- = \frac{i\sigma_y}{\sqrt{2}}(\Psi_L - i\Psi_R)$. The vector $\boldsymbol{\hat{S}} = (\hat{S}_x, \hat{S}_y, \hat{S}_z)$ contains the bilinears of the MZMs. The second term is nothing but a potential term that we will omit from now on. Separating the $U$ matrix from the spinors, the Kondo Hamiltonian evaluates to:
\begin{equation}
\hat{H}_K = \lambda \left(\psi_+^\dagger \hat U^\dagger \boldsymbol{\sigma} \hat U \psi_+^{\phantom{\dagger}} + \psi^\dagger_- \sigma_y \hat U^\dagger \sigma_y \boldsymbol{\sigma}\sigma_y \hat U \sigma_y \psi_-^{\phantom{\dagger}}\right)\cdot\boldsymbol{\hat{S}},
\end{equation}
where $\lambda = \frac{4t^2}{E_c}$, $\psi_+=\frac{1}{\sqrt{2}}\left(\psi_R + i\psi_L\right)$ and $\psi_-=\frac{i\sigma_y}{\sqrt{2}}\left(\psi_R - i\psi_L\right)$.

\subsection{Limit $E_z \gg E_s$}

In this section, we build upon the derivations from the previous section, assuming that $E_z \gg E_s$, and that $\hat{\bm{d}}$ only experiences weak fluctuations around the y-axis. Our strategy is to expand the unitary matrix $\hat U$ in terms of $\htheta$. We introduce a new representation of $\hat U$ using Pauli matrices to achieve this. We utilize the gauge freedom within $\hat U$ and redefine it as follows:
\begin{equation}
\hat U = \expp{-i\hphi/2 \sigma_y} \expp{i\htheta/2 \sigma_x} \expp{i\hphi/2 \sigma_y} = \expp{i \hat{W}/2},
\end{equation}
where
\begin{equation}
\hat{W} = \underbrace{\hat{\theta}\cos(\hphi)}_{\partial_\phi \delta \hat d_x}\sigma_x - \underbrace{\htheta\sin(\hphi)}_{\partial_\phi \delta \hat d_x}\sigma_z
\end{equation}
We can now expand $\hat U$ in powers of $\htheta$ since the fluctuations in this angle are small. Therefore, up to the first order, we obtain
\begin{equation}
\begin{split}
\hat U^\dagger \boldsymbol{\sigma} \hat U & \approx \left(\mathds{1} - \frac{i}{2}\hat{W}\right)\boldsymbol{\sigma}\left(\mathds{1} + \frac{i}{2}\hat{W}\right) \\ & = \boldsymbol{\sigma} + \partial_\phi \delta\hat{\bm d} \times \bm{\sigma} + \mathscr{O}(\htheta^2),
\end{split}
\end{equation}
where $\delta\hat{\bm{d}} = (\delta\hat d_x, 0, \delta\hat d_z)^\text{T}$. Similarly, we can calculate
\begin{equation}
\hat U^\dagger \sigma_y \bm{\sigma} \sigma_y \hat U \approx \sigma_y (\bm{\sigma} - \partial_\phi \delta\hat{\bm d}\times\bm{\sigma}) \sigma_y.
\end{equation}
The Hamiltonian can be expressed compactly as
\begin{equation}
\hat{H}_K = \lambda \sum_{a=\pm} \psi^\dagger_a \bm{\sigma} \psi_a \cdot \left(1 - a\,\partial_\phi \delta\hat{\bm d}\times\right)\hat{\bm{S}}.
\end{equation}
This is the origin of Eq.~\eqref{eq:KondoEzLarg} of the main text.

\subsection{Limit $E_z \ll E_s$}

In this limit, we also consider transition in higher excited states induced by angular momentum and orbital fluctuations. We chose the third row in Table~\ref{tab:spectrum} to be the ground state manifold (again setting $N_g^c = \frac{1}{2}$). The energy ground states can be written as 
\begin{align}
\begin{split}
    \ket{GS} & = \underbrace{\ket{N_c=\frac{1}{2}, j_y, J=3/4}_{s_y}}_{\text{condensate sector}}\otimes\underbrace{\ket{s_y}}_{\text{fermion sector}} \\
    & \equiv \ket{\frac{1}{2}, j_y, s_y},
\end{split}
\end{align}
where $j_y = \pm \nicefrac{1}{2}$ and $s_y = \pm \nicefrac{1}{2}$ are the eigenvalues of the two operator $\hat{J}_y$ and $\hat{S}_y$, respectively. As it will be shown, the ground state wavefunction of the condensate sector also depends on $s_y$ due to the implicit dependence of $\hat{\bm{J}}$ on $\hat{S}_y$. The two excited states can be written in a similar fashion as 
\begin{equation}
    \ket{l, k = \{0, 2\}} \equiv \ket{N_c=l, j_y = 0, J = 0}\otimes\{\ket{0}, \ket{\Gamma^\dagger_\uparrow}\Gamma^\dagger_\downarrow\ket{0}\}.
\end{equation}

The effective Hamiltonian after the Schrieffer-Wolf transformation reads 
\begin{equation}
   \Delta \hat{H} = -t^2 \sum_{\xi, \eta}\ket{GS}_\xi \bra{GS}_\xi \sum_{l, k} \frac{\hat{H}_t\ket{l, k}\bra{l, k}\hat{H}_t}{\Delta E_l}\ket{GS}_\eta \bra{GS}_\eta,
\end{equation}
where $\Delta E_l =  E_l - E_0$. $E_l$ is the energy of the first excited state, and $E_0$ is the ground state energy.  The projector onto a ground state can also compactly be written as
\begin{equation}
    \begin{split}
    \ket{GS}_{s_y}\bra{GS}_{s_y} & = \ket{\frac{1}{2}, j_y, s_y}\bra{\frac{1}{2}, j_y, s_y} \\ & = \ket{\frac{1}{2}, j_y}_{s_y}\bra{\frac{1}{2}, j_y}_{s_y}\otimes\underbrace{\ket{s_y}\bra{s_y}}_{\Pi_{s_y}}.
    \end{split}
\end{equation}
Now, we can re-write the effective Hamiltonian as
\begin{equation}
   \Delta \hat{H} = -t^2 \sum_{m, m'} \ket{\frac{1}{2}, m}\Delta H^{m,m'}\bra{\frac{1}{2}, m'}, 
\end{equation}
where 
\begin{align}
\begin{split}
    \, &\Delta H^{m,m'}  = -t^2   \sum_{l, k,\xi,\eta} \Pi_\xi \bra{\frac{1}{2}, m}_\xi \frac{\hat{H}_t\ket{l,k}\bra{l,k}\hat{H}_t}{\Delta E_l}\ket{\frac{1}{2}, m'}_\eta \Pi_\eta \\ 
    & = -t^2 \sum_{l,\xi,\eta
    } \Pi_\xi \bra{\frac{1}{2}, m}_\xi  \frac{\hat{H}_t\ket{l}\left(\sum_k\ket{k}\bra{k}\right)\bra{l}\hat{H}_t}{\Delta E_l}\ket{\frac{1}{2}, m'}_\eta \Pi_\eta,
\end{split}
\end{align}
where $\sum_k \ket{k}\bra{k} = \mathds{1}$ in the subspace of even electron parity.
The sum over the charge sector is equivalent to what was done in the last section. The effective Hamiltonian becomes
\begin{align}
\begin{split}
    \Delta H^{m,m'} & = - \frac{t^2}{\Delta E} \Pi_\xi \sum_{i,j} \left\{\bra{m}_\xi\gamma_j^\text{T} \hat U \psi_j\ket{0}\bra{0}\psi_i^\dagger \hat U^\dagger \gamma_i\ket{m'}_\eta \right.\\
    & \left. + \bra{m}_\xi \psi^\dagger_i \hat U^\dagger \gamma_i \ket{0}\bra{0} \gamma_j^T \hat U \psi_j \ket{m'}_\eta \right\} \Pi_\eta,
    \end{split}
    \label{eqqApx:HeffChargeIntegratedOut}
\end{align}
where $\Delta E = E_c - E_s\frac{3}{4}$.
In order to evaluate the matrix elements that appear in equation \eqref{eqqApx:HeffChargeIntegratedOut} we introduce a partition of unity of the form 
$\mathds{1} = \int_{0}^\theta \text{d}\theta \int_{0}^{2\pi}\text{d}\phi \, \frac{\sin(\theta)}{4\pi}\, \ket{\phi,\theta}\bra{\phi,\theta}$. Also we decompose the lead electrons in eigenvectors of the $\sigma_y$ matrix, that is $\psi_i = \sum_{n=\pm} \psi_n \bm{e}_n$, where $\bm{e}_n = \frac{1}{\sqrt{2}}(1, i n)^\text{T}$ and $n$ being the eigenvalue $n=\pm1$. Applying these two steps the Hamiltonian \eqref{eqqApx:HeffChargeIntegratedOut} becomes
\begin{align}
    \begin{split}
    \, \Delta H^{m, m'} & = -\frac{t^2}{\Delta E} \Pi_\xi \sum_{ij} \left\{\gamma_i^T U_\theta^{\xi m} \bm{e}^{\phantom{\dagger}}_m\psi^{\phantom{\dagger}}_{i,m}\psi^\dagger_{j,m}\bm{e}_{m'}^\dagger\left(U^{\eta m'}_\theta \right)^\dagger \gamma_{im'} \right. \\
    & + \left.\psi^\dagger_{i\bar{m}}\bm{e}_{\bar{m}}^\dagger \left(U^{\xi m}_\theta\right)^\dagger \gamma_i^{\phantom{\text{T}}} \gamma_j^\text{T} U^{\eta m'}_\theta \bm{e}_{\bar{m'}}\psi_{j,\bar{m'}} \right\} \Pi_\eta,
    \end{split}
\end{align}
where $\bar{m}=-m$ and
\begin{equation}
U_\theta^{\xi m} = \braket{m|_\xi 
\hat U|0} = \int\frac{\text{d}\theta\sin(\theta)}{2}\braket{m|_\xi\theta}\expp{i\theta/2 \sigma_x}.
\end{equation}
The projector in the orbital/Majorana space can explicitly written in terms of Majorana fermions as
\begin{equation}
    \Pi_\xi = \frac{1+2\xi \hat{S}_y}{2} = \frac{1}{2}\left(1 + \xi\frac{i}{2}\gamma^\text{T}_i\sigma_y\gamma_i\right),
\end{equation}
where $i$ can either be $R$ or $L$. Due to the parity constraint, both are equally good.
As we did for the lead electrons, we expand the Majorana fermions in terms of eigenvectors of $\sigma_y$. It is straightforward to show that
\begin{align}
    \bm{e}^\dagger_n  \cdot\gamma_i^\phd \Pi_\xi & =  \delta_{\xi n}\gamma_{in} ,\\
    \bm{e}^\dagger_n \cdot \Pi_\xi \gamma_i & = \delta_{\bar{\xi}n}\gamma_{in} ,\\
    \gamma_i^\text{T} \Pi_\xi \cdot \bm{e}_n^\phd & = \delta_{\xi\bar{n}}\gamma_{i\bar{n}}, \\
    \Pi_\xi \gamma_i^\text{T} \cdot \bm{e}_n & = \delta_{\bar{\xi}\bar{n}}\gamma_{i\bar{n}}.
\end{align}
Now, we are in a position to evaluate the sums of the shape
\begin{align}
\begin{split}
    \, & \sum_\xi \Pi_{\xi}\gamma^\text{T}_{in}\bm{e}_{n}^\phd\bm{e}^\dagger_{n'}U^{\xi m}_\theta \bm{e}_m  \\ & = \sum_\xi \int_0^\pi \frac{\sin(\theta)}{2} \braket{m|_\xi \theta} \underbrace{\bm{e}^\dagger_n \text{e}^{i\sigma_x \nicefrac{\theta}{2}}\bm{e}_m^\phd}_{\cos(\theta/2)\delta_{nm} - 2m\sin{\theta/2}\delta_{n\bar{m}} }\gamma_{i\bar n}\delta_{\bar \xi \bar n} \\
    & =  \int_0^\pi \frac{\sin(\theta)}{2} \Big(\bra{m}_m\ket{\theta}\cos(\theta/2)\delta_{nm}  \\
    &-2m\bra{m}_{\bar{m}} \ket{\theta}\sin(\theta/2)\delta_{n\bar{m}}\Big)\gamma_{i\bar n}.
\end{split}
    \label{eqApx:sum}
\end{align}
The two integrals are
\begin{equation}
    \int_0^\pi \text{d}\theta\, \frac{\sin(\theta)}{2} \sqrt{1 + \cos{\theta}}\cos(\theta/2) = \frac{1}{\sqrt{2}}
\end{equation}
and 
\begin{equation}
            \int_0^\pi \text{d}\theta\, \frac{\sin(\theta)}{2} \sqrt{1 - \cos{\theta}}\sin(\theta/2) = \frac{1}{\sqrt{2}}.
\end{equation}
After calculating the other three sums of the shape of sum \eqref{eqApx:sum} the effective Hamiltonian becomes
\begin{align}
    \begin{split}
    \, & \Delta H^{m.m'} = -\frac{t^2}{2 \Delta E}\sum_{i, j}\Big(\psi^\phd_{im}\psi^\dagger_{im'}(\gamma_{i\bar{m}}-2m\gamma_{im})  \\
    &(\gamma_{jm'} - 2m'\gamma_{j\bar{m'}}) \\ & +
     \psi^\dagger_{i\bar{m}}\psi^\phd_{j\bar{m'}}(\gamma_{i\bar{m}}+2m\gamma_{im})(\gamma_{jm'} + 2m'\gamma_{j\bar{m'}}) \Big).
    \end{split}
\end{align}
In the next step, we evaluate each matrix element in angular momentum space separately. That is, we set up the matrix
\begin{equation}
    \Delta H = \begin{pmatrix}
    \Delta H^{++} & \Delta H^{+-} \\
    \Delta H^{-+} & \Delta H^{--} 
    \end{pmatrix}.
\end{equation}
The matrix elements can be written in a compact way as
\begin{align}
    \begin{split}
    \Delta H^{++} & = -\frac{2t^2}{\Delta E} + \frac{t^2}{\Delta E}\psi^\dagger \sigma_y \psi +\frac{t^2}{\Delta}\psi^\dagger \sigma_y \tau_y \psi Z, \\
    \Delta H^{--} & = - \frac{2t^2}{\Delta E} - \frac{t^2}{\Delta E} \psi^\dagger \sigma_y \psi + \frac{t^2}{\Delta E} \psi^\dagger \sigma_y \tau_y \psi Z, \\
    \Delta H^{+-} & = \frac{2t^2}{\Delta E}\left( \psi^\dagger\sigma_-\psi Y - i\psi^\dagger\tau_y \sigma_-\psi\right), \\
    \Delta H^{-+} & = \frac{2t^2}{\Delta E}\left( \psi^\dagger\sigma_+\psi Y + i\psi^\dagger\tau_y \sigma_+\psi\right),
    \end{split}
\end{align}
where we introduced the spinor $\psi = (\psi_{L\uparrow},\psi_{L\downarrow},\psi_{R\uparrow}, \psi_{R\downarrow})^\text{T}$ and $\tau_i$ are Pauli matrices that act on the $R/L$-space. Furthermore, we introduced the matrices $\sigma_\pm = \frac{1}{2}(\sigma_z \pm i\sigma_x)$
The Kondo Hamiltonian can be found by decomposing the matrix in Pauli-matrices, which correspond to the components of $\bm{J}$ projected down onto the ground state manifold, that is
\begin{equation}
    \Delta\hat{H} = \sum_{i=0}^{3}\frac{1}{2}\text{Tr}\left[\Delta H J_i\right]J_i.
\end{equation}
Such a decomposition leads to (neglecting inessential constants)
\begin{align}
    \begin{split}
    \hat{H}_K & = \frac{t^2}{\Delta E}\left(\psi^\dagger \tau_y \sigma_y \psi Z + \psi^\dagger \sigma_y \psi J_y\right. \\ & +
    \left(\psi^\dagger\sigma_x\psi Y  + \psi^\dagger\tau_y\sigma_z X\right)J_x \\
    & + \left.\left(\psi^\dagger \sigma_z \psi Y + \psi^\dagger \tau_y \sigma_x \psi X \right)J_z\right) \\
    & = \frac{t^2}{\Delta E}\left( \psi\bm{\sigma}\psi + \psi^\dagger\bm{\sigma}\tau_y\psi Z J_y\right)\cdot\mathscr{S},
    \end{split}
\end{align}
where
\begin{equation}
    \mathscr{S} = \begin{pmatrix}
    YJ_x \\ J_y  \\ YJ_z
    \end{pmatrix}
\end{equation}
and $\bm{\sigma}$ is a vector with the Pauli matrices as components. Note that we chose the y-axis as the quantization axis, which means that in angular momentum space
\begin{equation}
    J_y = \begin{pmatrix}
    1 & 0 \\ 0 & -1
    \end{pmatrix},\quad
    J_z = 
    \begin{pmatrix}
    0 & 1\\ 1 & 0
    \end{pmatrix}\quad\text{and}\quad
    J_x = \begin{pmatrix}
    0 & -i \\ i & 0
    \end{pmatrix}.
\end{equation}

Further progress can be made by changing to a basis in which 
\begin{equation}
    -\tau_y = \begin{pmatrix}
    1 & 0 \\ 0 & -1
    \end{pmatrix}
\end{equation}
leaving us with the Hamiltonian 
\begin{equation}
    \hat H_K = \frac{2t^2}{\Delta E}\sum_{a=\pm} \psi^\dagger_a \bm{\sigma} \psi^\phd_a \cdot \frac{1 - aZ J_y}{2}\mathscr{S},
\end{equation}
where $\psi_+=\nicefrac{1}{\sqrt{2}}\left(\psi_L + i\psi_R\right)$ and $\psi_-=\nicefrac{i\sigma_y}{\sqrt{2}}\left(\psi_L - i\psi_R\right)$. Finally, we diagonalize the matrix structure of $\hat H_K$ with respect to projector $\frac{1-a Z J_y}{2}$ by applying the transformation $T=J_y\expp{i\frac{\pi}{4}X J_y}$ which yields the Kondo part of Hamiltonian \eqref{eq:KondoHamEzSmall} 
\begin{equation}
    \hat H_K = \frac{4t^2}{\Delta E} \sum_{a=\pm} \psi^\dagger_a \bm{\sigma} \psi_a^\phd \frac{1-aY}{2}\cdot\hat{\bm{J}},
    \label{eqApx:Kondo}
\end{equation}
where $\hat{\bm{J}} = (\nicefrac{J_x}{2}, \nicefrac{J_y}{2}, \nicefrac{J_z}{2})^\text{T}$ is a vector of angular momentum operator. 

The operator $\hat{H}_z = -E_z\hat{n}_y$ can be projected onto the ground state manifold as well. Again, we introduce the partition unity $\mathds{1}=\int_0^\pi \text{d}\theta\int_0^{2\pi} \text{d}\phi \frac{\sin(\theta)}{4 \pi}\ket{\phi,\theta}\bra{\phi, \theta}$. Applying one identity from both sides on $\hat H_z$ gives the matrix 
\begin{equation}
   \hat H_z = -E_z \begin{pmatrix}
    I_1 & I_3 & 0 & 0 \\
    I_3 & I_2 & 0 & 0 \\
    0 & 0 & I_2 & I_3 \\
    0 & 0 & I_3 & I_1 \\
    \end{pmatrix}
    = - \frac{E_z}{3} J_y Y,
\end{equation}
where 
\begin{equation}
    \begin{cases}
    I_1 = \int_0^\pi \frac{\text{d}\theta}{2} (1 + \cos(\theta))\cos{\theta} = \frac{1}{3}, & m=\xi=\eta \\
    I_2 = \int_0^\pi \frac{\text{d}\theta}{2} (1 - \cos(\theta))\cos{\theta} = -\frac{1}{3}, & m\neq\xi=\eta \\
    I_3 = \int_0^\pi \frac{\text{d}\theta}{2} \sin(\theta)\cos{\theta} = 0, & m=\xi\neq\eta.
    \end{cases}
\end{equation}
Applying the transformation $T$ yields
\begin{equation}
   \hat H_z =  - \frac{E_z}{3}Z.
   \label{eqApx:projectedHz}
\end{equation}
Adding the Hamiltonians \eqref{eqApx:projectedHz} and \eqref{eqApx:Kondo} gives rise the the effective Hamiltonian $\eqref{eq:KondoHamEzSmall}$.

\section{Poor Man's Scaling}
\label{app:PoorMan}

We employ a small coupling Renormalization Group (RG) scheme to Eq.~\eqref{eq:anisotropicKondoHam}. The high-energy states are integrated out in second-order perturbation theory. We assume the lead electrons can only access states with energies $D$ above and below the Fermi energy. One RG step reduces the bandwidth $D$ by $\delta D$.

\begin{figure}
    \centering
    \includegraphics{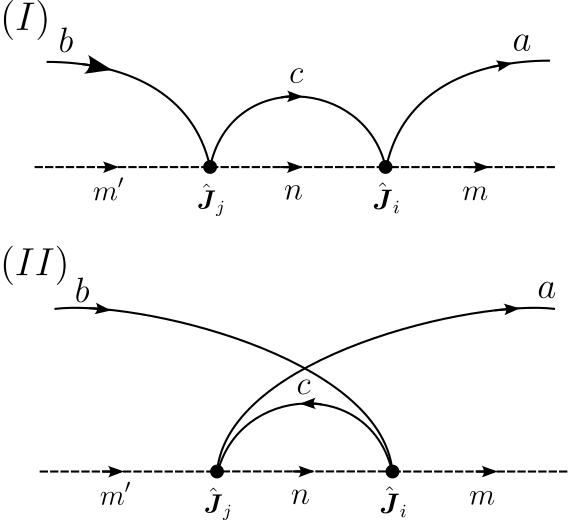}
    \caption{Scattering processes that contribute to the renormalization of the Kondo Hamiltonian $\Delta \hat{H}$ are as follows:
    (I) Particle excitation: In this process, a low-energy electron state $\ket{b}$ scatters at the impurity, denoted by $\hat{\bm{J}}$, into a high-energy state $\ket{c}$ which, in turn, scatters back into a low-energy state $\ket{a}$.
    (II) Hole excitation: In this scenario, a low-energy state $\ket{b}$ scatters into a high-energy hole state $\ket{c}$. The hole then scatters back into a low-energy electron state $\ket{a}$.}
\label{fig:diagrams}
\end{figure}

In second-order perturbation theory, the Hamiltonian, after integrating out the high-energy modes, takes the form:
\begin{equation}
\hat{H} = \hat{H}_0 + \hat{H}_K + \Delta \hat{H},
\end{equation}
where
\begin{equation}
\Delta \hat{H} = \ket{a}\underbrace{\bra{a}\Delta \hat{H} \ket{b}}_{\Delta H_{ab}}\bra{b}.
\end{equation}
Here, $\hat{H}_0$ is the free Hamiltonian of the lead electrons, and $\hat{H}_K$ is the Kondo Hamiltonian \eqref{eqApx:Kondo}. The states $\ket{a/b}$ are energy eigenstates with energies $E_{a/b}\in[-D+\delta D, D - \delta D]$. The matrix
\begin{equation}
\Delta H_{ab} = \bra{a}\frac{1}{2}\left(\hat{T}(E_a) + \hat{T}(E_b)\right)\ket{b} \stackrel{E_a\approx E_b = E}{=} \bra{a}\hat{T}(E)\ket{b}
\end{equation}
incorporates the T-matrix that describes the scattering events from low-lying energy states into the high-energy (integrated out) states. It can formally be written as
\begin{equation}
\hat{T}(E) = \hat{H}_K \hat{P}_H \hat{G}_0(E) \hat{P}_H \hat{H}_K,
\label{eqApx:TMatrix}
\end{equation}
where $\hat{P}_H$ is a projector into the Hilbert space of energetically high-lying states (i.e., $\epsilon\in[D-\delta D, D]$ or $\epsilon\in[-D, -D+\delta D]$), and $\hat{G}_0(E) = (E-\hat{H}_0)^{-1}$. The transition matrix $\hat{T}$ can be diagrammatically calculated by adding the two diagrams in Fig.~\ref{fig:diagrams}.

We will follow the strategy outlined above with an anisotropic version of Hamiltonian \eqref{eqApx:Kondo}, which reads:
\begin{equation}
\hat{H}_K = \frac{1}{2} \sum_a \left(\psi^\dagger_a \sigma_i \psi^\phd_a \left(\lambda^1_i - a\lambda^Y_i Y\right) \hat J_i \right),
\end{equation}
where we use the Einstein convention for an implicit sum over $i\in{x, y,z}$. The sum of both diagrams is
\begin{equation}
\begin{split}
\Delta\hat{H}_{ab} & = \hat{T}^{(I)}_{ab} + \hat{T}^{(II)}_{ab} \\
 & = -\frac{\nu \delta D}{2 D} \sum_a \left\{ \lambda_i^1\lambda_j^j [\sigma_i,\sigma_j]\hat J_i \hat J_j  - a\lambda_i^1\lambda_j^Y Y [\sigma_i,\sigma_j] \hat J_i \hat J_j \right. \\
& \left. - a\lambda^Y_i \lambda^1_j [\sigma_i, \sigma_j] Y \hat J_i \hat J_j + \lambda_i^Y\lambda_j^Y [\sigma_i,\sigma_j]\hat J_i \hat J_j \right\},
\end{split}
\end{equation}
where we approximate the integrals over energies that are integrated out by the factor $-\frac{\nu \delta D}{D}$. Using the relation
\begin{equation}
[\sigma_i, \sigma_j]\hat J_i \hat J_j = - 2\sigma_k\hat J_k\quad\text{with}\quad k\neq i,j,
\end{equation}
the result simplifies to
\begin{equation}
\begin{split}
\Delta \hat{H} & = -\frac{\nu\delta D}{D}\sum_a \left\{ (\lambda^1_i \lambda_j^1 + \lambda_i^Y\lambda_j^Y)\sigma_k \hat J_k \right. \\
& \left. - a (\lambda_i^1\lambda_j^Y + \lambda^Y_i\lambda^1_j)\sigma_k Y \hat J_k
\right\}.
\end{split}
\label{eqApx:DeltaHcompressed}
\end{equation}
From equation \eqref{eqApx:DeltaHcompressed}, we can see that the renormalized coupling constants take the form:
\begin{equation}
\begin{split}
\tilde{\lambda}^1_k  & = \lambda_k^1 + \frac{\nu \delta D}{D} (\lambda^1_i \lambda_j^1 + \lambda_i^Y\lambda_j^Y) \\
\tilde{\lambda}^Y_k & = \lambda_k^Y + \frac{\nu \delta D}{D} (\lambda_i^1\lambda_j^Y + \lambda^Y_i\lambda^1_j).
\end{split}
\end{equation}
These equations can be rewritten as a differential equation:
\begin{equation}
\begin{split}
\frac{d\lambda_k^1}{d D} & = \lim_{\delta D \rightarrow 0} \frac{\tilde{\lambda}k^1 - \lambda^1_k}{- \delta D} =  - \frac{\nu}{D}(\lambda^1_i \lambda_j^1 + \lambda_i^Y\lambda_j^Y) \\
\frac{d\lambda_k^Y}{d D} & = \lim_{\delta D \rightarrow 0} \frac{\tilde{\lambda}_k^Y - \lambda^Y_k}{-\delta D} = - \frac{\nu}{D}(\lambda_i^1\lambda_j^Y + \lambda^Y_i\lambda^1_j).
\end{split}
\end{equation}
Setting $\lambda^{1/Y}_x = \lambda^{1/Y}_z = \lambda_\perp$ and defining the RG-times $l=\ln(D_0/D)$, where $D_0$ is the initial bandwidth, the flow equations read
\begin{align}
    \frac{dg_\perp}{dl} & = g_\perp \left(g_y^1 +  g_y^Y\right),\label{eq:RGflow1} \\
    \frac{dg^1_y}{dl} & = 2g_\perp^2, \label{eq:RGflow2}\\
     \frac{dg^Y_y}{dl} & = 2g_\perp^2 .\label{eq:RGflow3}
\end{align}
Here, $g^{1/Y}_{\perp/y}=\nu \lambda^{1/Y}_{\perp/y}$ are dimensionless coupling constans. 
In order to visualize the tree dimensional RG flow, we define new parameters of the shape
\begin{equation}
    y^{1/Y}_{\perp, y} = \frac{g^{1/Y}_{\perp, y}}{1+g^{1/Y}_{\perp, y}},
\end{equation}
such that the $g^{1/Y}_{\perp, y} \rightarrow \infty$ corresponds to $y^{1/Y}_{\perp, y} \rightarrow 1$. The flow equation for the new parameters follows from equations \eqref{eq:RGflow1} - \eqref{eq:RGflow3} and read
\begin{align}
    \frac{\text d y_\perp}{\text d l} & = \frac{(y_\perp -1 )y_\perp\left(y^1_y(2y^{Y}_y - 1)-y^Y_y\right)}{(y^1_y -1 )(y^Y_y - 1)} \\
    \frac{\text d y_y^1}{\text d l} & = \frac{2(y^1_y-1)^2y_\perp^2}{(y_\perp-1)^2} \\
    \frac{\text d y_y^Y}{\text d l} & = \frac{2(y^Y_y-1)^2y_\perp^2}{(y_\perp-1)^2}.
\end{align}
This equation has been used to create Fig.~\ref{fig:RGflow} where we identified the point $(1, 1, 1)$ with the strong coupling fixed point, which we called $\infty$.

\section{Bosonization and Refermionization}
\label{app:EK}

This Appendix contains details for the Emery-Kivelson solution of Eq.~\eqref{eq:anisotropicKondoHam} and thereby is a supplement to Sec.~\ref{sec:BosonizationAndRefermionization}.

First, we define the spin operator of the lead electrons in terms of the chiral fermions $\tilde{\psi}_{R/L}$ as follows:
\begin{equation}
\bm{I}_a = \frac{\tilde{\psi}_a^\dagger\bm{\sigma}\tilde{\psi}_a^\phd}{2}, \quad\quad I^{\pm}_a = I^z_a \pm i I^x_a.
\end{equation}
Additionally, it is worth noting that
\begin{equation}
\begin{split}
\bm{I} \cdot \hat{\bm{J}} = \frac{1}{2}\left(I_a^+ \hat J_- + \text{H.c.}\right) + I^y_a \hat J_y.
\end{split}
\end{equation}
Now, the Hamiltonian can be rewritten in terms of these spin operators as follows:
\begin{align}
\begin{split}
\hat H_K & = \frac{\lambda_\perp}{2}\left(\left(I^+_+ + I^+_-\right)\hat J_- + \left(I^+_- - I^+_+\right)Y\hat J_- + \text{H.c.}\right) \\
& + \lambda^1_y\left(I^y_+ + I^y_-\right)\hat J_y + \lambda^Y_y (I^y_- - I^y_+)Y\hat J_y.
\end{split}
\end{align}
Here, we introduce an anisotropy in the Kondo coupling constants.
The lead electrons are bosonized in the following way:
\begin{equation}
\begin{split}
\tilde \psi_{a\sigma} & = \frac{F_{a\sigma}}{\sqrt{2\pi a_0}}\expp{i\sqrt{4\pi}\phi_{a\sigma}} \\ \Rightarrow & I^+_a = \psi^\dagger_{a\uparrow}\psi^\phd_{a\downarrow} = \frac{F^\dagger_{a\uparrow}F^\dagger_{a\downarrow}}{2\pi a_0}\expp{i\sqrt{4\pi}(\phi_{a\downarrow}-\phi_{a\uparrow})},
\end{split}
\end{equation}
and
\begin{equation}
I^y_a = \frac{1}{2}\left(\psi^\dagger_{a\uparrow}\psi^\phd_{a\uparrow} + \psi^\dagger_{a\downarrow}\psi^\phd_{a\downarrow}\right) = \frac{1}{2\sqrt{\pi}}\partial_x (\phi_{a\uparrow}(x) + \phi_{a\downarrow}(x))|_{x=0}.
\end{equation}
Using the rules to transform the operators into the $s/sf/c/cf$ basis as outlined in equations \eqref{eq:BosonMatrix} and \eqref{eq:basisChangeKleinFactors}, the bosonized Hamiltonian reads as follows:
\begin{align}
\begin{split}
\hat H_K & = \frac{\lambda_\perp}{\sqrt{8\pi a_0}}\left(\left(\chi_{sf}^\dagger + \chi^\phd_{sf}\right)F^\dagger_s \expp{-i\sqrt{4\pi}\phi_s}\hat J_-  \right. \\ 
+ & \left. \left(\chi^\phd_{sf} - \chi^\dagger_{sf}\right)F^\dagger_s \expp{-i\sqrt{4\pi}\phi_s}Y\hat J_+ + \text{H.c.} \right) \\
+ & \frac{\lambda_y^1}{\sqrt{\pi}}\partial\phi_s(x)|_{x=0}\hat J_y +\frac{\lambda_y^Y}{\sqrt{\pi}}\partial_x \phi_{sf}(x)|_{x=0} Y \hat J_y.
\end{split}
\end{align}
Here, we introduce the new fermion $\chi_a = \frac{F_s}{\sqrt{2\pi a_0}}\expp{i\sqrt{4\pi}\phi_{a}}$. We apply an Emery-Kivelson transformation $U_{\text{E.K.}}=\expp{i\sqrt{4\pi}\phi_s \hat J_y}$ which manages to decouple the $sf$ from the $s$ bosons/fermions. The operators affected by the Emery-Kivelson transformation transform as follows:
\begin{align}
\begin{split}
U^\dagger_{\text{E.K.}} \hat J_- U_{\text{E.K.}} & = \hat J_- \expp{i\sqrt{4\pi}\phi_s}, \\
U^\dagger_{\text{E.K.}} \partial_x \phi_s(x) U_{\text{E.K.}} & = \partial_x \phi_s(x) - \sqrt{\pi}\hat J_y\delta(x).
\label{eqApx:EmmeryKivelsonTrafo}
\end{split}
\end{align}
 The transformed and fully fermionized Hamiltonian reads as follows:
\begin{align}
\begin{split}
\hat H'_K & = \frac{\lambda_\perp}{\sqrt{4\pi a_0}} \left( (\chi_{sf}^\dagger + \chi^\phd_{sf})(d-d^\dagger)) + (\chi_{sf}^\phd - \chi^\dagger_{sf})(d+d^\dagger)Y \right) \\ & + \delta\lambda_y :\chi^\dagger_{sf}\chi^\phd_{sf}:(d^\dagger d -\frac{1}{2})  - \lambda^Y_y:\chi^\dagger_{sf}\chi^\phd_{sf}:Y(d^\dagger d -\frac{1}{2}),
\end{split}
\end{align}
where $F^\dagger_s \hat J_- = d$, $\hat J_y = (d^\dagger d - \frac{1}{2})$ and $\delta \lambda_y=\lambda_y^Y - 2\pi v_F$. The contribution of the Fermi velocity enters $v_F$ since the free Hamiltonian of the $\phi_s$ gets also transformed.  We also used the fact that the derivative of the Bose fields becomes in fermionic language the following:
\begin{align}
\partial_x \phi_a(x) = :\chi^\dagger_x \chi^\phd_x:,
\end{align}
where $: ... :$ denotes normal ordering. This is the origin of Eq.~\eqref{eq:fermionizedHam}.

\section{Observables}
\label{app:Observables}

\subsection{Green's function at the Toulouse point}
\label{secApx:GreensFunctions}

In this section, we calculate the Green's functions of the constituents in Hamiltonian $\eqref{eq:HamMajoranas}$ at the Toulouse point, which involves setting the coupling constants of the interaction terms to zero. In our case, this means we choose $\lambda_y^Y = 0$ and $\lambda_y^1 = 2\pi v_F$ (i.e. $\delta \lambda_y=0$).

First, we introduce some conventions. The retarded and imaginary time correlation functions of two observables, denoted as $A$ and $B$, are defined as
\begin{equation}
C^R_{AB}(t) = -i\theta(t) \braket{\left[A(t), B(t)\right]}
\end{equation}
and
\begin{equation}
C_{AB}(\tau) = - \braket{T_\tau A(\tau)B(\tau)},
\end{equation}
respectively. The transformation from imaginary time to Matsubara frequencies at zero temperature is given by:
\begin{align}
\, & \begin{split}
    C_{AB}(\tau) & = \frac{1}{\beta} \sum_m \expp{-i\omega_m\tau}C_{AB}(i\omega_m) \\
   & \stackrel{T\rightarrow0}{\longrightarrow}
     \frac{1}{2\pi} \int_{-\infty}^\infty \text{d}\omega_m \expp{-i\omega_m\tau} C_{AB}(i\omega_m) 
\end{split} 
\label{eqApx:ft}
\end{align} and
\begin{align}
\begin{split}
\,  C_{AB}(i\omega_m) & = \int_0^\beta \text{d}\tau \expp{i\omega_m \tau} C_{AB}(\tau) \\
& \stackrel{T\rightarrow0}{\longrightarrow} \frac{1}{2}\int_{-\infty}^\infty \text{d}\tau \expp{i\omega_m\tau} C_{AB}(\tau),
\end{split}
\end{align}
where $\beta$ is the inverse temperature. The retarded and imaginary time correlation functions are related by analytical continuation, which means:
\begin{equation}
C^{R}_{AB}(\omega) = \lim_{i\omega_m\rightarrow\omega + i0^+} C_{AB}(i\omega_m)
\end{equation}

Now we calculate the free local Green's function of the lead fermions: $G^{(0)}(\tau, x) = - \braket{T_\tau \chi^\phd_a(\tau, x) \chi_a^\dagger(0, 0)}$. In a diagonal basis, the Green's function reads:
\begin{equation}
G^{(0)}(i\omega_m)_k = \frac{1}{i\omega_m - \epsilon_k}.
\end{equation}
The local Green's function at $x=0$ is:
\begin{equation}
g^{(0)}(i\omega_m, x=0) = \int \frac{\text{d}k}{2\pi} \frac{1}{i\omega_m - \epsilon_k} = \int\text{d}\epsilon\frac{\rho(\epsilon)}{i\omega_m - \epsilon},
\end{equation}
where $\rho(\epsilon)=\nu$ is the density of states, which can be well modeled as being constant within the bandwidth $2D$, leading to:
\begin{equation}
g^{(0)}(i\omega_m) = \nu \int_{-D}^{D}\text{d}\epsilon \frac{1}{i\omega_m - \epsilon} = -i2\nu\arctan\left(\frac{D}{\omega}\right).
\end{equation}
In the limit $\omega_m \ll D$, the asymptotic behavior of the Green's function becomes:
\begin{equation}
g^{(0)}(i\omega_m) = -i\nu\pi\sign(\omega_m).
\end{equation}

To calculate the Green's function for the Majorana fermions $\eta$ and $\eta_{x, y, z}$ we introduce complex fermions which can be chosen in a very suggestive and physical way, that is
\begin{equation}
 f = \frac{i}{2}(\eta +i\eta_y), \quad s = \frac{1}{2}(\eta_z + i\eta_x) = \frac{\alpha}{2}(Z + iX) = \alpha \hat{S}^+.
\end{equation}
 The $f$ electron can be understood as a ladder operator within the Hilbert space that belongs to the order parameter angular momentum $\hat{\bm{J}}$. However, it acts differently on the subspaces defined by the MZMs. On the other hand, the $s$ fermion acts as a ladder operator in the orbital space.

The Hamiltonian \eqref{eq:HamMajoranas} can be expressed in terms of these new fermions and becomes
\begin{align}
 \hat{H}_\text{eff} & = \frac{\lambda_\perp}{\sqrt{2\pi a_0}}(\chi_{sf}^\dagger f + \text{H.c.}) + \lambda_y:\chi^\dagger_{sf}\chi_{sf}:(f^\dagger f - \frac{1}{2}) \nonumber \\ & + 2\delta\lambda_y :\chi^\dagger_{s}\chi_s:(f^\dagger f - \frac{1}{2})(s^\dagger s - \frac{1}{2}),
 \label{eq:fsHamiltonian}
 \end{align}
 where we ignored the Zeeman term for a moment. One can observe that in the case of no interaction (i.e., $\lambda_y=\delta\lambda_y=0$), the Hamiltonian \eqref{eq:fsHamiltonian} collapses to a resonant level model that describes a regular Kondo effect. However, the presence of the orbital space in which the $f$ fermion acts differently  distinguishes our system from ``just" a regular Kondo effect.

Now we can write down the local action for the Kondo Hamiltonian \eqref{eq:fsHamiltonian} at the Toulouse point in terms of the new fermions:
\begin{align}
\begin{split}
S_\text{K} & = \int\text{d}\omega_m
\begin{pmatrix}
\bar{\chi}_{sf} & \bar{f}
\end{pmatrix}
\begin{pmatrix}
-g^{(0)}(i\omega_m) & \tilde{\lambda} \\
\tilde{\lambda} & -i\omega_m
\end{pmatrix}
\begin{pmatrix}
\chi_{sf} \\ f
\end{pmatrix}
\\
& +
\int \text{d}\omega_m \bar{s}\,(-i\omega_m)\,s,
\end{split}
\end{align}
where $\tilde{\lambda} = \frac{\lambda_\perp}{\sqrt{2\pi a_0}}$. From the action, we can read off the Green's function:
\begin{equation}
\mathcal{G} = \left[
\begin{pmatrix}
\frac{i}{\nu \pi}\sign{\omega_m} & -\tilde{\lambda} \\
\tilde{\lambda} & i\omega_m
\end{pmatrix}
\right]^{-1} = \frac{-1}{|\omega_m| + \Gamma}
\begin{pmatrix}
i\omega_m \pi \nu & \tilde{\lambda}\pi\nu \\
\tilde{\lambda} \pi \nu & i\sign(\omega_m)
\end{pmatrix}
\end{equation}
where $\Gamma = \tilde{\lambda}^2 \nu\pi$. Collecting everything leads to the following list of Green's functions:
\begin{align}
G_0(i\omega_m) & = -i\pi\nu\sign(\omega_m)\quad\text{(Green's function $\chi_{s, c, cf}$)} \\
G(i\omega_m) & = \frac{-i\omega_m\pi\nu}{|\omega_m|+\Gamma}\quad\text{(Green's function $\chi_{sf}$)} \\
D_0(i\omega_m) & = \frac{1}{i\omega_m}\quad\text{(Green's function of $s$)} \\
D(i\omega_m) & = \frac{-i\sign(\omega_m)}{|\omega_m| + \Gamma}\quad\text{(Green's function $f$)} \\
F(i\omega_m) & = \frac{-\tilde{\lambda}\pi \nu}{|\omega_m|+\Gamma}\quad\text{($f\leftrightarrow\chi_{sf}$ matrix element)}\,.
\end{align}

We also find the Green's functions in imaginary time by applying the Fourier transform \eqref{eqApx:ft}. If we assume that $|\omega_m|\ll\Gamma$, the imaginary time Green's functions behave asymptotically at large $\tau$ as follows:
\begin{align}
G_0(\tau) & \approx - \frac{\nu}{\tau}, \label{eqApx:G0} \\
G(\tau) & \approx - \frac{2\nu}{\Gamma^2\tau^3}, \\
D_0(\tau) & \approx - \frac{\sign(\tau)}{2}, \\
D(\tau) & \approx - \frac{1}{\pi\Gamma\tau}, \\
F(\tau) & \approx - \frac{\tilde{\lambda}\nu}{\Gamma^2\tau^2}\,. \label{eqApx:F}
\end{align}

Furthermore, we deduce the Majorana Green's function from the fermionic one. We note that 
\begin{equation}
    \begin{split}
    \braket{T_\tau f(\tau) f(0)} & = \frac{1}{4}\braket{(\eta(\tau)+ i \eta_y(\tau))(\eta(0) - i \eta_y(0))} \\
    & = \frac{1}{4}\left(\braket{T_\tau \eta(0)\eta(\tau)}+\braket{T_\tau\eta_y(\tau)\eta_y(0)}\right) \\ & \equiv - D(\tau),
    \end{split}
    \label{AppEq:MajoranaGreensFunction}
\end{equation}
where it has been used that $\braket{T_\tau \eta(\tau)\eta_y(0)}=\braket{T_\tau \eta_y(\tau)\eta(0)}=0$ since that the Hamiltonian at the Toulouse point does not has terms which enable such processes. Also, the phase of the $f$ fermion is a Gauge degree of freedom from which we conclude that $\braket{T_\tau\eta_y(\tau)\eta_y(0)}=\braket{T_\tau \eta(0)\eta(\tau)}$. Hence, to fulfill equation \eqref{AppEq:MajoranaGreensFunction}
\begin{equation}
    \braket{T_\tau\eta_y(\tau)\eta_y(0)}=\braket{T_\tau \eta(0)\eta(\tau)}=-2D(\tau)
\end{equation}
must hold. It is straightforward to verify that similar relations also hold for $\xi_{s, sf, c, cf},\zeta_{s, sf, c, cf}, \eta_x$ and $\eta_y$.

\subsection{Correlation Functions at finite temperature}

Correlation functions of the operator at zero temperature with gap-less constituents show an algebraic decay in conformal field theories (e.g. equations \eqref{eqApx:G0}-\eqref{eqApx:F}). Using conformal transformations, the zero temperature correlation functions can be mapped onto their finite temperature counterparts as 
\begin{equation}
     \braket{T_\tau O(\tau)O(0)}\sim\left(\frac{1}{\tau}\right)^{2\Delta} \stackrel{\text{conf. trafo.}}{\longrightarrow} \left(\frac{\pi T}{\sin(\pi T \tau)}\right)^{2\Delta},
\end{equation}
where $\Delta$ is the scaling dimension of the operator $O(\tau)$ and $T$ is the temperature.
Integrals over these correlation functions 
\begin{equation}
    \begin{split}
        I(\Delta) = \int_{\tau_0}^{\beta-\tau_0}\text{d}\tau\,\left(\frac{\pi T}{\sin(\pi T \tau)}\right)^{2\Delta} = 2 \int_{\tau_0}^{\beta/2} \text{d}\tau \left( \frac{\pi T}{\sin(\pi T \tau)}\right)^{2\Delta} \\
    \end{split}
\end{equation}
can be expressed in a more convenient form by the coordinate transformation $x = \tan(\pi T \tau)$ and become
\begin{equation}
    I(\Delta) = 2(\pi T)^{2\Delta - 1}\int_{x_0}^\infty \text{d}x \frac{(1+x^2)^{\Delta - 1}}{x^{2\Delta}}
\end{equation}, which can be done exactly and evaluates to 
\begin{equation}
    I(\Delta) = \frac{2(\pi T)^{2\Delta - 1}}{x_0} \, {}_2 F_1\left(\frac{1}{2}, 1 - \Delta, \frac{3}{2}, -\frac{1}{x_0^2}\right),
\end{equation}
where $x_0 = \tan(\pi T \tau_0)$ and ${}_2 F_1$ is the hypergeometric function. We have introduced a regularization $\tau_0=\frac{1}{\Gamma}$ since the correlation functions are only valid for long imaginary times.  For further reference, we list a couple of special cases for $T\ll\Gamma$ (i.e. $x_0 \approx \pi T \tau_0$) to leading order in $T$:
\begin{enumerate}
    \item $\Delta = \nicefrac{1}{2}$
    \begin{equation}
        I\left(\frac{1}{2}\right) = 2\left(\ln\left(\frac{T}{\Gamma}\right) + \ln\left(\frac{2}{\pi}\right)\right),
    \end{equation}
    \item $\Delta = 1$
    \begin{equation}
        I(1) = 2\Gamma,
    \end{equation}
    \item $\Delta=\nicefrac{3}{2}$ \begin{equation}
        I\left(\frac{3}{2}\right) = \Gamma^2 - \pi^2 \ln(T)T^2 + \mathscr{O}(T^2),
    \end{equation}
    \item $\Delta=2$
    \begin{equation}
        I(2) = \frac{2}{3}\Gamma^3 + 2\pi^2\Gamma T^2 + \mathscr{O}(T^3),
    \end{equation}
    \item $\Delta>2$
    \begin{equation}
        I(\Delta) = \text{const.} + \mathscr{O}(T^2) .
    \end{equation}
\end{enumerate}

\subsection{Thermodynamics}

\subsubsection{Susceptibilities}

The local impurity susceptibility of $\hat{V}\in\{X, Y, Z, \hat{J}_x, \hat{J}_x, \hat{J}_x\}$ is defined as
\begin{equation}
\chi_{\hat{V}}(T, i\omega_m) = - \int^{\beta}_{0} \text{d}\tau \braket{T_\tau \hat{V}(\tau)\hat{V}(0)} \text e^{i\omega_m \tau},
\end{equation}
where $T$ and $\beta$ are the temperature and inverse temperature, respectively. Later on, we are only interested in the static susceptibility 
\begin{equation}
    \chi_{\hat{V}}(T) = \lim_{\omega_m\rightarrow 0}\chi_{\hat{V}}(T, \omega_m).
\end{equation}
However, for further reference, it will be beneficial to calculate the dynamical susceptibility for the $\hat{J_y}$ operator. First, we decompose the susceptibility in the static and dynamical part as

The calculation of the static susceptibilities is straightforward and relies on the decomposition into the four Majorana fermions $\eta, \eta_y, \eta_x, \eta_z$. One finds 
\begin{equation}
\hat{J}_y = \frac{i\alpha\beta}{2} = \frac{1}{2}\eta_x\eta_y\eta_z\eta.
\end{equation}
Thus, the time-dependent correlation function evaluates to
\begin{equation}
\begin{split}
 \braket{T_\tau \hat{J}_y(\tau)\hat{J}_y(0)} & = \frac{1}{4}\braket{T_\tau\wick{\c1 \eta_x^\tau\c2 \eta_y^\tau\c3\eta_z^\tau\c4\eta^\tau\c1\eta_x^0\c2\eta_y^0\c3\eta_z^0\c4\eta^0}} \\
& = \frac{1}{4}(-2D_0(\tau))^2(-2D(\tau))^2 = D(\tau)^2,
\end{split}
\end{equation}
where we used $\eta(\tau)=\eta^\tau$ as a shortcut notation. 
Using the former result, we obtain the static susceptibility
\begin{equation}
\begin{split}
\chi_{\hat J_y}(T) & = - \int_{\tau_0}^{\beta-\tau_0}\text{d}\tau D(\tau)^2 = - \left(\frac{1}{\pi\Gamma}\right)^2 I(1) = -\frac{2}{\pi^2\Gamma}
\end{split},
\end{equation}
where we introduced the cut of $\tau_0=\nicefrac{1}{\Gamma}$ to regularize the integral.

The dynamical susceptibility can be calculated along the same line, except that the integral reads
\begin{equation}
    \begin{split}
    \delta \chi_{\hat J_y}(T, i\omega_m) & = -\int_0^{\beta} \text d\tau D(\tau)^2 (\text e^{i\omega_m \tau} - 1) \\ & = - \left(\frac{1}{\pi \Gamma}\right)^2 \int_{0}^\beta \text d\tau (\pi T)^2\frac{\cos(\omega_m \tau)-1}{\sin^2(\pi T \tau)} \\ & = \frac{|\omega_m|}{2 \pi \Gamma^2}.
    \end{split} 
\end{equation}
Hence, the full dynamical susceptibility evaluates
\begin{equation}
    \chi_{\hat J_y}(T,i\omega_m) = -\frac{1}{\pi \Gamma}\left(\frac{2}{\pi}-\frac{|\omega_m|}{2\Gamma}\right).
\end{equation}

In the following, we only consider the static susceptibilities. 
For $\hat{J}_x$ and $\hat{J}_z$, the calculation is slightly more complicated since these operators are affected by the Emery-Kivelson transformation. According to equation \eqref{eqApx:EmmeryKivelsonTrafo}, we find that:
\begin{equation}
\begin{split}
U^\dagger_{\text{E.K.}} \hat{J}_z U^\phd_{\text{E.K.}} & = \frac{1}{2} U^\dagger_{\text{E.K.}} \left(J^+ + J^-\right) U^\phd_{\text{E.K.}} \\
& = \frac{1}{2}\left(\expp{-i\sqrt{4\pi}\phi_s}\hat{J}^+ + \expp{i\sqrt{4\pi}\phi_s}\hat{J}^-\right) \\
& = \sqrt{\frac{\pi a_0}{2}}\left(d^\dagger \chi_s + \text{H.c.}\right) \\
& = \sqrt{\frac{\pi a_0}{8}}\left(i\xi_s\eta + \zeta_s\eta_x\eta_y\eta_z\right),
\end{split}
\end{equation}
where we inserted the identity $\mathds{1}=F^\phd_s F^\dagger_s$. A similar calculation yields:
\begin{equation}
U^\dagger_\text{E.K.}\hat{J}_x U_\text{E.K.}^\phd = \sqrt{\frac{\pi a_0}{8}}\left(i\zeta_s\eta_y + \xi_s\eta_x\eta_y\eta_z\right).
\end{equation}
To calculate the susceptibility, we need the correlator of the transformed spin operators:
\begin{equation}
\begin{split}
\braket{T_\tau U^\dagger_\text{E.K.}\hat{J}_{x/z}(\tau)U^\phd_\text{E.K.}U^\dagger_\text{E.K.}\hat{J}_z(0)U^\phd_\text{E.K.}} = \pi a_0 G_0(\tau)D(\tau).
\end{split}
\end{equation}
Thus, the susceptibility can be calculated by
\begin{equation}
\begin{split}
\chi_{\hat{J}_{x, z}} & = - \pi a_0\int_{\tau_0}^{\beta-\tau_0}\text{d}\tau G_0(\tau)D(\tau) = \frac{a_0 \nu}{\Gamma} I(1) = 2a_0\nu.
\end{split}
\end{equation}

The second part of the impurity is given by the orbital degrees of freedom $X, Y, Z$ and follows the same strategy as for the condensate. Note that none of these operators are affected by the Emery-Kivelson transformation. 

\begin{equation}
\begin{split}
\chi_{Y}(T) & = - \int_{\tau_0}^{\beta-\tau_0} \text{d}\tau \braket{T_\tau Y(\tau)Y(0)} \\
& = \int_{\tau_0}^{\beta-\tau_0} \text{d}\tau \braket{T_\tau \eta_x(\tau)\eta_z(\tau)\eta_x(0)\eta_z(0)} \\
& = - \int_{\tau_0}^{\beta-\tau_0} \text{d}\tau (-2D_0)^2 = - \int_{\tau_0}^{\beta-\tau_0} \text{d}\tau \sim \frac{1}{T},
\end{split}
\end{equation}
where we introduced the cut of $\tau_0={\nicefrac{1}{\Gamma}}$ to regularize the integral. Furthermore:
\begin{equation}
\begin{split}
\chi_{Z}(T) & = - \int_{\tau_0}^{\beta-\tau_0} \text{d}\tau \braket{T_\tau Z(\tau)Z(0)} \\
& = \int_{\tau_0}^{\beta-\tau_0} \text{d}\tau \braket{T_\tau \eta_y(\tau)\eta_x(\tau)\eta_y(0)\eta_x(0)} \\
& = - \int_{\tau_0}^{\beta-\tau_0} \text{d}\tau (-2D(\tau))(-2D_0(\tau))\\
& = - \frac{1}{\pi \Gamma} I(\nicefrac{1}{2}) = \frac{2}{\pi\Gamma}\ln\left(\frac{\Gamma}{T}\right) - \frac{2}{\pi\Gamma}\ln\left(\frac{2}{\pi}\right).
\end{split}
\end{equation}
$\chi_{X}(T)$ can be calculated along the same line, which yields the same result as for $\chi_{Z}(T)$

\subsubsection{Finite Temperature Corrections to the Free Energy}

The scaling dimension of the interaction operators $O_{sf}$ and $O_s$ is determined by investigating the algebraic decay of their correlations:
\begin{equation}
\begin{split}
& \braket{T_\tau O_s(\tau)O_s(0)} \\
& = \braket{T_\tau \wick{ \c1 \xi_s^\tau \c2 \zeta_s^\tau \c3 \eta_x^\tau \c4 \eta_y^\tau \c5 \eta_z^\tau \c6 \eta^\tau \c1 \xi_s^0\c2 \zeta_s^0\c3 \eta_x^0\c4 \eta_y^0\c5 \eta_z^0 \c6 \eta^0}} \\
& = (-2G_0)^2(-2D_0)^2(-2D)^2 \sim \left( \frac{1}{\tau}\right)^4
\end{split}
\label{eqApx:OsContractions}
\end{equation}
and
\begin{equation}
\begin{split}
& \braket{T_\tau O_{sf}(\tau)O_{sf}(0)} = \braket{T_\tau\xi_{sf}^\tau\zeta_{sf}^\tau\eta_y^\tau\eta^\tau\xi_{sf}^0\zeta_{sf}^0\eta_y^0\eta^0} \\
& = \braket{T_\tau \wick{\c1 \xi_{sf}^\tau\c2 \zeta_{sf}^\tau\c3\eta_y^\tau\c4\eta^\tau\c1\xi_{sf}^0\c2\zeta_{sf}^0\c3\eta_y^0\c4\eta^0}} + \braket{T_\tau \wick{\c1\xi_{sf}^\tau\c2\zeta_{sf}^\tau\c3\eta_y^\tau\c4\eta^\tau\c4\xi_{sf}^0\c3\zeta_{sf}^0\c2\eta_y^0\c1\eta^0}} \\
& = (-2G)^2(-2D)^2 - (-2F)^4 \sim \left(\frac{1}{\tau}\right)^8,
\end{split}
\label{eqApx:OsfContractions}
\end{equation}
where we used $\xi_s^\tau = \xi_s(\tau)$ as a short-hand notation. Equations \eqref{eqApx:OsContractions} and \eqref{eqApx:OsfContractions} show that the scaling dimensions of $O_s$ and $O_{sf}$ are $\Delta_s = 2$ and $\Delta_{sf} = 4$, respectively.

Since both operators are irrelevant in an RG sense, we can incorporate them perturbatively into the calculation of the finite temperature corrections to the free energy. Since the operator $O_s$ has the lower scaling dimension, we will focus on that one.

In the leading order, the corrections to the free energy are given by the expression:
\begin{equation}
\begin{split}
\delta F(T) & = - \frac{1}{2} \int_{\tau_0}^{\beta-\tau_0} \text{d}\tau \braket{O_s(\tau)O_s(0)} \\
& = -\delta\lambda_y^2\left(\frac{\nu}{\pi\Gamma}\right)^2 I(2) \\
& = - \frac{(\delta\lambda_y \nu)^2}{\Gamma} T^2 - \frac{1}{3}\left(\frac{\delta\lambda_y\nu}{\pi}\right)^2 \Gamma.
\end{split}
\end{equation}
Thus, we obtain the final result:
\begin{equation}
\delta F(T) - \delta F(0) = - \frac{(\delta\lambda_y \nu)^2}{\Gamma} T^2.
\end{equation}
The correction leads directly to the expression for the thermodynamic entropy and the specific heat:
\begin{equation}
S(T) = - \frac{\partial}{\partial T}F(T) = 2\frac{(\delta\lambda_y \nu)^2}{\Gamma} T,
\end{equation}
and
\begin{equation}
C_v(T) = T \frac{\partial}{\partial T} S(T) = 2\frac{(\delta\lambda_y \nu)^2}{\Gamma} T.
\end{equation}

\subsection{Transport}

In the following chapter, we aim to calculate the conductance for both spin and charge currents. To achieve this, we employ the Kubo formula, which is expressed as:
\begin{equation}
G = \lim_{\omega \to 0} \text{Re}\frac{i}{\omega}C^R_{II},
\label{AppEq:KuboRaw}
\end{equation}
Here, the symbol $I$ represents the current operator, defined as:
\begin{equation}
I = \partial_t Q = -i[Q, H],
\label{eqApx:current}
\end{equation}
where the variable $Q$ can represent either the total charge or spin of the lead electrons. When expressed in terms of Matsubara frequencies, the Kubo formula takes the following form:
\begin{equation}
G = \lim_{\omega \to 0} \text{Re} \frac{1}{i\omega} \lim_{i\omega_m \to \omega + i0^+}\int_{-\beta/2}^{\beta/2} d\tau \expp{i\omega_m \tau} \braket{T_\tau I(\tau)I(0)}.
\label{eqApx:Kubo}
\end{equation}

\subsubsection{Charge and Spin current}

The total charge in the left/right wire is given by:
\begin{equation}
Q_{L/Rc} = -e \int \text{d}x \, \tilde{\psi}^\dagger_L(x) \tilde{\psi}^\phd_L(x)\,
\end{equation}
where, $\tilde{\psi}_L = (\tilde{\psi}_{L\uparrow},\tilde{\psi}_{L\downarrow})^\text{T}$ is a right-moving field that has been extended to the whole real axis. A basis change into the $\pm$ basis results in:
\begin{equation}
Q_{L/Rc} = -\frac{e}{2} \int \text{d}x \left(\tilde{\psi}^\dagger(x) \tilde{\psi}(x) \pm \tilde{\psi}^\dagger(x) \sigma_y \tilde{\tau}_y \tilde{\psi}(x)\right),
\end{equation}
where, $\tilde{\psi} = (\tilde{\psi}_+,\tilde{\psi}_-)^\text{T}$, and $\tilde{\tau}_i$ acts in $\pm$ space. After Bosonization and refermionization, the total charge expressed in terms of Majorana fermions becomes:
\begin{equation}
Q_{L/Rc} = -\frac{e}{2} \int \text{d}x \, (i\xi_c\zeta_c \pm i\xi_{sf}\xi_{cf}).
\end{equation}
Equation \eqref{eqApx:current} for the charge current yields:
\begin{align}
I_{L/Rc} = -\frac{e}{2}\left(\pm i\tilde{\lambda}\xi_{cf}\eta + \frac{\lambda^Y_y}{2}\xi_{cf}\xi_{sf}\eta_y\eta\right) \stackrel{\text{T.P.}}{=} \mp i\frac{\tilde{\lambda}e}{2}\xi_{cf}\eta,
\end{align}
where we used that we are evaluating the conductance at the Toulouse point (T.P.).

In full analogy to the charge current, we express the total spin of one wire first in chiral fermions with support on the whole axis. Thus, we have:
\begin{equation}
\begin{split}
Q_{L/Rs} & = \int \text{d}x \, \tilde{\psi}^\dagger_L(x) \sigma_y \tilde{\psi}^\phd_L(x) \\
& = \frac{1}{2} \int \text{d}x \left(\tilde{\psi}^\dagger(x) \sigma_y \tilde{\psi}(x) \pm \tilde{\psi}^\dagger\tilde{\tau}_y\tilde{\psi}(x)\right),
\end{split}
\end{equation}
where we changed into the $\pm$ basis after the second equal sign. The first part becomes:
\begin{equation}
\tilde{\psi}^\dagger(x)\sigma_y\tilde{\psi}(x) = \frac{2}{\sqrt{\pi}}\partial_x\phi_s(x)
\end{equation}
and, thus, is affected by the Emery-Kivelson transformation. The total magnetization becomes after the transformation:
\begin{equation}
\begin{split}
Q_{L/Rs} & = \frac{1}{2}\int\text{d}x\left(\frac{2}{\sqrt{\pi}}\partial_x\phi_s(x) \pm \tilde{\psi}^\dagger(x)\tilde{\tau}\tilde{\psi}(x)\right) - J^y \\
& = \frac{1}{2}\int\text{d}x \left(i\xi_s\zeta_s \pm i\zeta_{sf}\zeta_{cf}\right) - \frac{\eta_x\eta_y\eta_z\eta}{2}.
\end{split}
\end{equation}
With the corresponding current:
\begin{equation}
    I_{L/Rs} \stackrel{\text{T.P.}}{=}  \frac{\tilde\lambda}{2}\left(\pm i\zeta_{cf}\eta_y + \eta_x\eta_z\left(\xi_{sf} \eta_y - \zeta_{sf}\eta\right)\right),
\end{equation}
where the second term corresponds to $\dot{J}_{y}$.

\subsubsection{Charge Conductance in Strong Coupling Limit}

Knowing the charge currents, we can calculate the conductances $G_{ij}^c$, where $i, j \in {R, L}$. The current-current correlator is expressed as:
\begin{equation}
\begin{split}
& \braket{T_\tau I_{ic}(\tau)I_{jc}(0)} = -x_{ij}\left(\frac{\tilde{\lambda}e}{2}\right)^2\braket{T_\tau \wick{\c1 \xi_{cf} (\tau)
\c2 \eta(\tau) \c1 \xi_{cf}(0) \c2 \eta(0)}} \\
& = x_{ij}\left(\frac{\tilde{\lambda}e}{2}\right)^2\left(-2G_0(\tau)\right)\left(-2D(\tau)\right) = -x_{ij} (\tilde\lambda e)^2 G_0(\tau) D(\tau),
\end{split}
\end{equation}
where  $x_{ij}=(1-2\delta_{ij})$. Thus, the current correlation function expressed in Matsubara frequencies reads
\begin{equation}
    C_{II}(i\omega_m) = - x_{ij} \frac{(\tilde\lambda e)^2}{\beta} \sum_{n} G_0(i\omega_m - i\epsilon_n) D(i\epsilon_n), 
\end{equation}
where $\omega_m$ and $\epsilon_n$ are bosonic and fermionic Matsubara frequencies, respectively. The main task will be to evaluate the Mastubara sum
\begin{equation}
    \begin{split}
    \Sigma(i\omega_m) & = \frac{1}{\beta}\sum_n  G_0(i\omega_m - i\epsilon_n) D(i\epsilon_n) \\ & = \frac{-i\nu \pi}{\beta} \sum_{n} \frac{\sign(\omega_m - \epsilon_n)}{i\epsilon_m + i \Gamma \sign(\epsilon_n)}.
    \end{split}
\end{equation}
Note that this sum doesn't converge. However, we can regularize the sum by subtracting the $\omega_m = 0$ part, which gives an imaginary contribution and will, thus, vanish in the conductance anyway. Hence, we are left with the sum
\begin{equation}
    \begin{split}
    \delta\Sigma(i\omega_m) & = \frac{-i\nu\pi}{\beta} \sum_n \frac{\sign(\omega_m - \epsilon_n)+\sign(\epsilon_{n})}{i\epsilon_n + i \Gamma \sign(\epsilon_n)} \\ & = 
    \frac{-2\nu\pi}{\beta}\sum_{n=0}^{m-1}\frac{1}{\epsilon_n + \Gamma} \\ & =  -\nu\left(\psi\left(\frac{1}{2}+\frac{\omega_m + \Gamma}{2 \pi T}\right)-\psi\left(\frac{1}{2}+\frac{\Gamma}{2\pi T}\right)\right) \\  & = - \frac{\nu}{2 \pi T} \psi'\left(\frac{1}{2}+\frac{\Gamma}{2 \pi T}\right)\omega_m + \mathscr{O}(\omega_m^2),
    \end{split}
\end{equation}
where $\psi$ is the digamma function and $\psi'$ is it's first derivative. Since we are interested in the low-temperature correction, we expand the result up to the first leading order in the temperature and obtain 
\begin{equation}
    \delta\Sigma(i\omega_m) \approx -\frac{\nu}{\Gamma}\left(1-\frac{\pi^2}{3}\left(\frac{T}{\Gamma}^2\right)\right)\omega_m.
\end{equation}
Therefore, we find the current correlation function to be
\begin{equation}
    \begin{split}
    C_{II}(i\omega_m) & = -x_{ij} (\tilde \lambda e)^2 \delta\Sigma(i\omega_m)  \\ & = x_{ij}\frac{e^2}{\pi}\left(1 - \frac{\pi^2}{3}\left(\frac{T}{\Gamma}\right)^2\right)\omega_m.
    \end{split}
\end{equation}
Inserting this result into the Kubo formula \eqref{AppEq:KuboRaw} leads to the final result for the DC conductance is
\begin{equation}
    \begin{split}
    G^c_{ij} & = \text{Re}\lim_{\omega\rightarrow 0} \frac{1}{i\omega} \lim_{i\omega_m \rightarrow \omega + i0^+} x_{ij}\frac{e^2}{\pi}\omega_m\left(1-\frac{\pi}{3}\left(\frac{T}{\Gamma}\right)\right) \\ & = \text{Re} \lim_{\omega\rightarrow 0} \frac{1}{i\omega} x_{ij}\frac{e}{\pi}(-i\omega) \times \left(1-\frac{\pi}{3} \left(\frac{T}{\Gamma}\right)\right) \\ & = -x_{ij}G_0^c\underbrace{2 \left(1 - \frac{\pi}{3}\left(\frac{T}{\Gamma}\right)^2\right)}_{G(T)},
    \end{split}
\end{equation}
where $G^c_0 = e^2/h$ is the perfect conductance. Note that $\hbar=1$ and, thus, $h=2\pi$.

\subsubsection{Spin Conductance}

The calculation of the spin conductance is analogous to that of charge conductance, with the key difference being the more intricate shape of the spin current. To compute $G^s_{ij}$, we first introduce a new notation:
\begin{equation}
I_{L/Rs} = i_{L/Rs} - \dot{J}_y,
\end{equation}
where $i_{L/Rs} = \pm i\frac{\tilde{\lambda}}{2}\zeta_{cf} \eta_y$. The current-current correlator essentially involves a sum over three distinct correlators: $\braket{T_\tau i_{L/Rs}(\tau)i_{L/Rs}(0)}$, $\braket{T_\tau i_{L/Rs}(\tau)\dot{J}y(0)}$, and $\braket{T\tau \dot{J}_y(\tau)\dot{J}_y(0)}$. It is straightforward to show that:
\begin{equation}
\braket{T_\tau i_{is}(\tau)i_{js}(0)} = \frac{\braket{T_\tau I_{ic}(\tau)I_{jc}(0)}}{e^2}
\end{equation}
and that $\braket{T_\tau i_{L/Rs}(\tau)\dot{J}_y(0)} = 0$. Thus, only the last correlator $\braket{T_\tau \dot{J}_y(\tau)\dot{J}_y(0)}$ remains to be calculated. 

First, we express the correlator in terms of Matsubara frequencies 
\begin{equation}
    \begin{split}
    C_{\dot{J}_y \dot J_y}(i\omega_m) & = - \int_0^\beta \text d\tau \text d\tau' \text e^{i\omega_m (\tau - \tau')} \braket{\dot J_y(\tau) \dot J_y(\tau')} \\ & = - \int_0^\beta \text d\tau \text d\tau' \text e^{i\omega_m (\tau - \tau')}\partial_\tau \partial_{\tau'}\delta \chi_{\hat{J_y}}(\tau,\tau') \\ & = - \omega_m^2 \delta \chi_{\hat{J}_y}(i\omega_m) = \frac{\omega_m^2|\omega_m|}{2\pi\Gamma^2}.
    \end{split}
\end{equation}
Here we implicitly subtracted the $C_{\dot{J}_y \dot J_y}(0)$ by using $\delta \chi_{\hat{J}_y}(i\omega_m)$ instead of $\chi_{\hat{J}_y}(i\omega_m)$. The third equality is achieved by partial integration. We see that the $C_{\dot{J}_y \dot J_y}$ correlator is of third order in $\omega$ and, thus, does not contribute to the DC conductance since the term vanishes in the limit $\omega\rightarrow0$.

\subsubsection{Conductance at Weak Coupling}

Finally, we aim to determine the temperature dependence of conductance at weak coupling (i.e., temperatures above the Kondo temperature). To achieve this, we express the Kondo Hamiltonian in a slightly different form:
\begin{equation}
\hat{H}_K = \frac{\lambda}{2}\psi^\dagger\left(\mathds{1} - \tilde{\tau}_z Y\right)\bm{\sigma}\psi\cdot\hat{\bm{J}},
\end{equation}
where $\psi=(\psi_+, \psi_-)^\text{T}$. The charge current expressed in terms of the $\pm$ fermions is given by:
\begin{equation}
I_{L/Rc} = \mp \frac{e\lambda}{2}\psi^\dagger \left(\varepsilon_{2nm}\tilde{\tau}_y\sigma_m - \delta_{2n}\tilde{\tau}_x Y\right)\psi \hat{J}_n
\end{equation}
The current correlator evaluates as follows:
\begin{equation}
\braket{T_\tau I_{ic}(\tau)I_{jc}(0)} = 3\,x_{ij}\,(\lambda e)^2 G_0(\tau)G_0(-\tau),
\end{equation}
where the free Green's function of the electrons at finite temperature is:
\begin{equation}
G_0(\tau) = -\braket{\psi_{\gamma}^\phd(\tau)\psi^\dagger_{\gamma}(0)} = -\nu \frac{\pi T}{\sin(\pi T \tau)}.
\end{equation}
Note that $\braket{\psi_{\gamma}^\phd(\tau)\psi^\dagger_{\sigma}(0)}=0$ if $\gamma \neq \sigma$. Inserting the current-current correlator into the Kubo formula leads to the integral:
\begin{equation}
\lim_{i\omega_n\rightarrow \omega + i0^+} \int_{0}^\beta \text{d}\tau\,G_0(\tau)G_0(-\tau) \expp{i\omega_n\tau} = -i\pi\nu^2\omega,
\end{equation}
which is valid for small $\omega$. This result yields the conductance at weak coupling:
\begin{equation}
G_{ij}^c = -6x_{ij}\,\frac{e^2}{h}(\lambda\pi\nu)^2.
\end{equation}

The spin current in the basis of the $\pm$ lead fermions is defined as:
\begin{equation}
I_{L/Rs} = \frac{\lambda}{2}\left\{\varepsilon_{2nm}\psi^\dagger\sigma_m\psi \hat{J}_n - \psi^\dagger(\varepsilon_{2nm}\tilde{\tau}_z\sigma_m \pm \tilde{\tau}_x\sigma_n)\psi Y\hat{J}_n\right\}.
\end{equation}
For this case, the current-current correlator reads:
\begin{equation}
\braket{T_\tau I_{is}(\tau)I_{js}(0)} = -  \lambda^2 (4- 3x_{ij})G_0(\tau)G_0(-\tau).
\end{equation}
This results in the spin conductance:
\begin{equation}
G^s_{ij} = (4-3x_{ij}) \frac{2}{h}(\lambda\pi\nu)^2.
\end{equation}

\bibliography{SpinFullTopoKondo.bib}

\begin{thebibliography}{80}%
\makeatletter
\providecommand \@ifxundefined [1]{%
 \@ifx{#1\undefined}
}%
\providecommand \@ifnum [1]{%
 \ifnum #1\expandafter \@firstoftwo
 \else \expandafter \@secondoftwo
 \fi
}%
\providecommand \@ifx [1]{%
 \ifx #1\expandafter \@firstoftwo
 \else \expandafter \@secondoftwo
 \fi
}%
\providecommand \natexlab [1]{#1}%
\providecommand \enquote  [1]{``#1''}%
\providecommand \bibnamefont  [1]{#1}%
\providecommand \bibfnamefont [1]{#1}%
\providecommand \citenamefont [1]{#1}%
\providecommand \href@noop [0]{\@secondoftwo}%
\providecommand \href [0]{\begingroup \@sanitize@url \@href}%
\providecommand \@href[1]{\@@startlink{#1}\@@href}%
\providecommand \@@href[1]{\endgroup#1\@@endlink}%
\providecommand \@sanitize@url [0]{\catcode `\\12\catcode `\$12\catcode
  `\&12\catcode `\#12\catcode `\^12\catcode `\_12\catcode `\%12\relax}%
\providecommand \@@startlink[1]{}%
\providecommand \@@endlink[0]{}%
\providecommand \url  [0]{\begingroup\@sanitize@url \@url }%
\providecommand \@url [1]{\endgroup\@href {#1}{\urlprefix }}%
\providecommand \urlprefix  [0]{URL }%
\providecommand \Eprint [0]{\href }%
\providecommand \doibase [0]{https://doi.org/}%
\providecommand \selectlanguage [0]{\@gobble}%
\providecommand \bibinfo  [0]{\@secondoftwo}%
\providecommand \bibfield  [0]{\@secondoftwo}%
\providecommand \translation [1]{[#1]}%
\providecommand \BibitemOpen [0]{}%
\providecommand \bibitemStop [0]{}%
\providecommand \bibitemNoStop [0]{.\EOS\space}%
\providecommand \EOS [0]{\spacefactor3000\relax}%
\providecommand \BibitemShut  [1]{\csname bibitem#1\endcsname}%
\let\auto@bib@innerbib\@empty
\bibitem [{\citenamefont {Chiu}\ \emph {et~al.}(2016)\citenamefont {Chiu},
  \citenamefont {Teo}, \citenamefont {Schnyder},\ and\ \citenamefont
  {Ryu}}]{ChiuRyu2016}%
  \BibitemOpen
  \bibfield  {author} {\bibinfo {author} {\bibfnamefont {C.-K.}\ \bibnamefont
  {Chiu}}, \bibinfo {author} {\bibfnamefont {J.~C.~Y.}\ \bibnamefont {Teo}},
  \bibinfo {author} {\bibfnamefont {A.~P.}\ \bibnamefont {Schnyder}},\ and\
  \bibinfo {author} {\bibfnamefont {S.}~\bibnamefont {Ryu}},\ }\bibfield
  {title} {\bibinfo {title} {Classification of topological quantum matter with
  symmetries},\ }\href {https://doi.org/10.1103/RevModPhys.88.035005}
  {\bibfield  {journal} {\bibinfo  {journal} {Rev. Mod. Phys.}\ }\textbf
  {\bibinfo {volume} {88}},\ \bibinfo {pages} {035005} (\bibinfo {year}
  {2016})}\BibitemShut {NoStop}%
\bibitem [{\citenamefont {Volovik}(2003)}]{Volovik2003}%
  \BibitemOpen
  \bibfield  {author} {\bibinfo {author} {\bibfnamefont {G.~E.}\ \bibnamefont
  {Volovik}},\ }\href@noop {} {\emph {\bibinfo {title} {The universe in a
  helium droplet}}},\ Vol.\ \bibinfo {volume} {117}\ (\bibinfo  {publisher}
  {OUP Oxford},\ \bibinfo {year} {2003})\BibitemShut {NoStop}%
\bibitem [{\citenamefont {Vollhardt}\ and\ \citenamefont
  {Wolfle}(2013)}]{VollhardtWoelfle2013}%
  \BibitemOpen
  \bibfield  {author} {\bibinfo {author} {\bibfnamefont {D.}~\bibnamefont
  {Vollhardt}}\ and\ \bibinfo {author} {\bibfnamefont {P.}~\bibnamefont
  {Wolfle}},\ }\href@noop {} {\emph {\bibinfo {title} {The superfluid phases of
  helium 3}}}\ (\bibinfo  {publisher} {Courier Corporation},\ \bibinfo {year}
  {2013})\BibitemShut {NoStop}%
\bibitem [{\citenamefont {Joynt}\ and\ \citenamefont
  {Taillefer}(2002)}]{JoyntTaillefer2002}%
  \BibitemOpen
  \bibfield  {author} {\bibinfo {author} {\bibfnamefont {R.}~\bibnamefont
  {Joynt}}\ and\ \bibinfo {author} {\bibfnamefont {L.}~\bibnamefont
  {Taillefer}},\ }\bibfield  {title} {\bibinfo {title} {The superconducting
  phases of $\text{{UPt}}_3$},\ }\href
  {https://journals.aps.org/rmp/abstract/10.1103/RevModPhys.74.235} {\bibfield
  {journal} {\bibinfo  {journal} {Reviews of Modern Physics}\ }\textbf
  {\bibinfo {volume} {74}},\ \bibinfo {pages} {235} (\bibinfo {year}
  {2002})}\BibitemShut {NoStop}%
\bibitem [{\citenamefont {Jiao}\ \emph {et~al.}(2020)\citenamefont {Jiao},
  \citenamefont {Howard}, \citenamefont {Ran}, \citenamefont {Wang},
  \citenamefont {Rodriguez}, \citenamefont {Sigrist}, \citenamefont {Wang},
  \citenamefont {Butch},\ and\ \citenamefont {Madhavan}}]{JiaoMadhavan2020}%
  \BibitemOpen
  \bibfield  {author} {\bibinfo {author} {\bibfnamefont {L.}~\bibnamefont
  {Jiao}}, \bibinfo {author} {\bibfnamefont {S.}~\bibnamefont {Howard}},
  \bibinfo {author} {\bibfnamefont {S.}~\bibnamefont {Ran}}, \bibinfo {author}
  {\bibfnamefont {Z.}~\bibnamefont {Wang}}, \bibinfo {author} {\bibfnamefont
  {J.~O.}\ \bibnamefont {Rodriguez}}, \bibinfo {author} {\bibfnamefont
  {M.}~\bibnamefont {Sigrist}}, \bibinfo {author} {\bibfnamefont
  {Z.}~\bibnamefont {Wang}}, \bibinfo {author} {\bibfnamefont {N.~P.}\
  \bibnamefont {Butch}},\ and\ \bibinfo {author} {\bibfnamefont
  {V.}~\bibnamefont {Madhavan}},\ }\bibfield  {title} {\bibinfo {title} {Chiral
  superconductivity in heavy-{F}ermion metal {U}{T}e$_2$},\ }\href
  {https://www.nature.com/articles/s41586-020-2122-2} {\bibfield  {journal}
  {\bibinfo  {journal} {Nature}\ }\textbf {\bibinfo {volume} {579}},\ \bibinfo
  {pages} {523} (\bibinfo {year} {2020})}\BibitemShut {NoStop}%
\bibitem [{\citenamefont {Aoki}\ \emph {et~al.}(2022)\citenamefont {Aoki},
  \citenamefont {Brison}, \citenamefont {Flouquet}, \citenamefont {Ishida},
  \citenamefont {Knebel}, \citenamefont {Tokunaga},\ and\ \citenamefont
  {Yanase}}]{AokiYanase2022}%
  \BibitemOpen
  \bibfield  {author} {\bibinfo {author} {\bibfnamefont {D.}~\bibnamefont
  {Aoki}}, \bibinfo {author} {\bibfnamefont {J.-P.}\ \bibnamefont {Brison}},
  \bibinfo {author} {\bibfnamefont {J.}~\bibnamefont {Flouquet}}, \bibinfo
  {author} {\bibfnamefont {K.}~\bibnamefont {Ishida}}, \bibinfo {author}
  {\bibfnamefont {G.}~\bibnamefont {Knebel}}, \bibinfo {author} {\bibfnamefont
  {Y.}~\bibnamefont {Tokunaga}},\ and\ \bibinfo {author} {\bibfnamefont
  {Y.}~\bibnamefont {Yanase}},\ }\bibfield  {title} {\bibinfo {title}
  {Unconventional superconductivity in $\text{{UT}e}_2$},\ }\href
  {https://iopscience.iop.org/article/10.1088/1361-648X/ac5863} {\bibfield
  {journal} {\bibinfo  {journal} {Journal of Physics: Condensed Matter}\
  }\textbf {\bibinfo {volume} {34}},\ \bibinfo {pages} {243002} (\bibinfo
  {year} {2022})}\BibitemShut {NoStop}%
\bibitem [{\citenamefont {Lake}\ \emph {et~al.}(2022)\citenamefont {Lake},
  \citenamefont {Patri},\ and\ \citenamefont {Senthil}}]{LakeSenthil2022}%
  \BibitemOpen
  \bibfield  {author} {\bibinfo {author} {\bibfnamefont {E.}~\bibnamefont
  {Lake}}, \bibinfo {author} {\bibfnamefont {A.~S.}\ \bibnamefont {Patri}},\
  and\ \bibinfo {author} {\bibfnamefont {T.}~\bibnamefont {Senthil}},\
  }\bibfield  {title} {\bibinfo {title} {Pairing symmetry of twisted bilayer
  graphene: A phenomenological synthesis},\ }\href
  {https://doi.org/10.1103/PhysRevB.106.104506} {\bibfield  {journal} {\bibinfo
   {journal} {Phys. Rev. B}\ }\textbf {\bibinfo {volume} {106}},\ \bibinfo
  {pages} {104506} (\bibinfo {year} {2022})}\BibitemShut {NoStop}%
\bibitem [{\citenamefont {Fidkowski}\ and\ \citenamefont
  {Kitaev}(2011)}]{FidkowskiKitaev2011}%
  \BibitemOpen
  \bibfield  {author} {\bibinfo {author} {\bibfnamefont {L.}~\bibnamefont
  {Fidkowski}}\ and\ \bibinfo {author} {\bibfnamefont {A.}~\bibnamefont
  {Kitaev}},\ }\bibfield  {title} {\bibinfo {title} {Topological phases of
  {F}ermions in one dimension},\ }\href
  {https://doi.org/10.1103/PhysRevB.83.075103} {\bibfield  {journal} {\bibinfo
  {journal} {Phys. Rev. B}\ }\textbf {\bibinfo {volume} {83}},\ \bibinfo
  {pages} {075103} (\bibinfo {year} {2011})}\BibitemShut {NoStop}%
\bibitem [{\citenamefont {Rachel}(2018)}]{Rachel2018}%
  \BibitemOpen
  \bibfield  {author} {\bibinfo {author} {\bibfnamefont {S.}~\bibnamefont
  {Rachel}},\ }\bibfield  {title} {\bibinfo {title} {Interacting topological
  insulators: a review},\ }\href
  {https://iopscience.iop.org/article/10.1088/1361-6633/aad6a6} {\bibfield
  {journal} {\bibinfo  {journal} {Reports on Progress in Physics}\ }\textbf
  {\bibinfo {volume} {81}},\ \bibinfo {pages} {116501} (\bibinfo {year}
  {2018})}\BibitemShut {NoStop}%
\bibitem [{\citenamefont {Fu}(2010)}]{Fu2010}%
  \BibitemOpen
  \bibfield  {author} {\bibinfo {author} {\bibfnamefont {L.}~\bibnamefont
  {Fu}},\ }\bibfield  {title} {\bibinfo {title} {Electron teleportation via
  {M}ajorana bound states in a mesoscopic superconductor},\ }\href
  {https://journals.aps.org/prl/abstract/10.1103/PhysRevLett.104.056402}
  {\bibfield  {journal} {\bibinfo  {journal} {Phys. Rev. Lett.}\ }\textbf
  {\bibinfo {volume} {104}},\ \bibinfo {pages} {056402} (\bibinfo {year}
  {2010})}\BibitemShut {NoStop}%
\bibitem [{\citenamefont {Kitaev}(2001)}]{Kitaev2001}%
  \BibitemOpen
  \bibfield  {author} {\bibinfo {author} {\bibfnamefont {A.~Y.}\ \bibnamefont
  {Kitaev}},\ }\bibfield  {title} {\bibinfo {title} {Unpaired {M}ajorana
  {F}ermions in quantum wires},\ }\href
  {https://iopscience.iop.org/article/10.1070/1063-7869/44/10S/S29} {\bibfield
  {journal} {\bibinfo  {journal} {Physics-uspekhi}\ }\textbf {\bibinfo {volume}
  {44}},\ \bibinfo {pages} {131} (\bibinfo {year} {2001})}\BibitemShut
  {NoStop}%
\bibitem [{\citenamefont {Alicea}(2012)}]{Alicea2012}%
  \BibitemOpen
  \bibfield  {author} {\bibinfo {author} {\bibfnamefont {J.}~\bibnamefont
  {Alicea}},\ }\bibfield  {title} {\bibinfo {title} {New directions in the
  pursuit of {M}ajorana {F}ermions in solid state systems},\ }\href
  {https://iopscience.iop.org/article/10.1088/0034-4885/75/7/076501} {\bibfield
   {journal} {\bibinfo  {journal} {Reports on progress in physics}\ }\textbf
  {\bibinfo {volume} {75}},\ \bibinfo {pages} {076501} (\bibinfo {year}
  {2012})}\BibitemShut {NoStop}%
\bibitem [{\citenamefont {Oreg}\ and\ \citenamefont
  {Von~Oppen}(2020)}]{OregvonOppen2020}%
  \BibitemOpen
  \bibfield  {author} {\bibinfo {author} {\bibfnamefont {Y.}~\bibnamefont
  {Oreg}}\ and\ \bibinfo {author} {\bibfnamefont {F.}~\bibnamefont
  {Von~Oppen}},\ }\bibfield  {title} {\bibinfo {title} {{M}ajorana zero modes
  in networks of cooper-pair boxes: topologically ordered states and
  topological quantum computation},\ }\href
  {https://www.annualreviews.org/doi/abs/10.1146/annurev-conmatphys-031218-013618}
  {\bibfield  {journal} {\bibinfo  {journal} {Annual Review of Condensed Matter
  Physics}\ }\textbf {\bibinfo {volume} {11}},\ \bibinfo {pages} {397}
  (\bibinfo {year} {2020})}\BibitemShut {NoStop}%
\bibitem [{\citenamefont {Terhal}\ \emph {et~al.}(2012)\citenamefont {Terhal},
  \citenamefont {Hassler},\ and\ \citenamefont
  {DiVincenzo}}]{TerhalDiVincenzo2012}%
  \BibitemOpen
  \bibfield  {author} {\bibinfo {author} {\bibfnamefont {B.~M.}\ \bibnamefont
  {Terhal}}, \bibinfo {author} {\bibfnamefont {F.}~\bibnamefont {Hassler}},\
  and\ \bibinfo {author} {\bibfnamefont {D.~P.}\ \bibnamefont {DiVincenzo}},\
  }\bibfield  {title} {\bibinfo {title} {From {M}ajorana {F}ermions to
  topological order},\ }\href {https://doi.org/10.1103/PhysRevLett.108.260504}
  {\bibfield  {journal} {\bibinfo  {journal} {Phys. Rev. Lett.}\ }\textbf
  {\bibinfo {volume} {108}},\ \bibinfo {pages} {260504} (\bibinfo {year}
  {2012})}\BibitemShut {NoStop}%
\bibitem [{\citenamefont {Sagi}\ \emph {et~al.}(2019)\citenamefont {Sagi},
  \citenamefont {Ebisu}, \citenamefont {Tanaka}, \citenamefont {Stern},\ and\
  \citenamefont {Oreg}}]{SagiOreg2019}%
  \BibitemOpen
  \bibfield  {author} {\bibinfo {author} {\bibfnamefont {E.}~\bibnamefont
  {Sagi}}, \bibinfo {author} {\bibfnamefont {H.}~\bibnamefont {Ebisu}},
  \bibinfo {author} {\bibfnamefont {Y.}~\bibnamefont {Tanaka}}, \bibinfo
  {author} {\bibfnamefont {A.}~\bibnamefont {Stern}},\ and\ \bibinfo {author}
  {\bibfnamefont {Y.}~\bibnamefont {Oreg}},\ }\bibfield  {title} {\bibinfo
  {title} {Spin liquids from {M}ajorana zero modes in a cooper-pair box},\
  }\href {https://doi.org/10.1103/PhysRevB.99.075107} {\bibfield  {journal}
  {\bibinfo  {journal} {Phys. Rev. B}\ }\textbf {\bibinfo {volume} {99}},\
  \bibinfo {pages} {075107} (\bibinfo {year} {2019})}\BibitemShut {NoStop}%
\bibitem [{\citenamefont {Ziesen}\ \emph {et~al.}(2019)\citenamefont {Ziesen},
  \citenamefont {Hassler},\ and\ \citenamefont {Roy}}]{ZiesenRoy2019}%
  \BibitemOpen
  \bibfield  {author} {\bibinfo {author} {\bibfnamefont {A.}~\bibnamefont
  {Ziesen}}, \bibinfo {author} {\bibfnamefont {F.}~\bibnamefont {Hassler}},\
  and\ \bibinfo {author} {\bibfnamefont {A.}~\bibnamefont {Roy}},\ }\bibfield
  {title} {\bibinfo {title} {Topological ordering in the {M}ajorana toric
  code},\ }\href {https://doi.org/10.1103/PhysRevB.100.104508} {\bibfield
  {journal} {\bibinfo  {journal} {Phys. Rev. B}\ }\textbf {\bibinfo {volume}
  {100}},\ \bibinfo {pages} {104508} (\bibinfo {year} {2019})}\BibitemShut
  {NoStop}%
\bibitem [{\citenamefont {K\"onig}\ \emph {et~al.}(2020)\citenamefont
  {K\"onig}, \citenamefont {Coleman},\ and\ \citenamefont
  {Tsvelik}}]{KoenigTsvelik2020b}%
  \BibitemOpen
  \bibfield  {author} {\bibinfo {author} {\bibfnamefont {E.~J.}\ \bibnamefont
  {K\"onig}}, \bibinfo {author} {\bibfnamefont {P.}~\bibnamefont {Coleman}},\
  and\ \bibinfo {author} {\bibfnamefont {A.~M.}\ \bibnamefont {Tsvelik}},\
  }\bibfield  {title} {\bibinfo {title} {Soluble limit and criticality of
  {F}ermions in $\mathbf{Z}_{2}$ gauge theories},\ }\href
  {https://doi.org/10.1103/PhysRevB.102.155143} {\bibfield  {journal} {\bibinfo
   {journal} {Phys. Rev. B}\ }\textbf {\bibinfo {volume} {102}},\ \bibinfo
  {pages} {155143} (\bibinfo {year} {2020})}\BibitemShut {NoStop}%
\bibitem [{\citenamefont {B{\'e}ri}\ and\ \citenamefont
  {Cooper}(2012)}]{BeriCooper2012}%
  \BibitemOpen
  \bibfield  {author} {\bibinfo {author} {\bibfnamefont {B.}~\bibnamefont
  {B{\'e}ri}}\ and\ \bibinfo {author} {\bibfnamefont {N.}~\bibnamefont
  {Cooper}},\ }\bibfield  {title} {\bibinfo {title} {Topological {K}ondo effect
  with {M}ajorana {F}ermions},\ }\href
  {https://www.google.com/search?client=firefox-b-e&q=opological+{K}ondo+effect+with+{M}ajorana+{F}ermions}
  {\bibfield  {journal} {\bibinfo  {journal} {Physical Review Letters}\
  }\textbf {\bibinfo {volume} {109}},\ \bibinfo {pages} {156803} (\bibinfo
  {year} {2012})}\BibitemShut {NoStop}%
\bibitem [{\citenamefont {Altland}\ \emph
  {et~al.}(2014{\natexlab{a}})\citenamefont {Altland}, \citenamefont {B\'eri},
  \citenamefont {Egger},\ and\ \citenamefont {Tsvelik}}]{AltlandTsvelik2014}%
  \BibitemOpen
  \bibfield  {author} {\bibinfo {author} {\bibfnamefont {A.}~\bibnamefont
  {Altland}}, \bibinfo {author} {\bibfnamefont {B.}~\bibnamefont {B\'eri}},
  \bibinfo {author} {\bibfnamefont {R.}~\bibnamefont {Egger}},\ and\ \bibinfo
  {author} {\bibfnamefont {A.~M.}\ \bibnamefont {Tsvelik}},\ }\bibfield
  {title} {\bibinfo {title} {Multichannel {K}ondo impurity dynamics in a
  {M}ajorana device},\ }\href
  {https://journals.aps.org/prl/abstract/10.1103/PhysRevLett.113.076401}
  {\bibfield  {journal} {\bibinfo  {journal} {Phys. Rev. Lett.}\ }\textbf
  {\bibinfo {volume} {113}},\ \bibinfo {pages} {076401} (\bibinfo {year}
  {2014}{\natexlab{a}})}\BibitemShut {NoStop}%
\bibitem [{\citenamefont {Altland}\ \emph
  {et~al.}(2014{\natexlab{b}})\citenamefont {Altland}, \citenamefont
  {B{\'e}ri}, \citenamefont {Egger},\ and\ \citenamefont
  {Tsvelik}}]{AltlandTsvelik2014b}%
  \BibitemOpen
  \bibfield  {author} {\bibinfo {author} {\bibfnamefont {A.}~\bibnamefont
  {Altland}}, \bibinfo {author} {\bibfnamefont {B.}~\bibnamefont {B{\'e}ri}},
  \bibinfo {author} {\bibfnamefont {R.}~\bibnamefont {Egger}},\ and\ \bibinfo
  {author} {\bibfnamefont {A.}~\bibnamefont {Tsvelik}},\ }\bibfield  {title}
  {\bibinfo {title} {Bethe ansatz solution of the topological {K}ondo model},\
  }\href {https://iopscience.iop.org/article/10.1088/1751-8113/47/26/265001}
  {\bibfield  {journal} {\bibinfo  {journal} {Journal of Physics A:
  Mathematical and Theoretical}\ }\textbf {\bibinfo {volume} {47}},\ \bibinfo
  {pages} {265001} (\bibinfo {year} {2014}{\natexlab{b}})}\BibitemShut
  {NoStop}%
\bibitem [{\citenamefont {Buccheri}\ \emph {et~al.}(2015)\citenamefont
  {Buccheri}, \citenamefont {Babujian}, \citenamefont {Korepin}, \citenamefont
  {Sodano},\ and\ \citenamefont {Trombettoni}}]{Buccheri2015}%
  \BibitemOpen
  \bibfield  {author} {\bibinfo {author} {\bibfnamefont {F.}~\bibnamefont
  {Buccheri}}, \bibinfo {author} {\bibfnamefont {H.}~\bibnamefont {Babujian}},
  \bibinfo {author} {\bibfnamefont {V.~E.}\ \bibnamefont {Korepin}}, \bibinfo
  {author} {\bibfnamefont {P.}~\bibnamefont {Sodano}},\ and\ \bibinfo {author}
  {\bibfnamefont {A.}~\bibnamefont {Trombettoni}},\ }\bibfield  {title}
  {\bibinfo {title} {Thermodynamics of the topological {K}ondo model},\ }\href
  {https://www.sciencedirect.com/science/article/pii/S0550321315001406}
  {\bibfield  {journal} {\bibinfo  {journal} {Nuclear Physics B}\ }\textbf
  {\bibinfo {volume} {896}},\ \bibinfo {pages} {52} (\bibinfo {year}
  {2015})}\BibitemShut {NoStop}%
\bibitem [{\citenamefont {Gau}\ \emph {et~al.}(2018)\citenamefont {Gau},
  \citenamefont {Plugge},\ and\ \citenamefont {Egger}}]{GauEgger2018}%
  \BibitemOpen
  \bibfield  {author} {\bibinfo {author} {\bibfnamefont {M.}~\bibnamefont
  {Gau}}, \bibinfo {author} {\bibfnamefont {S.}~\bibnamefont {Plugge}},\ and\
  \bibinfo {author} {\bibfnamefont {R.}~\bibnamefont {Egger}},\ }\bibfield
  {title} {\bibinfo {title} {Quantum transport in coupled {M}ajorana box
  systems},\ }\href {https://doi.org/10.1103/PhysRevB.97.184506} {\bibfield
  {journal} {\bibinfo  {journal} {Phys. Rev. B}\ }\textbf {\bibinfo {volume}
  {97}},\ \bibinfo {pages} {184506} (\bibinfo {year} {2018})}\BibitemShut
  {NoStop}%
\bibitem [{\citenamefont {Papaj}\ \emph {et~al.}(2019)\citenamefont {Papaj},
  \citenamefont {Zhu},\ and\ \citenamefont {Fu}}]{PapajFu2019}%
  \BibitemOpen
  \bibfield  {author} {\bibinfo {author} {\bibfnamefont {M.}~\bibnamefont
  {Papaj}}, \bibinfo {author} {\bibfnamefont {Z.}~\bibnamefont {Zhu}},\ and\
  \bibinfo {author} {\bibfnamefont {L.}~\bibnamefont {Fu}},\ }\bibfield
  {title} {\bibinfo {title} {Multichannel charge {K}ondo effect and
  non-{F}ermi-liquid fixed points in conventional and topological
  superconductor islands},\ }\href {https://doi.org/10.1103/PhysRevB.99.014512}
  {\bibfield  {journal} {\bibinfo  {journal} {Phys. Rev. B}\ }\textbf {\bibinfo
  {volume} {99}},\ \bibinfo {pages} {014512} (\bibinfo {year}
  {2019})}\BibitemShut {NoStop}%
\bibitem [{\citenamefont {Li}\ \emph {et~al.}(2023{\natexlab{a}})\citenamefont
  {Li}, \citenamefont {Oreg},\ and\ \citenamefont
  {V\"ayrynen}}]{LiVayrynen2023b}%
  \BibitemOpen
  \bibfield  {author} {\bibinfo {author} {\bibfnamefont {G.}~\bibnamefont
  {Li}}, \bibinfo {author} {\bibfnamefont {Y.}~\bibnamefont {Oreg}},\ and\
  \bibinfo {author} {\bibfnamefont {J.~I.}\ \bibnamefont {V\"ayrynen}},\
  }\bibfield  {title} {\bibinfo {title} {Multichannel topological {K}ondo
  effect},\ }\href {https://doi.org/10.1103/PhysRevLett.130.066302} {\bibfield
  {journal} {\bibinfo  {journal} {Phys. Rev. Lett.}\ }\textbf {\bibinfo
  {volume} {130}},\ \bibinfo {pages} {066302} (\bibinfo {year}
  {2023}{\natexlab{a}})}\BibitemShut {NoStop}%
\bibitem [{\citenamefont {Wauters}\ \emph {et~al.}(2023)\citenamefont
  {Wauters}, \citenamefont {Chung}, \citenamefont {Maffi},\ and\ \citenamefont
  {Burrello}}]{WautersBurrello2023}%
  \BibitemOpen
  \bibfield  {author} {\bibinfo {author} {\bibfnamefont {M.~M.}\ \bibnamefont
  {Wauters}}, \bibinfo {author} {\bibfnamefont {C.-M.}\ \bibnamefont {Chung}},
  \bibinfo {author} {\bibfnamefont {L.}~\bibnamefont {Maffi}},\ and\ \bibinfo
  {author} {\bibfnamefont {M.}~\bibnamefont {Burrello}},\ }\bibfield  {title}
  {\bibinfo {title} {The topological {K}ondo model out of equilibrium},\ }\href
  {https://arxiv.org/abs/2307.03773} {\bibfield  {journal} {\bibinfo  {journal}
  {arXiv:2307.03773}\ } (\bibinfo {year} {2023})}\BibitemShut {NoStop}%
\bibitem [{\citenamefont {Andrei}(1980)}]{Andrei1980}%
  \BibitemOpen
  \bibfield  {author} {\bibinfo {author} {\bibfnamefont {N.}~\bibnamefont
  {Andrei}},\ }\bibfield  {title} {\bibinfo {title} {Diagonalization of the
  {K}ondo hamiltonian},\ }\href
  {https://journals.aps.org/prl/abstract/10.1103/PhysRevLett.45.379} {\bibfield
   {journal} {\bibinfo  {journal} {Phys. Rev. Lett.}\ }\textbf {\bibinfo
  {volume} {45}},\ \bibinfo {pages} {379} (\bibinfo {year} {1980})}\BibitemShut
  {NoStop}%
\bibitem [{\citenamefont {Vigman}(1980)}]{Vigman1980}%
  \BibitemOpen
  \bibfield  {author} {\bibinfo {author} {\bibfnamefont {P.}~\bibnamefont
  {Vigman}},\ }\bibfield  {title} {\bibinfo {title} {Exact solution of
  $\text{s}.\text{d}$ exchange model at {T} = 0},\ }\href
  {https://iopscience.iop.org/article/10.1088/0022-3719/14/10/014} {\bibfield
  {journal} {\bibinfo  {journal} {Soviet Journal of Experimental and
  Theoretical Physics Letters}\ }\textbf {\bibinfo {volume} {31}},\ \bibinfo
  {pages} {364} (\bibinfo {year} {1980})}\BibitemShut {NoStop}%
\bibitem [{\citenamefont {Andrei}\ and\ \citenamefont
  {Destri}(1984)}]{AndreiDestri1984}%
  \BibitemOpen
  \bibfield  {author} {\bibinfo {author} {\bibfnamefont {N.}~\bibnamefont
  {Andrei}}\ and\ \bibinfo {author} {\bibfnamefont {C.}~\bibnamefont
  {Destri}},\ }\bibfield  {title} {\bibinfo {title} {Solution of the
  multichannel {K}ondo problem},\ }\href
  {https://journals.aps.org/prl/abstract/10.1103/PhysRevLett.52.364} {\bibfield
   {journal} {\bibinfo  {journal} {Physical review letters}\ }\textbf {\bibinfo
  {volume} {52}},\ \bibinfo {pages} {364} (\bibinfo {year} {1984})}\BibitemShut
  {NoStop}%
\bibitem [{\citenamefont {Tsvelick}\ and\ \citenamefont
  {Wiegmann}(1985)}]{TsvelickWiegmann1985}%
  \BibitemOpen
  \bibfield  {author} {\bibinfo {author} {\bibfnamefont {A.}~\bibnamefont
  {Tsvelick}}\ and\ \bibinfo {author} {\bibfnamefont {P.}~\bibnamefont
  {Wiegmann}},\ }\bibfield  {title} {\bibinfo {title} {Exact solution of the
  multichannel {K}ondo problem, scaling, and integrability},\ }\href
  {https://link.springer.com/article/10.1007/BF01017853} {\bibfield  {journal}
  {\bibinfo  {journal} {Journal of Statistical Physics}\ }\textbf {\bibinfo
  {volume} {38}},\ \bibinfo {pages} {125} (\bibinfo {year} {1985})}\BibitemShut
  {NoStop}%
\bibitem [{\citenamefont {Affleck}\ and\ \citenamefont
  {Ludwig}(1991{\natexlab{a}})}]{AffleckLudwig1991a}%
  \BibitemOpen
  \bibfield  {author} {\bibinfo {author} {\bibfnamefont {I.}~\bibnamefont
  {Affleck}}\ and\ \bibinfo {author} {\bibfnamefont {A.~W.}\ \bibnamefont
  {Ludwig}},\ }\bibfield  {title} {\bibinfo {title} {Critical theory of
  overscreened {K}ondo fixed points},\ }\href
  {https://www.sciencedirect.com/science/article/pii/055032139190419X}
  {\bibfield  {journal} {\bibinfo  {journal} {Nuclear Physics B}\ }\textbf
  {\bibinfo {volume} {360}},\ \bibinfo {pages} {641} (\bibinfo {year}
  {1991}{\natexlab{a}})}\BibitemShut {NoStop}%
\bibitem [{\citenamefont {Affleck}\ and\ \citenamefont
  {Ludwig}(1991{\natexlab{b}})}]{AffleckLudwig1991b}%
  \BibitemOpen
  \bibfield  {author} {\bibinfo {author} {\bibfnamefont {I.}~\bibnamefont
  {Affleck}}\ and\ \bibinfo {author} {\bibfnamefont {A.~W.}\ \bibnamefont
  {Ludwig}},\ }\bibfield  {title} {\bibinfo {title} {The {K}ondo effect,
  conformal field theory and fusion rules},\ }\href
  {https://www.sciencedirect.com/science/article/pii/055032139190109B}
  {\bibfield  {journal} {\bibinfo  {journal} {Nuclear Physics B}\ }\textbf
  {\bibinfo {volume} {352}},\ \bibinfo {pages} {849} (\bibinfo {year}
  {1991}{\natexlab{b}})}\BibitemShut {NoStop}%
\bibitem [{\citenamefont {Affleck}\ and\ \citenamefont
  {Ludwig}(1993)}]{AffleckLudwig1993}%
  \BibitemOpen
  \bibfield  {author} {\bibinfo {author} {\bibfnamefont {I.}~\bibnamefont
  {Affleck}}\ and\ \bibinfo {author} {\bibfnamefont {A.~W.}\ \bibnamefont
  {Ludwig}},\ }\bibfield  {title} {\bibinfo {title} {Exact
  conformal-field-theory results on the multichannel {K}ondo effect:
  Single-{F}ermion green’s function, self-energy, and resistivity},\ }\href
  {https://journals.aps.org/prb/abstract/10.1103/PhysRevB.48.7297} {\bibfield
  {journal} {\bibinfo  {journal} {Physical Review B}\ }\textbf {\bibinfo
  {volume} {48}},\ \bibinfo {pages} {7297} (\bibinfo {year}
  {1993})}\BibitemShut {NoStop}%
\bibitem [{\citenamefont {Jerez}\ \emph {et~al.}(1998)\citenamefont {Jerez},
  \citenamefont {Andrei},\ and\ \citenamefont {Zar{\'a}nd}}]{JerezZarand1998}%
  \BibitemOpen
  \bibfield  {author} {\bibinfo {author} {\bibfnamefont {A.}~\bibnamefont
  {Jerez}}, \bibinfo {author} {\bibfnamefont {N.}~\bibnamefont {Andrei}},\ and\
  \bibinfo {author} {\bibfnamefont {G.}~\bibnamefont {Zar{\'a}nd}},\ }\bibfield
   {title} {\bibinfo {title} {Solution of the multichannel coqblin-schrieffer
  impurity model and application to multilevel systems},\ }\href
  {https://journals.aps.org/prb/abstract/10.1103/PhysRevB.58.3814} {\bibfield
  {journal} {\bibinfo  {journal} {Physical Review B}\ }\textbf {\bibinfo
  {volume} {58}},\ \bibinfo {pages} {3814} (\bibinfo {year}
  {1998})}\BibitemShut {NoStop}%
\bibitem [{\citenamefont {Keller}\ \emph {et~al.}(2014)\citenamefont {Keller},
  \citenamefont {Amasha}, \citenamefont {Weymann}, \citenamefont {Moca},
  \citenamefont {Rau}, \citenamefont {Katine}, \citenamefont {Shtrikman},
  \citenamefont {Zar{\'a}nd},\ and\ \citenamefont
  {Goldhaber-Gordon}}]{KellerGoldhaberGordon2014}%
  \BibitemOpen
  \bibfield  {author} {\bibinfo {author} {\bibfnamefont {A.}~\bibnamefont
  {Keller}}, \bibinfo {author} {\bibfnamefont {S.}~\bibnamefont {Amasha}},
  \bibinfo {author} {\bibfnamefont {I.}~\bibnamefont {Weymann}}, \bibinfo
  {author} {\bibfnamefont {C.}~\bibnamefont {Moca}}, \bibinfo {author}
  {\bibfnamefont {I.}~\bibnamefont {Rau}}, \bibinfo {author} {\bibfnamefont
  {J.}~\bibnamefont {Katine}}, \bibinfo {author} {\bibfnamefont
  {H.}~\bibnamefont {Shtrikman}}, \bibinfo {author} {\bibfnamefont
  {G.}~\bibnamefont {Zar{\'a}nd}},\ and\ \bibinfo {author} {\bibfnamefont
  {D.}~\bibnamefont {Goldhaber-Gordon}},\ }\bibfield  {title} {\bibinfo {title}
  {Emergent ${SU}(4)$ {K}ondo physics in a spin--charge-entangled double
  quantum dot},\ }\href {https://www.nature.com/articles/nphys2844} {\bibfield
  {journal} {\bibinfo  {journal} {Nature Physics}\ }\textbf {\bibinfo {volume}
  {10}},\ \bibinfo {pages} {145} (\bibinfo {year} {2014})}\BibitemShut
  {NoStop}%
\bibitem [{\citenamefont {Keller}\ \emph {et~al.}(2015)\citenamefont {Keller},
  \citenamefont {Peeters}, \citenamefont {Moca}, \citenamefont {Weymann},
  \citenamefont {Mahalu}, \citenamefont {Umansky}, \citenamefont {Zar{\'a}nd},\
  and\ \citenamefont {Goldhaber-Gordon}}]{KellerGoldhaberGordon2015}%
  \BibitemOpen
  \bibfield  {author} {\bibinfo {author} {\bibfnamefont {A.}~\bibnamefont
  {Keller}}, \bibinfo {author} {\bibfnamefont {L.}~\bibnamefont {Peeters}},
  \bibinfo {author} {\bibfnamefont {C.}~\bibnamefont {Moca}}, \bibinfo {author}
  {\bibfnamefont {I.}~\bibnamefont {Weymann}}, \bibinfo {author} {\bibfnamefont
  {D.}~\bibnamefont {Mahalu}}, \bibinfo {author} {\bibfnamefont
  {V.}~\bibnamefont {Umansky}}, \bibinfo {author} {\bibfnamefont
  {G.}~\bibnamefont {Zar{\'a}nd}},\ and\ \bibinfo {author} {\bibfnamefont
  {D.}~\bibnamefont {Goldhaber-Gordon}},\ }\bibfield  {title} {\bibinfo {title}
  {Universal {F}ermi liquid crossover and quantum criticality in a mesoscopic
  system},\ }\href {https://www.nature.com/articles/nature15261} {\bibfield
  {journal} {\bibinfo  {journal} {Nature}\ }\textbf {\bibinfo {volume} {526}},\
  \bibinfo {pages} {237} (\bibinfo {year} {2015})}\BibitemShut {NoStop}%
\bibitem [{\citenamefont {Pouse}\ \emph {et~al.}(2023)\citenamefont {Pouse},
  \citenamefont {Peeters}, \citenamefont {Hsueh}, \citenamefont {Gennser},
  \citenamefont {Cavanna}, \citenamefont {Kastner}, \citenamefont {Mitchell},\
  and\ \citenamefont {Goldhaber-Gordon}}]{PouseGoldhaberGordon2021}%
  \BibitemOpen
  \bibfield  {author} {\bibinfo {author} {\bibfnamefont {W.}~\bibnamefont
  {Pouse}}, \bibinfo {author} {\bibfnamefont {L.}~\bibnamefont {Peeters}},
  \bibinfo {author} {\bibfnamefont {C.~L.}\ \bibnamefont {Hsueh}}, \bibinfo
  {author} {\bibfnamefont {U.}~\bibnamefont {Gennser}}, \bibinfo {author}
  {\bibfnamefont {A.}~\bibnamefont {Cavanna}}, \bibinfo {author} {\bibfnamefont
  {M.~A.}\ \bibnamefont {Kastner}}, \bibinfo {author} {\bibfnamefont {A.~K.}\
  \bibnamefont {Mitchell}},\ and\ \bibinfo {author} {\bibfnamefont
  {D.}~\bibnamefont {Goldhaber-Gordon}},\ }\bibfield  {title} {\bibinfo {title}
  {Quantum simulation of an exotic quantum critical point in a two-site charge
  {K}ondo circuit},\ }\href {https://doi.org/10.1038/s41567-022-01905-4}
  {\bibfield  {journal} {\bibinfo  {journal} {Nature Physics}\ }\textbf
  {\bibinfo {volume} {19}},\ \bibinfo {pages} {492} (\bibinfo {year}
  {2023})}\BibitemShut {NoStop}%
\bibitem [{\citenamefont {Lopes}\ \emph {et~al.}(2020)\citenamefont {Lopes},
  \citenamefont {Affleck},\ and\ \citenamefont {Sela}}]{LopesSela2020}%
  \BibitemOpen
  \bibfield  {author} {\bibinfo {author} {\bibfnamefont {P.~L.~S.}\
  \bibnamefont {Lopes}}, \bibinfo {author} {\bibfnamefont {I.}~\bibnamefont
  {Affleck}},\ and\ \bibinfo {author} {\bibfnamefont {E.}~\bibnamefont
  {Sela}},\ }\bibfield  {title} {\bibinfo {title} {Anyons in multichannel
  {K}ondo systems},\ }\href
  {https://journals.aps.org/prb/abstract/10.1103/PhysRevB.101.085141}
  {\bibfield  {journal} {\bibinfo  {journal} {Phys. Rev. B}\ }\textbf {\bibinfo
  {volume} {101}},\ \bibinfo {pages} {085141} (\bibinfo {year}
  {2020})}\BibitemShut {NoStop}%
\bibitem [{\citenamefont {Komijani}(2020)}]{Komijani2020}%
  \BibitemOpen
  \bibfield  {author} {\bibinfo {author} {\bibfnamefont {Y.}~\bibnamefont
  {Komijani}},\ }\bibfield  {title} {\bibinfo {title} {Isolating {K}ondo anyons
  for topological quantum computation},\ }\href
  {https://journals.aps.org/prb/abstract/10.1103/PhysRevB.101.235131}
  {\bibfield  {journal} {\bibinfo  {journal} {Physical Review B}\ }\textbf
  {\bibinfo {volume} {101}},\ \bibinfo {pages} {235131} (\bibinfo {year}
  {2020})}\BibitemShut {NoStop}%
\bibitem [{\citenamefont {Gabay}\ \emph {et~al.}(2022)\citenamefont {Gabay},
  \citenamefont {Han}, \citenamefont {Lopes}, \citenamefont {Affleck},\ and\
  \citenamefont {Sela}}]{GabaySela2022}%
  \BibitemOpen
  \bibfield  {author} {\bibinfo {author} {\bibfnamefont {D.}~\bibnamefont
  {Gabay}}, \bibinfo {author} {\bibfnamefont {C.}~\bibnamefont {Han}}, \bibinfo
  {author} {\bibfnamefont {P.~L.~S.}\ \bibnamefont {Lopes}}, \bibinfo {author}
  {\bibfnamefont {I.}~\bibnamefont {Affleck}},\ and\ \bibinfo {author}
  {\bibfnamefont {E.}~\bibnamefont {Sela}},\ }\bibfield  {title} {\bibinfo
  {title} {Multi-impurity chiral {K}ondo model: Correlation functions and anyon
  fusion rules},\ }\href
  {https://journals.aps.org/prb/abstract/10.1103/PhysRevB.105.035151}
  {\bibfield  {journal} {\bibinfo  {journal} {Phys. Rev. B}\ }\textbf {\bibinfo
  {volume} {105}},\ \bibinfo {pages} {035151} (\bibinfo {year}
  {2022})}\BibitemShut {NoStop}%
\bibitem [{\citenamefont {Lotem}\ \emph {et~al.}(2022)\citenamefont {Lotem},
  \citenamefont {Sela},\ and\ \citenamefont {Goldstein}}]{LotemGoldstein2022}%
  \BibitemOpen
  \bibfield  {author} {\bibinfo {author} {\bibfnamefont {M.}~\bibnamefont
  {Lotem}}, \bibinfo {author} {\bibfnamefont {E.}~\bibnamefont {Sela}},\ and\
  \bibinfo {author} {\bibfnamefont {M.}~\bibnamefont {Goldstein}},\ }\bibfield
  {title} {\bibinfo {title} {Manipulating non-abelian anyons in a chiral
  multichannel {K}ondo model},\ }\href
  {https://doi.org/10.1103/PhysRevLett.129.227703} {\bibfield  {journal}
  {\bibinfo  {journal} {Phys. Rev. Lett.}\ }\textbf {\bibinfo {volume} {129}},\
  \bibinfo {pages} {227703} (\bibinfo {year} {2022})}\BibitemShut {NoStop}%
\bibitem [{\citenamefont {Lotem}\ \emph {et~al.}(2023)\citenamefont {Lotem},
  \citenamefont {Sela},\ and\ \citenamefont {Goldstein}}]{LotemGoldstein2022b}%
  \BibitemOpen
  \bibfield  {author} {\bibinfo {author} {\bibfnamefont {M.}~\bibnamefont
  {Lotem}}, \bibinfo {author} {\bibfnamefont {E.}~\bibnamefont {Sela}},\ and\
  \bibinfo {author} {\bibfnamefont {M.}~\bibnamefont {Goldstein}},\ }\bibfield
  {title} {\bibinfo {title} {Chiral numerical renormalization group},\ }\href
  {https://doi.org/10.1103/PhysRevB.107.155417} {\bibfield  {journal} {\bibinfo
   {journal} {Phys. Rev. B}\ }\textbf {\bibinfo {volume} {107}},\ \bibinfo
  {pages} {155417} (\bibinfo {year} {2023})}\BibitemShut {NoStop}%
\bibitem [{\citenamefont {Potok}\ \emph {et~al.}(2007)\citenamefont {Potok},
  \citenamefont {Rau}, \citenamefont {Shtrikman}, \citenamefont {Oreg},\ and\
  \citenamefont {Goldhaber-Gordon}}]{PotokGoldhaberGordon2007}%
  \BibitemOpen
  \bibfield  {author} {\bibinfo {author} {\bibfnamefont {R.}~\bibnamefont
  {Potok}}, \bibinfo {author} {\bibfnamefont {I.}~\bibnamefont {Rau}}, \bibinfo
  {author} {\bibfnamefont {H.}~\bibnamefont {Shtrikman}}, \bibinfo {author}
  {\bibfnamefont {Y.}~\bibnamefont {Oreg}},\ and\ \bibinfo {author}
  {\bibfnamefont {D.}~\bibnamefont {Goldhaber-Gordon}},\ }\bibfield  {title}
  {\bibinfo {title} {Observation of the two-channel {K}ondo effect},\ }\href
  {https://www.nature.com/articles/nature05556} {\bibfield  {journal} {\bibinfo
   {journal} {Nature}\ }\textbf {\bibinfo {volume} {446}},\ \bibinfo {pages}
  {167} (\bibinfo {year} {2007})}\BibitemShut {NoStop}%
\bibitem [{\citenamefont {Iftikhar}\ \emph {et~al.}(2015)\citenamefont
  {Iftikhar}, \citenamefont {Jezouin}, \citenamefont {Anthore}, \citenamefont
  {Gennser}, \citenamefont {Parmentier}, \citenamefont {Cavanna},\ and\
  \citenamefont {Pierre}}]{IftikharPierre2015}%
  \BibitemOpen
  \bibfield  {author} {\bibinfo {author} {\bibfnamefont {Z.}~\bibnamefont
  {Iftikhar}}, \bibinfo {author} {\bibfnamefont {S.}~\bibnamefont {Jezouin}},
  \bibinfo {author} {\bibfnamefont {A.}~\bibnamefont {Anthore}}, \bibinfo
  {author} {\bibfnamefont {U.}~\bibnamefont {Gennser}}, \bibinfo {author}
  {\bibfnamefont {F.}~\bibnamefont {Parmentier}}, \bibinfo {author}
  {\bibfnamefont {A.}~\bibnamefont {Cavanna}},\ and\ \bibinfo {author}
  {\bibfnamefont {F.}~\bibnamefont {Pierre}},\ }\bibfield  {title} {\bibinfo
  {title} {Two-channel {K}ondo effect and renormalization flow with macroscopic
  quantum charge states},\ }\href {https://www.nature.com/articles/nature15384}
  {\bibfield  {journal} {\bibinfo  {journal} {Nature}\ }\textbf {\bibinfo
  {volume} {526}},\ \bibinfo {pages} {233} (\bibinfo {year}
  {2015})}\BibitemShut {NoStop}%
\bibitem [{\citenamefont {Iftikhar}\ \emph {et~al.}(2018)\citenamefont
  {Iftikhar}, \citenamefont {Anthore}, \citenamefont {Mitchell}, \citenamefont
  {Parmentier}, \citenamefont {Gennser}, \citenamefont {Ouerghi}, \citenamefont
  {Cavanna}, \citenamefont {Mora}, \citenamefont {Simon},\ and\ \citenamefont
  {Pierre}}]{IftikharPierre2018}%
  \BibitemOpen
  \bibfield  {author} {\bibinfo {author} {\bibfnamefont {Z.}~\bibnamefont
  {Iftikhar}}, \bibinfo {author} {\bibfnamefont {A.}~\bibnamefont {Anthore}},
  \bibinfo {author} {\bibfnamefont {A.}~\bibnamefont {Mitchell}}, \bibinfo
  {author} {\bibfnamefont {F.}~\bibnamefont {Parmentier}}, \bibinfo {author}
  {\bibfnamefont {U.}~\bibnamefont {Gennser}}, \bibinfo {author} {\bibfnamefont
  {A.}~\bibnamefont {Ouerghi}}, \bibinfo {author} {\bibfnamefont
  {A.}~\bibnamefont {Cavanna}}, \bibinfo {author} {\bibfnamefont
  {C.}~\bibnamefont {Mora}}, \bibinfo {author} {\bibfnamefont {P.}~\bibnamefont
  {Simon}},\ and\ \bibinfo {author} {\bibfnamefont {F.}~\bibnamefont
  {Pierre}},\ }\bibfield  {title} {\bibinfo {title} {Tunable quantum
  criticality and super-ballistic transport in a “charge” {K}ondo
  circuit},\ }\href {https://www.science.org/doi/10.1126/science.aan5592}
  {\bibfield  {journal} {\bibinfo  {journal} {Science}\ }\textbf {\bibinfo
  {volume} {360}},\ \bibinfo {pages} {1315} (\bibinfo {year}
  {2018})}\BibitemShut {NoStop}%
\bibitem [{\citenamefont {Nozieres}\ and\ \citenamefont
  {Blandin}(1980)}]{NozieresBlandin1980}%
  \BibitemOpen
  \bibfield  {author} {\bibinfo {author} {\bibfnamefont {P.}~\bibnamefont
  {Nozieres}}\ and\ \bibinfo {author} {\bibfnamefont {A.}~\bibnamefont
  {Blandin}},\ }\bibfield  {title} {\bibinfo {title} {{K}ondo effect in real
  metals},\ }\href {https://hal.science/jpa-00209235} {\bibfield  {journal}
  {\bibinfo  {journal} {Journal de Physique}\ }\textbf {\bibinfo {volume}
  {41}},\ \bibinfo {pages} {193} (\bibinfo {year} {1980})}\BibitemShut
  {NoStop}%
\bibitem [{\citenamefont {Mitchell}\ \emph {et~al.}(2021)\citenamefont
  {Mitchell}, \citenamefont {Liberman}, \citenamefont {Sela},\ and\
  \citenamefont {Affleck}}]{MitchellAffleck2021}%
  \BibitemOpen
  \bibfield  {author} {\bibinfo {author} {\bibfnamefont {A.~K.}\ \bibnamefont
  {Mitchell}}, \bibinfo {author} {\bibfnamefont {A.}~\bibnamefont {Liberman}},
  \bibinfo {author} {\bibfnamefont {E.}~\bibnamefont {Sela}},\ and\ \bibinfo
  {author} {\bibfnamefont {I.}~\bibnamefont {Affleck}},\ }\bibfield  {title}
  {\bibinfo {title} {{SO}(5) non-{F}ermi liquid in a coulomb box device},\
  }\href {https://journals.aps.org/prl/abstract/10.1103/PhysRevLett.126.147702}
  {\bibfield  {journal} {\bibinfo  {journal} {Phys. Rev. Lett.}\ }\textbf
  {\bibinfo {volume} {126}},\ \bibinfo {pages} {147702} (\bibinfo {year}
  {2021})}\BibitemShut {NoStop}%
\bibitem [{\citenamefont {Liberman}\ \emph {et~al.}(2021)\citenamefont
  {Liberman}, \citenamefont {Mitchell}, \citenamefont {Affleck},\ and\
  \citenamefont {Sela}}]{LibermanSela2021}%
  \BibitemOpen
  \bibfield  {author} {\bibinfo {author} {\bibfnamefont {A.}~\bibnamefont
  {Liberman}}, \bibinfo {author} {\bibfnamefont {A.~K.}\ \bibnamefont
  {Mitchell}}, \bibinfo {author} {\bibfnamefont {I.}~\bibnamefont {Affleck}},\
  and\ \bibinfo {author} {\bibfnamefont {E.}~\bibnamefont {Sela}},\ }\bibfield
  {title} {\bibinfo {title} {{SO}(5) critical point in a spin-flavor {K}ondo
  device: Bosonization and re{F}ermionization solution},\ }\href
  {https://journals.aps.org/prb/abstract/10.1103/PhysRevB.103.195131}
  {\bibfield  {journal} {\bibinfo  {journal} {Phys. Rev. B}\ }\textbf {\bibinfo
  {volume} {103}},\ \bibinfo {pages} {195131} (\bibinfo {year}
  {2021})}\BibitemShut {NoStop}%
\bibitem [{\citenamefont {Li}\ \emph {et~al.}(2023{\natexlab{b}})\citenamefont
  {Li}, \citenamefont {K\"onig},\ and\ \citenamefont
  {V\"ayrynen}}]{LiVayrynen2023}%
  \BibitemOpen
  \bibfield  {author} {\bibinfo {author} {\bibfnamefont {G.}~\bibnamefont
  {Li}}, \bibinfo {author} {\bibfnamefont {E.~J.}\ \bibnamefont {K\"onig}},\
  and\ \bibinfo {author} {\bibfnamefont {J.~I.}\ \bibnamefont {V\"ayrynen}},\
  }\bibfield  {title} {\bibinfo {title} {Topological symplectic {K}ondo
  effect},\ }\href {https://doi.org/10.1103/PhysRevB.107.L201401} {\bibfield
  {journal} {\bibinfo  {journal} {Phys. Rev. B}\ }\textbf {\bibinfo {volume}
  {107}},\ \bibinfo {pages} {L201401} (\bibinfo {year}
  {2023}{\natexlab{b}})}\BibitemShut {NoStop}%
\bibitem [{\citenamefont {K{\"o}nig}\ and\ \citenamefont
  {Tsvelik}(2023)}]{KoenigTsvelik2023}%
  \BibitemOpen
  \bibfield  {author} {\bibinfo {author} {\bibfnamefont {E.~J.}\ \bibnamefont
  {K{\"o}nig}}\ and\ \bibinfo {author} {\bibfnamefont {A.~M.}\ \bibnamefont
  {Tsvelik}},\ }\bibfield  {title} {\bibinfo {title} {Exact solution of the
  topological symplectic {K}ondo problem},\ }\href
  {https://www.sciencedirect.com/science/article/pii/S0003491623000167}
  {\bibfield  {journal} {\bibinfo  {journal} {Annals of Physics}\ ,\ \bibinfo
  {pages} {169231}} (\bibinfo {year} {2023})}\BibitemShut {NoStop}%
\bibitem [{\citenamefont {Ren}\ \emph {et~al.}(2023)\citenamefont {Ren},
  \citenamefont {K{\"o}nig},\ and\ \citenamefont {Tsvelik}}]{RenTsvelik2023}%
  \BibitemOpen
  \bibfield  {author} {\bibinfo {author} {\bibfnamefont {T.}~\bibnamefont
  {Ren}}, \bibinfo {author} {\bibfnamefont {E.~J.}\ \bibnamefont {K{\"o}nig}},\
  and\ \bibinfo {author} {\bibfnamefont {A.~M.}\ \bibnamefont {Tsvelik}},\
  }\bibfield  {title} {\bibinfo {title} {Topological quantum computation on a
  chiral {K}ondo chain},\ }\href {https://arxiv.org/abs/2309.03010} {\bibfield
  {journal} {\bibinfo  {journal} {arXiv preprint arXiv:2309.03010}\ } (\bibinfo
  {year} {2023})}\BibitemShut {NoStop}%
\bibitem [{\citenamefont {Kimura}(2021)}]{Kimura2021}%
  \BibitemOpen
  \bibfield  {author} {\bibinfo {author} {\bibfnamefont {T.}~\bibnamefont
  {Kimura}},\ }\bibfield  {title} {\bibinfo {title} {Abcd of {K}ondo effect},\
  }\href {https://journals.jps.jp/doi/full/10.7566/JPSJ.90.024708?mobileUi=0}
  {\bibfield  {journal} {\bibinfo  {journal} {Journal of the Physical Society
  of Japan}\ }\textbf {\bibinfo {volume} {90}},\ \bibinfo {pages} {024708}
  (\bibinfo {year} {2021})}\BibitemShut {NoStop}%
\bibitem [{\citenamefont {Haim}\ and\ \citenamefont
  {Oreg}(2019)}]{HaimOreg2019}%
  \BibitemOpen
  \bibfield  {author} {\bibinfo {author} {\bibfnamefont {A.}~\bibnamefont
  {Haim}}\ and\ \bibinfo {author} {\bibfnamefont {Y.}~\bibnamefont {Oreg}},\
  }\bibfield  {title} {\bibinfo {title} {Time-reversal-invariant topological
  superconductivity in one and two dimensions},\ }\href
  {https://www.sciencedirect.com/science/article/pii/S0370157319302613#sec2}
  {\bibfield  {journal} {\bibinfo  {journal} {Physics Reports}\ }\textbf
  {\bibinfo {volume} {825}},\ \bibinfo {pages} {1} (\bibinfo {year}
  {2019})}\BibitemShut {NoStop}%
\bibitem [{\citenamefont {Wong}\ and\ \citenamefont {Law}(2012)}]{WangLaw2012}%
  \BibitemOpen
  \bibfield  {author} {\bibinfo {author} {\bibfnamefont {C.~L.~M.}\
  \bibnamefont {Wong}}\ and\ \bibinfo {author} {\bibfnamefont {K.~T.}\
  \bibnamefont {Law}},\ }\bibfield  {title} {\bibinfo {title} {{M}ajorana
  kramers doublets in ${d}_{{x}^{2}\ensuremath{-}{y}^{2}}$-wave superconductors
  with rashba spin-orbit coupling},\ }\href
  {https://doi.org/10.1103/PhysRevB.86.184516} {\bibfield  {journal} {\bibinfo
  {journal} {Phys. Rev. B}\ }\textbf {\bibinfo {volume} {86}},\ \bibinfo
  {pages} {184516} (\bibinfo {year} {2012})}\BibitemShut {NoStop}%
\bibitem [{\citenamefont {Nakosai}\ \emph {et~al.}(2013)\citenamefont
  {Nakosai}, \citenamefont {Budich}, \citenamefont {Tanaka}, \citenamefont
  {Trauzettel},\ and\ \citenamefont {Nagaosa}}]{NakoseiNagaosa2013}%
  \BibitemOpen
  \bibfield  {author} {\bibinfo {author} {\bibfnamefont {S.}~\bibnamefont
  {Nakosai}}, \bibinfo {author} {\bibfnamefont {J.~C.}\ \bibnamefont {Budich}},
  \bibinfo {author} {\bibfnamefont {Y.}~\bibnamefont {Tanaka}}, \bibinfo
  {author} {\bibfnamefont {B.}~\bibnamefont {Trauzettel}},\ and\ \bibinfo
  {author} {\bibfnamefont {N.}~\bibnamefont {Nagaosa}},\ }\bibfield  {title}
  {\bibinfo {title} {{M}ajorana bound states and nonlocal spin correlations in
  a quantum wire on an unconventional superconductor},\ }\href
  {https://doi.org/10.1103/PhysRevLett.110.117002} {\bibfield  {journal}
  {\bibinfo  {journal} {Phys. Rev. Lett.}\ }\textbf {\bibinfo {volume} {110}},\
  \bibinfo {pages} {117002} (\bibinfo {year} {2013})}\BibitemShut {NoStop}%
\bibitem [{\citenamefont {Keselman}\ \emph {et~al.}(2013)\citenamefont
  {Keselman}, \citenamefont {Fu}, \citenamefont {Stern},\ and\ \citenamefont
  {Berg}}]{KeselmanBerg2013}%
  \BibitemOpen
  \bibfield  {author} {\bibinfo {author} {\bibfnamefont {A.}~\bibnamefont
  {Keselman}}, \bibinfo {author} {\bibfnamefont {L.}~\bibnamefont {Fu}},
  \bibinfo {author} {\bibfnamefont {A.}~\bibnamefont {Stern}},\ and\ \bibinfo
  {author} {\bibfnamefont {E.}~\bibnamefont {Berg}},\ }\bibfield  {title}
  {\bibinfo {title} {Inducing time-reversal-invariant topological
  superconductivity and {F}ermion parity pumping in quantum wires},\ }\href
  {https://doi.org/10.1103/PhysRevLett.111.116402} {\bibfield  {journal}
  {\bibinfo  {journal} {Phys. Rev. Lett.}\ }\textbf {\bibinfo {volume} {111}},\
  \bibinfo {pages} {116402} (\bibinfo {year} {2013})}\BibitemShut {NoStop}%
\bibitem [{\citenamefont {Zhang}\ \emph {et~al.}(2013)\citenamefont {Zhang},
  \citenamefont {Kane},\ and\ \citenamefont {Mele}}]{ZhangMele2013}%
  \BibitemOpen
  \bibfield  {author} {\bibinfo {author} {\bibfnamefont {F.}~\bibnamefont
  {Zhang}}, \bibinfo {author} {\bibfnamefont {C.~L.}\ \bibnamefont {Kane}},\
  and\ \bibinfo {author} {\bibfnamefont {E.~J.}\ \bibnamefont {Mele}},\
  }\bibfield  {title} {\bibinfo {title} {Time-reversal-invariant topological
  superconductivity and {M}ajorana kramers pairs},\ }\href
  {https://doi.org/10.1103/PhysRevLett.111.056402} {\bibfield  {journal}
  {\bibinfo  {journal} {Phys. Rev. Lett.}\ }\textbf {\bibinfo {volume} {111}},\
  \bibinfo {pages} {056402} (\bibinfo {year} {2013})}\BibitemShut {NoStop}%
\bibitem [{\citenamefont {Schrade}\ \emph {et~al.}(2015)\citenamefont
  {Schrade}, \citenamefont {Zyuzin}, \citenamefont {Klinovaja},\ and\
  \citenamefont {Loss}}]{SchradeLoss2015}%
  \BibitemOpen
  \bibfield  {author} {\bibinfo {author} {\bibfnamefont {C.}~\bibnamefont
  {Schrade}}, \bibinfo {author} {\bibfnamefont {A.~A.}\ \bibnamefont {Zyuzin}},
  \bibinfo {author} {\bibfnamefont {J.}~\bibnamefont {Klinovaja}},\ and\
  \bibinfo {author} {\bibfnamefont {D.}~\bibnamefont {Loss}},\ }\bibfield
  {title} {\bibinfo {title} {Proximity-induced $\ensuremath{\pi}$ josephson
  junctions in topological insulators and kramers pairs of {M}ajorana
  {F}ermions},\ }\href {https://doi.org/10.1103/PhysRevLett.115.237001}
  {\bibfield  {journal} {\bibinfo  {journal} {Phys. Rev. Lett.}\ }\textbf
  {\bibinfo {volume} {115}},\ \bibinfo {pages} {237001} (\bibinfo {year}
  {2015})}\BibitemShut {NoStop}%
\bibitem [{\citenamefont {Kim}\ \emph {et~al.}(2016)\citenamefont {Kim},
  \citenamefont {Liu}, \citenamefont {Gaidamauskas}, \citenamefont {Paaske},
  \citenamefont {Flensberg},\ and\ \citenamefont {Lutchyn}}]{KimLutchyn2016}%
  \BibitemOpen
  \bibfield  {author} {\bibinfo {author} {\bibfnamefont {Y.}~\bibnamefont
  {Kim}}, \bibinfo {author} {\bibfnamefont {D.~E.}\ \bibnamefont {Liu}},
  \bibinfo {author} {\bibfnamefont {E.}~\bibnamefont {Gaidamauskas}}, \bibinfo
  {author} {\bibfnamefont {J.}~\bibnamefont {Paaske}}, \bibinfo {author}
  {\bibfnamefont {K.}~\bibnamefont {Flensberg}},\ and\ \bibinfo {author}
  {\bibfnamefont {R.~M.}\ \bibnamefont {Lutchyn}},\ }\bibfield  {title}
  {\bibinfo {title} {Signatures of {M}ajorana kramers pairs in
  superconductor-luttinger liquid and superconductor-quantum dot-normal lead
  junctions},\ }\href {https://doi.org/10.1103/PhysRevB.94.075439} {\bibfield
  {journal} {\bibinfo  {journal} {Phys. Rev. B}\ }\textbf {\bibinfo {volume}
  {94}},\ \bibinfo {pages} {075439} (\bibinfo {year} {2016})}\BibitemShut
  {NoStop}%
\bibitem [{\citenamefont {Camjayi}\ \emph {et~al.}(2017)\citenamefont
  {Camjayi}, \citenamefont {Arrachea}, \citenamefont {Aligia},\ and\
  \citenamefont {von Oppen}}]{CamjayivonOppen2017}%
  \BibitemOpen
  \bibfield  {author} {\bibinfo {author} {\bibfnamefont {A.}~\bibnamefont
  {Camjayi}}, \bibinfo {author} {\bibfnamefont {L.}~\bibnamefont {Arrachea}},
  \bibinfo {author} {\bibfnamefont {A.}~\bibnamefont {Aligia}},\ and\ \bibinfo
  {author} {\bibfnamefont {F.}~\bibnamefont {von Oppen}},\ }\bibfield  {title}
  {\bibinfo {title} {Fractional spin and josephson effect in
  time-reversal-invariant topological superconductors},\ }\href
  {https://doi.org/10.1103/PhysRevLett.119.046801} {\bibfield  {journal}
  {\bibinfo  {journal} {Phys. Rev. Lett.}\ }\textbf {\bibinfo {volume} {119}},\
  \bibinfo {pages} {046801} (\bibinfo {year} {2017})}\BibitemShut {NoStop}%
\bibitem [{\citenamefont {Schrade}\ and\ \citenamefont
  {Fu}(2018)}]{SchradeFu2018}%
  \BibitemOpen
  \bibfield  {author} {\bibinfo {author} {\bibfnamefont {C.}~\bibnamefont
  {Schrade}}\ and\ \bibinfo {author} {\bibfnamefont {L.}~\bibnamefont {Fu}},\
  }\bibfield  {title} {\bibinfo {title} {Parity-controlled $2\ensuremath{\pi}$
  josephson effect mediated by {M}ajorana kramers pairs},\ }\href
  {https://doi.org/10.1103/PhysRevLett.120.267002} {\bibfield  {journal}
  {\bibinfo  {journal} {Phys. Rev. Lett.}\ }\textbf {\bibinfo {volume} {120}},\
  \bibinfo {pages} {267002} (\bibinfo {year} {2018})}\BibitemShut {NoStop}%
\bibitem [{\citenamefont {Aligia}\ and\ \citenamefont
  {Arrachea}(2018)}]{AligiaArrachea2018}%
  \BibitemOpen
  \bibfield  {author} {\bibinfo {author} {\bibfnamefont {A.~A.}\ \bibnamefont
  {Aligia}}\ and\ \bibinfo {author} {\bibfnamefont {L.}~\bibnamefont
  {Arrachea}},\ }\bibfield  {title} {\bibinfo {title} {Entangled end states
  with fractionalized spin projection in a time-reversal-invariant topological
  superconducting wire},\ }\href {https://doi.org/10.1103/PhysRevB.98.174507}
  {\bibfield  {journal} {\bibinfo  {journal} {Phys. Rev. B}\ }\textbf {\bibinfo
  {volume} {98}},\ \bibinfo {pages} {174507} (\bibinfo {year}
  {2018})}\BibitemShut {NoStop}%
\bibitem [{\citenamefont {Arrachea}\ \emph {et~al.}(2019)\citenamefont
  {Arrachea}, \citenamefont {Camjayi}, \citenamefont {Aligia},\ and\
  \citenamefont {Gru\~neiro}}]{ArracheaGruneiro2019}%
  \BibitemOpen
  \bibfield  {author} {\bibinfo {author} {\bibfnamefont {L.}~\bibnamefont
  {Arrachea}}, \bibinfo {author} {\bibfnamefont {A.}~\bibnamefont {Camjayi}},
  \bibinfo {author} {\bibfnamefont {A.~A.}\ \bibnamefont {Aligia}},\ and\
  \bibinfo {author} {\bibfnamefont {L.}~\bibnamefont {Gru\~neiro}},\ }\bibfield
   {title} {\bibinfo {title} {Catalog of andreev spectra and josephson effects
  in structures with time-reversal-invariant topological superconductor
  wires},\ }\href {https://doi.org/10.1103/PhysRevB.99.085431} {\bibfield
  {journal} {\bibinfo  {journal} {Phys. Rev. B}\ }\textbf {\bibinfo {volume}
  {99}},\ \bibinfo {pages} {085431} (\bibinfo {year} {2019})}\BibitemShut
  {NoStop}%
\bibitem [{\citenamefont {Knapp}\ \emph {et~al.}(2020)\citenamefont {Knapp},
  \citenamefont {Chew},\ and\ \citenamefont {Alicea}}]{KnappAlicea2020}%
  \BibitemOpen
  \bibfield  {author} {\bibinfo {author} {\bibfnamefont {C.}~\bibnamefont
  {Knapp}}, \bibinfo {author} {\bibfnamefont {A.}~\bibnamefont {Chew}},\ and\
  \bibinfo {author} {\bibfnamefont {J.}~\bibnamefont {Alicea}},\ }\bibfield
  {title} {\bibinfo {title} {Fragility of the fractional josephson effect in
  time-reversal-invariant topological superconductors},\ }\href
  {https://doi.org/10.1103/PhysRevLett.125.207002} {\bibfield  {journal}
  {\bibinfo  {journal} {Phys. Rev. Lett.}\ }\textbf {\bibinfo {volume} {125}},\
  \bibinfo {pages} {207002} (\bibinfo {year} {2020})}\BibitemShut {NoStop}%
\bibitem [{\citenamefont {Schrade}\ and\ \citenamefont
  {Fu}(2022)}]{SchradeFu2022}%
  \BibitemOpen
  \bibfield  {author} {\bibinfo {author} {\bibfnamefont {C.}~\bibnamefont
  {Schrade}}\ and\ \bibinfo {author} {\bibfnamefont {L.}~\bibnamefont {Fu}},\
  }\bibfield  {title} {\bibinfo {title} {Quantum computing with {M}ajorana
  kramers pairs},\ }\href {https://doi.org/10.1103/PhysRevLett.129.227002}
  {\bibfield  {journal} {\bibinfo  {journal} {Phys. Rev. Lett.}\ }\textbf
  {\bibinfo {volume} {129}},\ \bibinfo {pages} {227002} (\bibinfo {year}
  {2022})}\BibitemShut {NoStop}%
\bibitem [{\citenamefont {Mohammadi}\ and\ \citenamefont
  {Kargarian}(2022)}]{FatemehKargarian2022}%
  \BibitemOpen
  \bibfield  {author} {\bibinfo {author} {\bibfnamefont {F.}~\bibnamefont
  {Mohammadi}}\ and\ \bibinfo {author} {\bibfnamefont {M.}~\bibnamefont
  {Kargarian}},\ }\bibfield  {title} {\bibinfo {title} {Designing
  ${\mathbb{z}}_{2}$ and
  ${\mathbb{z}}_{2}\ifmmode\times\else\texttimes\fi{}{\mathbb{z}}_{2}$
  topological orders in networks of {M}ajorana bound states},\ }\href
  {https://doi.org/10.1103/PhysRevB.105.165107} {\bibfield  {journal} {\bibinfo
   {journal} {Phys. Rev. B}\ }\textbf {\bibinfo {volume} {105}},\ \bibinfo
  {pages} {165107} (\bibinfo {year} {2022})}\BibitemShut {NoStop}%
\bibitem [{\citenamefont {Bao}\ and\ \citenamefont
  {Zhang}(2017)}]{BaoZhang2017}%
  \BibitemOpen
  \bibfield  {author} {\bibinfo {author} {\bibfnamefont {Z.-q.}\ \bibnamefont
  {Bao}}\ and\ \bibinfo {author} {\bibfnamefont {F.}~\bibnamefont {Zhang}},\
  }\bibfield  {title} {\bibinfo {title} {Topological {M}ajorana two-channel
  {K}ondo effect},\ }\href {https://doi.org/10.1103/PhysRevLett.119.187701}
  {\bibfield  {journal} {\bibinfo  {journal} {Phys. Rev. Lett.}\ }\textbf
  {\bibinfo {volume} {119}},\ \bibinfo {pages} {187701} (\bibinfo {year}
  {2017})}\BibitemShut {NoStop}%
\bibitem [{\citenamefont {Rampp}\ \emph {et~al.}(2022)\citenamefont {Rampp},
  \citenamefont {K\"onig},\ and\ \citenamefont
  {Schmalian}}]{RamppSchmalian2022}%
  \BibitemOpen
  \bibfield  {author} {\bibinfo {author} {\bibfnamefont {M.~A.}\ \bibnamefont
  {Rampp}}, \bibinfo {author} {\bibfnamefont {E.~J.}\ \bibnamefont {K\"onig}},\
  and\ \bibinfo {author} {\bibfnamefont {J.}~\bibnamefont {Schmalian}},\
  }\bibfield  {title} {\bibinfo {title} {Topologically enabled
  superconductivity},\ }\href {https://doi.org/10.1103/PhysRevLett.129.077001}
  {\bibfield  {journal} {\bibinfo  {journal} {Phys. Rev. Lett.}\ }\textbf
  {\bibinfo {volume} {129}},\ \bibinfo {pages} {077001} (\bibinfo {year}
  {2022})}\BibitemShut {NoStop}%
\bibitem [{\citenamefont {Else}(2021)}]{Else2021}%
  \BibitemOpen
  \bibfield  {author} {\bibinfo {author} {\bibfnamefont {D.~V.}\ \bibnamefont
  {Else}},\ }\bibfield  {title} {\bibinfo {title} {Topological goldstone phases
  of matter},\ }\href {https://doi.org/10.1103/PhysRevB.104.115129} {\bibfield
  {journal} {\bibinfo  {journal} {Phys. Rev. B}\ }\textbf {\bibinfo {volume}
  {104}},\ \bibinfo {pages} {115129} (\bibinfo {year} {2021})}\BibitemShut
  {NoStop}%
\bibitem [{\citenamefont {Kane}\ \emph {et~al.}(2017)\citenamefont {Kane},
  \citenamefont {Stern},\ and\ \citenamefont {Halperin}}]{KaneHalperin2017}%
  \BibitemOpen
  \bibfield  {author} {\bibinfo {author} {\bibfnamefont {C.~L.}\ \bibnamefont
  {Kane}}, \bibinfo {author} {\bibfnamefont {A.}~\bibnamefont {Stern}},\ and\
  \bibinfo {author} {\bibfnamefont {B.~I.}\ \bibnamefont {Halperin}},\
  }\bibfield  {title} {\bibinfo {title} {Pairing in luttinger liquids and
  quantum hall states},\ }\href {https://doi.org/10.1103/PhysRevX.7.031009}
  {\bibfield  {journal} {\bibinfo  {journal} {Phys. Rev. X}\ }\textbf {\bibinfo
  {volume} {7}},\ \bibinfo {pages} {031009} (\bibinfo {year}
  {2017})}\BibitemShut {NoStop}%
\bibitem [{\citenamefont {Lapa}(2020)}]{Lapa2020}%
  \BibitemOpen
  \bibfield  {author} {\bibinfo {author} {\bibfnamefont {M.~F.}\ \bibnamefont
  {Lapa}},\ }\bibfield  {title} {\bibinfo {title} {Topology of superconductors
  beyond mean-field theory},\ }\href
  {https://doi.org/10.1103/PhysRevResearch.2.033309} {\bibfield  {journal}
  {\bibinfo  {journal} {Phys. Rev. Res.}\ }\textbf {\bibinfo {volume} {2}},\
  \bibinfo {pages} {033309} (\bibinfo {year} {2020})}\BibitemShut {NoStop}%
\bibitem [{\citenamefont {Lapa}\ and\ \citenamefont
  {Levin}(2020)}]{LapaLevin2020}%
  \BibitemOpen
  \bibfield  {author} {\bibinfo {author} {\bibfnamefont {M.~F.}\ \bibnamefont
  {Lapa}}\ and\ \bibinfo {author} {\bibfnamefont {M.}~\bibnamefont {Levin}},\
  }\bibfield  {title} {\bibinfo {title} {Rigorous results on topological
  superconductivity with particle number conservation},\ }\href
  {https://doi.org/10.1103/PhysRevLett.124.257002} {\bibfield  {journal}
  {\bibinfo  {journal} {Phys. Rev. Lett.}\ }\textbf {\bibinfo {volume} {124}},\
  \bibinfo {pages} {257002} (\bibinfo {year} {2020})}\BibitemShut {NoStop}%
\bibitem [{\citenamefont {Schrieffer}\ and\ \citenamefont
  {Wolff}(1966)}]{SchriefferWolff1966}%
  \BibitemOpen
  \bibfield  {author} {\bibinfo {author} {\bibfnamefont {J.~R.}\ \bibnamefont
  {Schrieffer}}\ and\ \bibinfo {author} {\bibfnamefont {P.~A.}\ \bibnamefont
  {Wolff}},\ }\bibfield  {title} {\bibinfo {title} {Relation between the
  anderson and {K}ondo hamiltonians},\ }\href
  {https://doi.org/10.1103/PhysRev.149.491} {\bibfield  {journal} {\bibinfo
  {journal} {Phys. Rev.}\ }\textbf {\bibinfo {volume} {149}},\ \bibinfo {pages}
  {491} (\bibinfo {year} {1966})}\BibitemShut {NoStop}%
\bibitem [{\citenamefont {Emery}\ and\ \citenamefont
  {Kivelson}(1992)}]{EmeryKivelson1992}%
  \BibitemOpen
  \bibfield  {author} {\bibinfo {author} {\bibfnamefont {V.~J.}\ \bibnamefont
  {Emery}}\ and\ \bibinfo {author} {\bibfnamefont {S.}~\bibnamefont
  {Kivelson}},\ }\bibfield  {title} {\bibinfo {title} {Mapping of the
  two-channel {K}ondo problem to a resonant-level model},\ }\href
  {https://journals.aps.org/prb/abstract/10.1103/PhysRevB.46.10812} {\bibfield
  {journal} {\bibinfo  {journal} {Phys. Rev. B}\ }\textbf {\bibinfo {volume}
  {46}},\ \bibinfo {pages} {10812} (\bibinfo {year} {1992})}\BibitemShut
  {NoStop}%
\bibitem [{\citenamefont {Anderson}(1970)}]{Anderson1970}%
  \BibitemOpen
  \bibfield  {author} {\bibinfo {author} {\bibfnamefont {P.~W.}\ \bibnamefont
  {Anderson}},\ }\bibfield  {title} {\bibinfo {title} {A poor
  man{\textquotesingle}s derivation of scaling laws for the {K}ondo problem},\
  }\href {https://doi.org/10.1088/0022-3719/3/12/008} {\bibfield  {journal}
  {\bibinfo  {journal} {Journal of Physics C: Solid State Physics}\ }\textbf
  {\bibinfo {volume} {3}},\ \bibinfo {pages} {2436} (\bibinfo {year}
  {1970})}\BibitemShut {NoStop}%
\bibitem [{\citenamefont {Gogolin}\ \emph {et~al.}(2004)\citenamefont
  {Gogolin}, \citenamefont {Nersesyan},\ and\ \citenamefont
  {Tsvelik}}]{Gogolin2004}%
  \BibitemOpen
  \bibfield  {author} {\bibinfo {author} {\bibfnamefont {A.~O.}\ \bibnamefont
  {Gogolin}}, \bibinfo {author} {\bibfnamefont {A.~A.}\ \bibnamefont
  {Nersesyan}},\ and\ \bibinfo {author} {\bibfnamefont {A.~M.}\ \bibnamefont
  {Tsvelik}},\ }\href@noop {} {\emph {\bibinfo {title} {Bosonization and
  strongly correlated systems}}}\ (\bibinfo  {publisher} {Cambridge University
  Press},\ \bibinfo {address} {Cambridge, England},\ \bibinfo {year}
  {2004})\BibitemShut {NoStop}%
\bibitem [{\citenamefont {Zar{\'{a}}nd}\ and\ \citenamefont {von
  Delft}(2000)}]{Zarnd2000}%
  \BibitemOpen
  \bibfield  {author} {\bibinfo {author} {\bibfnamefont {G.}~\bibnamefont
  {Zar{\'{a}}nd}}\ and\ \bibinfo {author} {\bibfnamefont {J.}~\bibnamefont {von
  Delft}},\ }\bibfield  {title} {\bibinfo {title} {Analytical calculation of
  the finite-size crossover spectrum of the anisotropic two-channel {K}ondo
  model},\ }\href {https://doi.org/10.1103/physrevb.61.6918} {\bibfield
  {journal} {\bibinfo  {journal} {Physical Review B}\ }\textbf {\bibinfo
  {volume} {61}},\ \bibinfo {pages} {6918} (\bibinfo {year}
  {2000})}\BibitemShut {NoStop}%
\bibitem [{\citenamefont {von Delft}\ \emph {et~al.}(1998)\citenamefont {von
  Delft}, \citenamefont {Zar\'and},\ and\ \citenamefont
  {Fabrizio}}]{vonDelft1998}%
  \BibitemOpen
  \bibfield  {author} {\bibinfo {author} {\bibfnamefont {J.}~\bibnamefont {von
  Delft}}, \bibinfo {author} {\bibfnamefont {G.}~\bibnamefont {Zar\'and}},\
  and\ \bibinfo {author} {\bibfnamefont {M.}~\bibnamefont {Fabrizio}},\
  }\bibfield  {title} {\bibinfo {title} {Finite-size bosonization of 2-channel
  {K}ondo model: A bridge between numerical renormalization group and conformal
  field theory},\ }\href {https://doi.org/10.1103/PhysRevLett.81.196}
  {\bibfield  {journal} {\bibinfo  {journal} {Phys. Rev. Lett.}\ }\textbf
  {\bibinfo {volume} {81}},\ \bibinfo {pages} {196} (\bibinfo {year}
  {1998})}\BibitemShut {NoStop}%
\bibitem [{\citenamefont {Toulouse}\ and\ \citenamefont
  {Seances}(1969)}]{ToulouseSeances1969}%
  \BibitemOpen
  \bibfield  {author} {\bibinfo {author} {\bibfnamefont {G.}~\bibnamefont
  {Toulouse}}\ and\ \bibinfo {author} {\bibfnamefont {C.~R.}\ \bibnamefont
  {Seances}},\ }\href@noop {} {\bibfield  {journal} {\bibinfo  {journal} {Acad.
  Sci., Ser. B}\ }\textbf {\bibinfo {volume} {268}},\ \bibinfo {pages} {1200}
  (\bibinfo {year} {1969})}\BibitemShut {NoStop}%
\bibitem [{\citenamefont {Friedan}\ and\ \citenamefont
  {Konechny}(2004)}]{FriedanKonechy2004}%
  \BibitemOpen
  \bibfield  {author} {\bibinfo {author} {\bibfnamefont {D.}~\bibnamefont
  {Friedan}}\ and\ \bibinfo {author} {\bibfnamefont {A.}~\bibnamefont
  {Konechny}},\ }\bibfield  {title} {\bibinfo {title} {Boundary entropy of
  one-dimensional quantum systems at low temperature},\ }\href
  {https://doi.org/10.1103/PhysRevLett.93.030402} {\bibfield  {journal}
  {\bibinfo  {journal} {Phys. Rev. Lett.}\ }\textbf {\bibinfo {volume} {93}},\
  \bibinfo {pages} {030402} (\bibinfo {year} {2004})}\BibitemShut {NoStop}%
\bibitem [{\citenamefont {Shnir}(2006)}]{Shnir2006}%
  \BibitemOpen
  \bibfield  {author} {\bibinfo {author} {\bibfnamefont {Y.}~\bibnamefont
  {Shnir}},\ }\href {https://books.google.de/books?id=G5huCG0k2dAC} {\emph
  {\bibinfo {title} {Magnetic Monopoles}}},\ Theoretical and Mathematical
  Physics\ (\bibinfo  {publisher} {Springer Berlin Heidelberg},\ \bibinfo
  {year} {2006})\BibitemShut {NoStop}%
\end{thebibliography}%

\end{document}